\documentclass{aa}
\usepackage{graphics}
\usepackage{rotating}
\usepackage{natbib}
\bibpunct{(}{)}{;}{a}{}{,}

\DeclareRobustCommand{\ion}[2]{%
\relax\ifmmode
\ifx\testbx\f@series
{\mathbf{#1\,\mathsc{#2}}}\else
{\mathrm{#1\,\mathsc{#2}}}\fi
\else\textup{#1\,{\mdseries\textsc{#2}}}%
\fi}

\title{Relating dust, gas and the rate of star formation in M~31 }

\author{ F. S. Tabatabaei and E.\,M. Berkhuijsen }
\institute{Max-Planck Institut f\"ur Radioastronomie, Auf dem H\"ugel 69, 53121 Bonn, Germany   }
\offprints{F. Tabatabaei\\ tabataba@mpifr-bonn.mpg.de}

\begin{document}

\titlerunning{Dust, gas and SFR in  M~31}
\authorrunning{Tabatabaei et al.}
\abstract
{}
{Investigation of relationships between dust and gas, and study of the star formation law in M~31. }
{We derive distributions of dust temperature and dust 
opacity across M\,31 at 45$\arcsec$ resolution using the Spitzer data. 
With the opacity map and a standard dust model we de-redden the 
H$\alpha$ emission yielding the first de-reddened H$\alpha$ map of M~31. We compare the emissions from dust, H$\alpha$, HI 
and H$_2$ by means of radial distributions, pixel-to-pixel correlations 
and wavelet cross-correlations. We calculate the star formation rate 
and star formation efficiency from the de-reddened H$\alpha$ emission.
}
{The dust temperature steeply decreases from 30\,K near the center
to 15\,K at large radii. The mean dust optical depth at the H$\alpha$
wavelength along the line of sight is about 0.7.  The 
radial decrease of the dust-to-gas ratio is similar to that of 
the oxygen abundance. Extinction
is about linearly correlated with the total gas surface density
within limited radial intervals. On scales $<$\,2\,kpc, cold dust
emission is best correlated with that of neutral gas and
warm dust emission  with that of ionized gas.
H$\alpha$ emission is slightly better correlated with  emission
at 70\,$\mu$m than at 24\,$\mu$m. The star formation 
rate in M\,31 is low. In the area 6\,kpc\,$<$\,$R$\,$<$\,17\,kpc, the total SFR is $\simeq$\,0.3\,${\rm M}_{\odot} {\rm yr}^{-1}$. A linear 
relationship exists between surface densities of SFR and H$_2$.
The Kennicutt-Schmidt law between SFR and total gas has a power-law index of 1.30$\pm$0.05  in the radial range of $R$\,=\,7-11\,kpc increasing by about 0.3 for $R$\,=\,11-13\,kpc. }
{
The better 70\,$\mu$m--H$\alpha$ than 24\,$\mu$m--H$\alpha$ correlation plus an excess in the 24\,$\mu$m/70\,$\mu$m intensity ratio indicates that other sources than dust grains, e.g. of stellar origin, contribute to the 24\,$\mu$m emission.  The lack of H$_2$ in the central region
could be related to the lack of HI and the low opacity/high
temperature of the dust.      
Since neither SFR nor SFE is well correlated
with the surface density of H$_2$ or total gas, other factors than
gas density must play an important role in  the formation of massive stars in M~31.  The molecular depletion 
time scale of 1.1\,Gyr indicates that M~31 is about three times less
efficient in forming young massive stars than M~33.
}
\maketitle

%________________________________________________________________

\section{Introduction}
Dust, neutral gas and ionized gas are the major components
of the interstellar medium (ISM) in galaxies. Observations of 
their properties and inter-relationships can give important
clues to the physics governing star formation.
Relationships between components in the ISM are to be expected. Observations have shown that in the Galaxy dust and neutral gas are well mixed. In dense clouds of molecular gas mixed with cold dust most of the stars are formed. They subsequently heat the dust and gas in their surroundings and ionize the atomic gas. As the major coolants of the ISM are continuum emission and line emission at various frequencies, a close comparison of these emissions could shed light on spatial and physical connections between the emitting components. Present-day IR and radio telescopes have produced sensitive high resolution maps of several nearby galaxies, which are ideal laboratories to study the interplay between the ISM and star formation  \citep[e.g. ][]{Kennicutt_07,Bigiel_08,Verley_09}. 

The spiral galaxy nearest to us, the Andromeda Nebula 
(NGC224), is a highly inclined Sb galaxy of low surface brightness.
Table 1 lists the positional data on M~31. Its proximity
and large extent on the sky ($> 5^{\circ} \times 1^{\circ}$) enable detailed studies
of the ISM over a large radial range.

Surveys of M~31 at high angular resolution ($< 1\arcmin$) are available at
many wavelengths. In the HI line the galaxy was mapped by \cite{Brinks} 
at $24\arcsec \times 36\arcsec$ resolution, the northeastern half by
\cite{Braun_90} at 10$\arcsec$ resolution and, most recently,
the entire galaxy with high sensitivity by \cite{Braun_09} at 
a resolution of 15$\arcsec$. \cite{Nieten} made a survey in the 
$^{12}$CO(1-0) line at a resolution of 23$\arcsec$. \cite{Devereux_etal_94b} observed M~31 in the H$\alpha$ line to obtain the distribution of the ionized gas. The dust emission from M~31 was recently observed by the multiband imaging photometer Spitzer \citep[MIPS, ][]{Rieke} with high sensitivity at 24\,$\mu$m, 70\,$\mu$m, and 160\,$\mu$m at resolutions~$\leq 40\arcsec$. 
 
The relationships between gas and dust as well as between gas and 
star formation in M~31 have been studied in the past at resolutions 
of several arcminutes. \cite{Walterbos_87} derived a nearly
constant dust temperature across M~31 using the IRAS 100\,$\mu$m and
60\,$\mu$m maps. They also found a strong increase in the atomic gas-
to-dust surface density ratio with increasing radius. This increase 
was confirmed by \cite{Walterbos_88} who used optical extinction as dust
tracer, and by \cite{Nieten} using the ISO map at 175\,$\mu$m 
\citep{Haas}. Interestingly, the latter authors did not find 
a radial increase in the molecular gas-to-dust ratio. 

The dependence of star formation on HI surface density in M~31 has been 
studied by a number of authors \citep{Emerson_74,Berkhuijsen_77,Tenjes,
Unwin,Nakai_82,Nakai_84} using the number density of HII regions or of OB 
stars as star formation tracers. They obtained power-law indices between 
0.5 and 2, possibly depending on the region in M~31, the star formation 
tracer and the angular resolution. \cite{Braun_09} plotted the star formation 
density derived from the brightnesses at 8$\mu$m, 24$\mu$m and UV against 
the surface densities of molecular gas, HI and total gas, but did not fit 
power laws to their data. 
 
The high-resolution data available for M~31 show the morphologies of the emission from dust and gas components in detail. We apply a 2-D wavelet analysis technique \citep{Frick_etal_01} to the MIPS IR data \citep{Gordon_06} and the gas (HI, H$_2$, and H$\alpha$) maps to study the scale distribution of emission power and to separate the diffuse emission components from compact sources. We then compare the wavelet-decomposed maps at various spatial scales. We also use pixel-to-pixel (Pearson) correlations to derive quantitative relations not only between different ISM components but also between them and the present-day star formation rate.

Following \cite{Walterbos_87} and \cite{Haas}, we derive the dust temperature 
assuming a $\lambda^{-2}$ emissivity law for the MIPS bands at which the emission 
from the big grains and hence the LTE condition is relevant, and present a map of 
the dust color temperature. We also obtain  the distribution of the optical 
depth and analyze the gas-to-dust surface-density ratio at a resolution
of 45$\arcsec$ (170~pc\,$\times$\,660~pc  along the major and minor axis,
respectively, in the galaxy plane), 9 times higher than 
before \citep{Walterbos_87}. We use the optical depth map to 
de-redden the H$\alpha$  emission observed by \cite{Devereux_etal_94b}  
yielding the distribution of the absorption-free emission from the ionized gas, 
and use this as an indicator of massive star formation. We compare it with the 
distributions of neutral gas to obtain the dependence of the star 
formation rate on gas surface density. 

The paper is organized as follows: The relevant data sets are described
in Sect. 2. In Sect. 3 we derive maps of the dust color temperature and 
optical depth, and correct the H$\alpha$ emission for absorption 
by dust. Radial profiles of the dust and gas emission and of the various 
gas-to-dust ratios are obtained in Sect. 4. Sect. 5 is devoted to
wavelet decompositions and wavelet spectra of the dust and gas distributions, 
and their cross correlations. Complementary, we discuss in 
Sect. 6 classical correlations between gas and dust. In Sect. 7 
the dependence of the star formation rate on the gas surface density is presented. 
Finally, in Sect. 8 we summarize our results. 
\begin{table}
\begin{center}
\caption{Positional data adopted for M~31.}
\begin{tabular}{ l l } 
\hline
\hline
Position of nucleus    & RA\,=\,$00^{h}42^{m}45.97^{s}$      \\
  \,\,\,(J2000)  &  DEC\,=\,$41^{\circ}16\arcmin11.64\arcsec$\\
Position angle of major axis   & 37$^{\circ}$ \\
Inclination$^{1}$   & 75$^{\circ}$  (0$^{\circ}$=face on)\\
Distance$^{2}$   & 780$\pm$40\,kpc$^3$\\
\hline
\noalign {\medskip}
\multicolumn{2}{l}{$^{1}$ \cite{Berkhuijsen_77} and \cite{Braun_91}}\\
\multicolumn{2}{l}{$^{2}$ \cite{Stanek} }\\
\multicolumn{2}{l}{$^{3}$ 1$\arcmin$=\,227$\pm$12\,pc along major axis}\\
\end{tabular}
\end{center}
\end{table}
\section{Data}
Table 2 summarizes the data used in this work.
M~31 was mapped in IR (at 24\,$\mu$m, 70\,$\mu$m, and 160\,$\mu$m) by MIPS  in August 2004 covering a region of about 1$^{\circ} \times 3^{\circ}$ \citep{Gordon_06}.  The
basic data reduction and mosaicing was performed using the MIPS instrument team Data
Analysis Tool versions 2.90 \citep{Gordon_05}.  

M~31 was observed in the $^{12}$CO(1-0) line with the IRAM telescope
by  \cite{Nieten} at a resolution of 23$\arcsec$.  They derived the distribution of the molecular gas using a constant conversion factor of X$_{\rm CO} =\,1.9\,\times \, 10^{20}$\,mol.\,K$^{-1}$\,km$^{-1}$\,s. The galaxy was observed in the 21-cm HI
line with the Westerbork interferometer by \cite{Brinks}
at a resolution of 24$\arcsec \times 36\arcsec$. The HI survey has been corrected
for missing spacings. The H$\alpha$ observations of \cite{Devereux_etal_94b} were carried out on the Case Western Burell-Schmidt telescope
at the Kitt Peak National Observatory, providing a 2$^{\circ} \times 2^{\circ}$ field
of view.

Although the resolution of 40$\arcsec$ of the 160\,$\mu$m image is the lowest
of the data listed in Table 2, we smoothed all maps to a Gaussian
beam with a half-power width of 45$\arcsec$ for a comparison with radio 
continuum data at 20~cm \citep{Hoernes_etal_98} in a forthcoming study 
(Tabatabaei et al. in prep.). As the point spread function (PSF) of 
the MIPS data is not Gaussian, we convolved the MIPS images using 
custom kernels created with Fast Fourier transforms to account for 
the detailed structure of the PSFs. Details of the kernel creation 
can be found in \cite{Gordon_07}. 

\begin{table*}
\begin{center}
\caption{M~31 data used in this study. }
\begin{tabular}{ l l l l} 
\hline
Wavelength & Resolution & Telescope  &   Ref. \\
\hline
160\,$\mu$m             &  $40\arcsec$ & Spitzer& \cite{Gordon_06}\\
70\,$\mu$m               &  $18\arcsec$ & Spitzer&\cite{Gordon_06}\\
24\,$\mu$m             &  $6\arcsec$ & Spitzer & \cite{Gordon_06} \\
2.6\,mm\,$^{12}$CO(1-0)           &  $23\arcsec$ &IRAM 30-m & \cite{Nieten} \\
21\,cm\,HI               &  24$\arcsec \, \times \,30\arcsec$ & WSRT &\cite{Brinks}\\
6570\AA{}\,H$\alpha$   &  $2\arcsec$ (pixel size)  & KPNO& \cite{Devereux_etal_94b}\\
\hline
\end{tabular}
\end{center}
\end{table*}
After convolution, the maps were transformed to the same grid of 15$\arcsec$ width 
with the reference coordinates and position angle of the major axis 
given in Table 1. Finally, they were cut to a common extent 
of 110$\arcmin \times 38.\arcmin5$, for which most data sets are complete.
The field is not centred on the nucleus of M~31, but extends to 56.$\arcmin25$
along the northern major axis (corresponding to a radius of $R$\,=\,12.8\,kpc) and to 
53.$\arcmin75$ along the southern major axis ($R$\,=\,12.2\,kpc). The H$_2$ map of Nieten et
al. (2006) extends to 48.$\arcmin5$ along the southern major axis ($R$\,=\,11.0\,kpc).
With an extent of 19.$\arcmin25$ along the minor axis in both directions, the field
covers radii of $R <$\,16.9\,kpc in the plane of M~31. Hence, radial profiles derived by 
averageing the data in circular rings in the plane of the galaxy (equivalent to
elliptical rings in the plane of the sky) are incomplete at $R>$\,12\,kpc because
of missing data near the major axis.

\section{Dust temperature and opacity}
\cite{Walterbos_87} extensively studied the distributions
of the dust temperature and opacity in M~31 using the IRAS data at 
60\,$\mu$m and 100\,$\mu$m smoothed to a resolution of 4.$\arcmin3\times  6.\arcmin9$. 
Assuming a $\lambda^{-2}$ emissivity law, they found a remarkably constant dust temperature (21-22\,K) across the disk between 2\,kpc and 15\,kpc radius. Using this temperature, they obtained the opacity distribution at 100\,$\mu$m. Below we apply a similar method to the
70\,$\mu$m and 160\,$\mu$m MIPS maps to derive the distributions of dust
temperature and optical depth at the H$\alpha$ wavelength at higher
resolution and sensitivity.  

\subsection{Dust temperature}
We derived the color temperature of the dust, $T_{\rm d}$, between 70\,$\mu$m 
and 160\,$\mu$m assuming a $\lambda^{-2}$ emissivity law that should be
appropriate for interstellar grains emitting at these wavelengths
\citep{andriesse,Draine}.  The resulting dust temperature
map  (Fig.~1) and a histogram of the temperatures (Fig.~\ref{fig:thistring}a)
show that $T_{\rm d}$ varies between  15.6\,$\pm$\,0.8\,K and 30.8\,$\pm$\,0.3\,K. The mean value of 18.7\,$\pm$\,1.4\,K\,(standard deviation) is lower than that obtained by \cite{Walterbos_87} and close to the ISO measurements (16\,$\pm$\,2\,K) of \cite{Haas}.   
Figure~1 shows that dust of {\bf $\sim$\,18\,K} exists all over M~31.
Warmer dust with $T_{\rm d}>20$\,K dominates in star forming regions and in an 
extended area around the center of the galaxy, while cooler dust 
dominates in interarm regions.

Figure~\ref{fig:thistring}b  shows the dust temperature averaged in rings of 0.2~kpc width in the plane of M~31 against radius R\footnote{Because the spiral structure is different in the northern and southern half (northeast and southwest of the minor axis, i.e. left and right of the minor axis, respectively), we present all radial profiles for each half separately.}. On both sides of the center the dust temperature falls very fast from about 30\,K near 
the nucleus to 19~K at $R\simeq$\,4.5\,kpc. To the outer parts of the galaxy, 
it then stays within a small range of about 17\,K-19\,K in the north and 
16\,K-19\,K in the south. This indicates different radiation characteristics 
between the inner 4~kpc and beyond. In {\bf the ring of bright emission, the so called `10~kpc ring',} the temperature 
is clearly enhanced, especially in the northern half. 
Thus, in contrast to the finding of \cite{Walterbos_87}, $T_{\rm d}$ is not constant in the range $R$\,=\,2-15\,kpc but varies between 22.5\,$\pm$\,0.5\,K and 17.2\,$\pm$\,0.7\,K. 
%We note that the temperatures in Figs. 1, 2 are about 3\,K lower than those derived by \cite{Walterbos_87}, because the emission at 170$\mu$m traces cooler dust than that at 100$\mu$m used by them.

It is interesting that the mean dust temperature obtained between 
70\,$\mu$m and 160\,$\mu$m is about 3\,K lower in M~31 than in M~33 \citep{Tabatabaei_3_07}. The emission from cold dust in M~31 is stronger than in
M~33, which can also be inferred from the total emission spectra based
on IRAS and ISO observations \citep[see ][]{Haas,Hippelein}.
\begin{figure*}
\begin{center}
\resizebox{\hsize}{!}{\includegraphics*{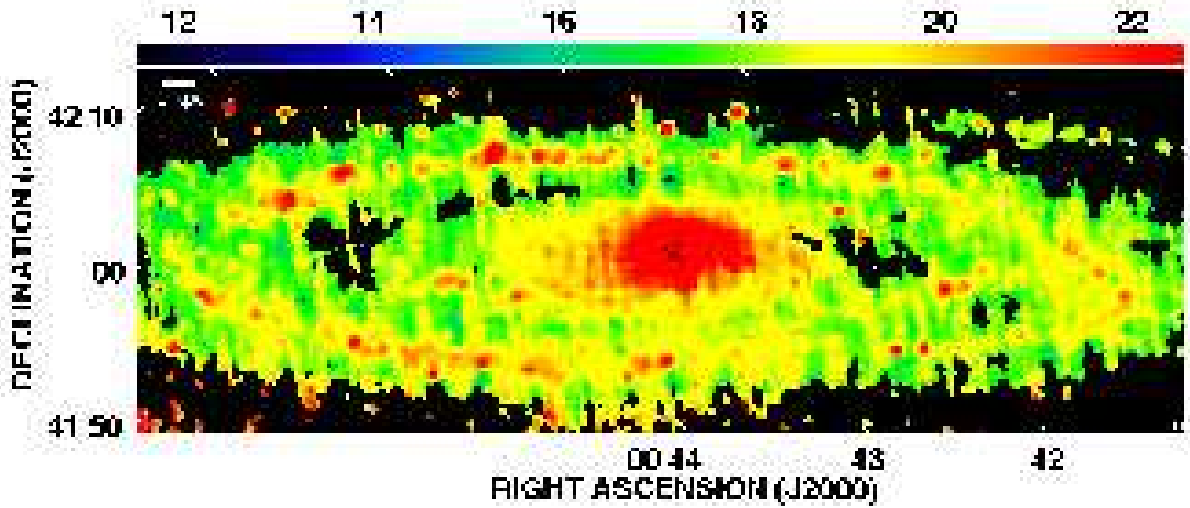}}
\caption{Dust temperature in M~31 obtained from the ratio I$_{70\mu m}$/I$_{160\mu m}$ based on the Spitzer MIPS data. Only pixels with intensity above 3$\times$ noise level were used. The angular resolution of 45$\arcsec$ is shown in the lower right-hand corner of the map. The cross indicates the location of the center. The bar at the top gives the dust temperature in Kelvin.}
\end{center}
\label{fig:tempere}
\end{figure*}

\subsection{Dust opacity distribution}
\label{sec:tau}
The total dust optical depth  along the line of sight $\tau_{160}$ 
was obtained from the dust intensity at 160\,$\mu$m and the derived temperature. 
 Following \cite{Tabatabaei_3_07}, $\tau_{160}$ was converted into the dust 
optical depth at the wavelength of the H$\alpha$ line, $\tau_{{\rm H}\alpha}$, by
multiplying it by the ratio of the dust extinction coefficient per unit mass at the 
corresponding wavelengths, $\tau_{{\rm H}{\alpha}} \simeq 2200\, \tau_{160}$ \citep[see e.g. 
Figure 12.8 of][]{krugel}.  Figure~3 shows the distribution of $\tau_{{\rm H}\alpha}$ 
across the disk of M~31 at  an angular resolution of 45$\arcsec$. Regions with 
considerable dust opacity ($\tau_{{\rm H}\alpha}>\,0.6$) follow the spiral arms, 
even the inner arms which are either weak or not detected in H$\alpha$ emission. 
The high opacity clumps ($\tau_{{\rm H}\alpha}>\,1.5$), however, only occur in the 
arms at $\simeq$\,5\,kpc  and in the `10\,kpc ring'. On average, the optical depth 
is largest in the '10\,kpc ring'. Hence, in this ring, dust has the highest density like the atomic gas \citep{Brinks}.  
Fig.~\ref{fig:tauhistring}a shows the histogram of $\tau_{{\rm H}\alpha}$, indicating 
a most probable value of 0.5 and a mean value of 0.7\,$\pm$\,0.4 across the galaxy. This
agrees with the value of $\tau_{{\rm H}\alpha}$ = 0.5\,$\pm$\,0.4 that follows from the mean
total extinction obtained by Barmby et al (2000) towards 314 globular clusters.

The variation of the mean dust optical depth $\tau_{{\rm H}\alpha}$  with galactocentric 
radius is shown  in Fig.~\ref{fig:tauhistring}b. 
In the north, $\tau_{{\rm H}\alpha}$ peaks not only in the `10\,kpc ring' (with two 
maxima at $R=$\,9.9 and 10.9\,kpc) but also near 5\,kpc (with two maxima at $R=$\,4.3 
and 5.9\,kpc). Beyond 11\,kpc $\tau_{{\rm H}\alpha}$ drops with an exponential scale 
length of 2.48$\pm$0.07\,kpc in the north and 5.06$\pm$0.22\,kpc in the south (Sect.~4, 
eq.~1). 

In Fig.~\ref{fig:extinction} we compare the radial variation of $\tau_{{\rm H}\alpha}$ for the
total area with earlier determinations. The various estimates agree well given the 
large uncertainties. \cite{Xu_96} derived $\tau_{\rm V}$ from high-resolution IRAS data
using a dust heating/cooling model and a sandwich configuration of dust and stars.
Although they left out the discrete sources, their values may be too high because
they did not include inter-arm regions in their study. \cite{Montalto}
calculated the extinction $A_{\rm FUV}$ from the total infrared TIR-to-FUV intensity ratio
and a sandwich model for stars and dust. They note that at $R <$\,8\,kpc the geometry
of M~31 may differ from the sandwich model due to the stars in the bulge, making 
the inner points less reliable. This would also affect the results of \cite{Xu_96} at $R <$\,8\,kpc. The TIR-to-FUV ratio is applicable if the dust is mainly heated
by young stars, but in M~31 about 70\% of the cold dust is heated by the ISRF
\citep{Xu_96}. Therefore, $A_{\rm FUV}$ is overestimated, as was also found for M33
\citep{Verley_09}. The curve of \cite{Tempel} closely agrees with our
data. They derived $\tau_{\rm V}$ from MIPS data and a star-dust model. Their smooth curve
underestimates $\tau_{\rm V}$ in the brightest regions by about 0.1 and may overestimate
$\tau_{\rm V}$ in regions of low brightness.

\begin{figure*}
\begin{center}
%\resizebox{7.3cm}{!}{\includegraphics*{m33-T.40arc.ps}}
\resizebox{\hsize}{!}{\includegraphics*{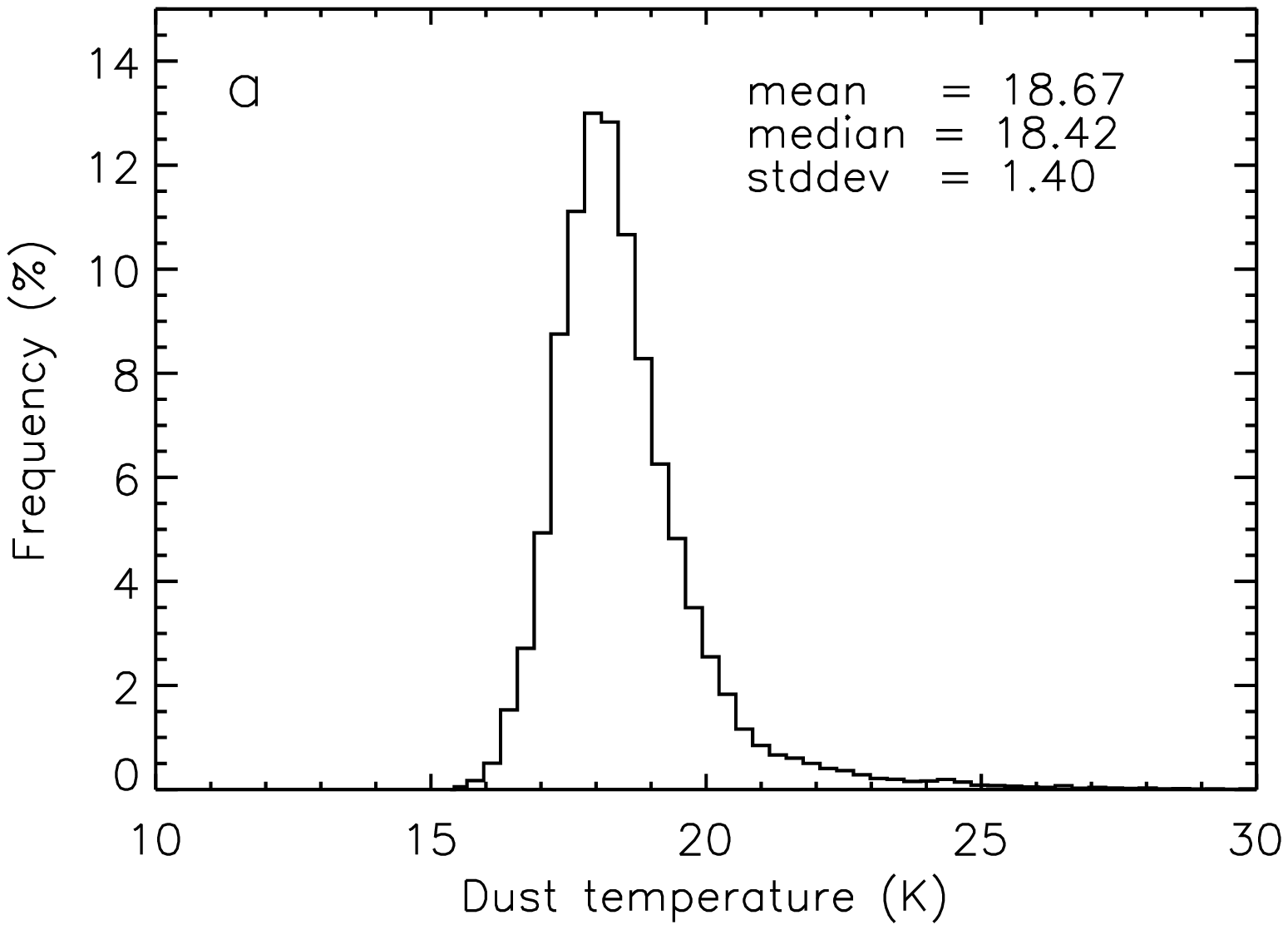}
\includegraphics*{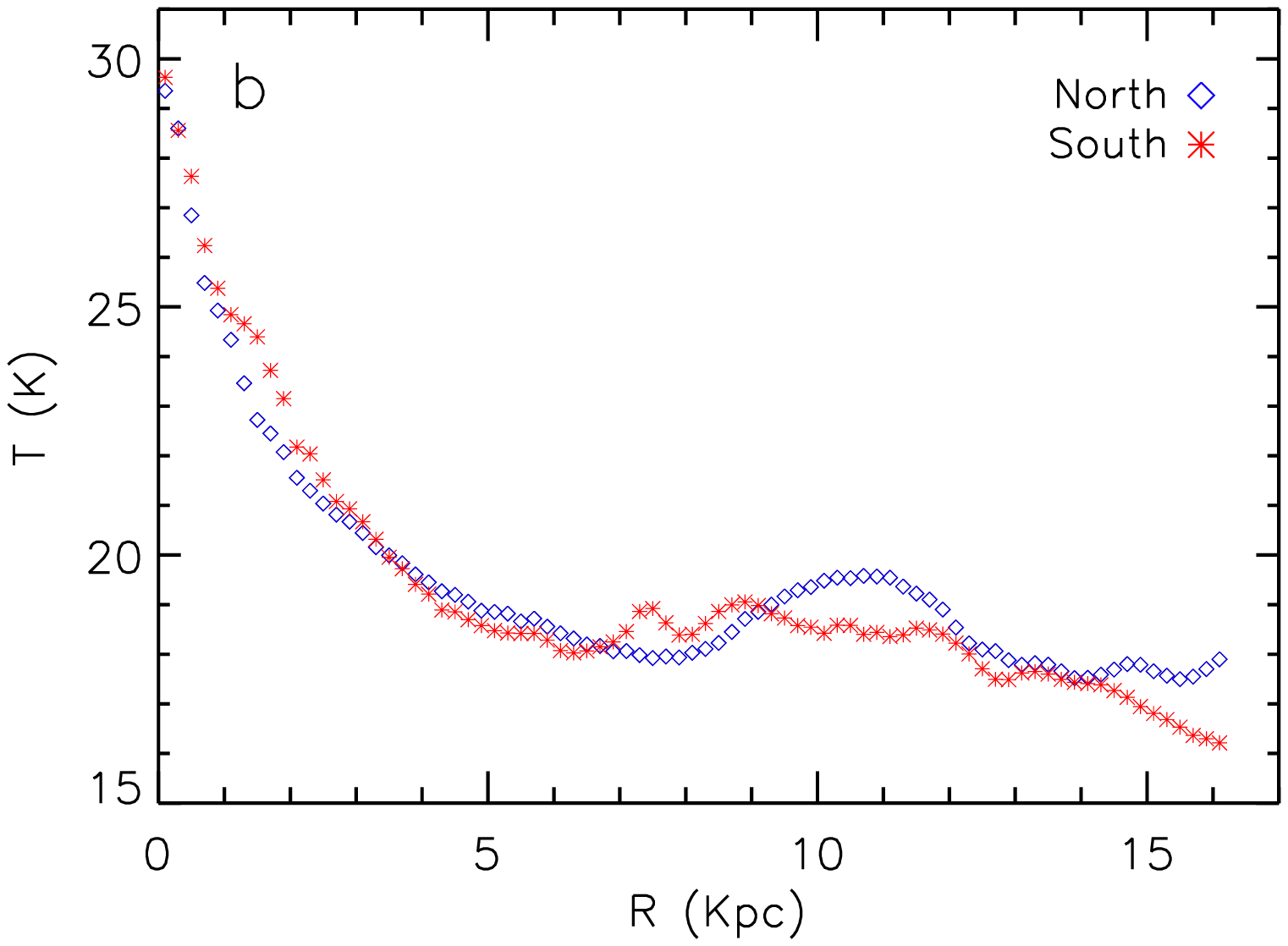}}
\caption[]{{\it a}) Histogram of the dust temperature shown in Fig.~1. {\it b}) Distribution of the dust temperature in rings of 0.2\,kpc in the galactic plane in the northern  and southern halves of M~31.}
\label{fig:thistring}
\end{center}
\end{figure*}

The opacity map in Figure~3 can be used to correct the H$\alpha$ emission for the extinction by dust. In general, extinction depends on the relative distribution of emitting regions and dust along the line of sight and changes with the geometry \citep[e.g. well mixed diffuse medium
or shell-like in HII regions ][]{Witt}. In this study, individual HII regions are rarely resolved and the geometry is close to a mixed diffuse medium.  
Furthermore, there is no information about the relative position of emittors and absorbers along the line of sight.
For the Milky Way, \cite{Dickinson} found indications of a non-uniform mixing by comparing the z-distribution of atomic gas and dust. They adopted one third of the total dust optical depth as the effective extinction as a first-order approach. 
This is also in agreement with \cite{Kruegel_09} taking scattering into account.   Moreover, \cite{Magnier} found that on average the extinction comes from dust associated with only one-third of average N(HI) in their study of OB associations along the eastern spiral arm regions of M~31. Therefore,  we use an effective optical depth $\tau_{\rm eff} = 0.33\,\times\, \tau_{{\rm H}\alpha}$ in this paper. The attenuation factor for the H$\alpha$ intensity then is $e^{-\tau_{\rm eff}}$  and we derive the intrinsic H$\alpha$ intensity $I_0$ from the observed H$\alpha$ intensity  $I=I_0\,e^{-\tau_{\rm eff}}$. Integration of the H$\alpha$ map out to a radius of 16\,kpc yields a ratio of corrected-to-observed total H$\alpha$ flux density of 1.29, thus about 30\% of the total H$\alpha$ emission is obscured by dust within M~31. The corrected H$\alpha$ map is shown in Fig.~\ref{fig:wave}a.

\begin{figure*}
\begin{center}
\resizebox{\hsize}{!}{\includegraphics*{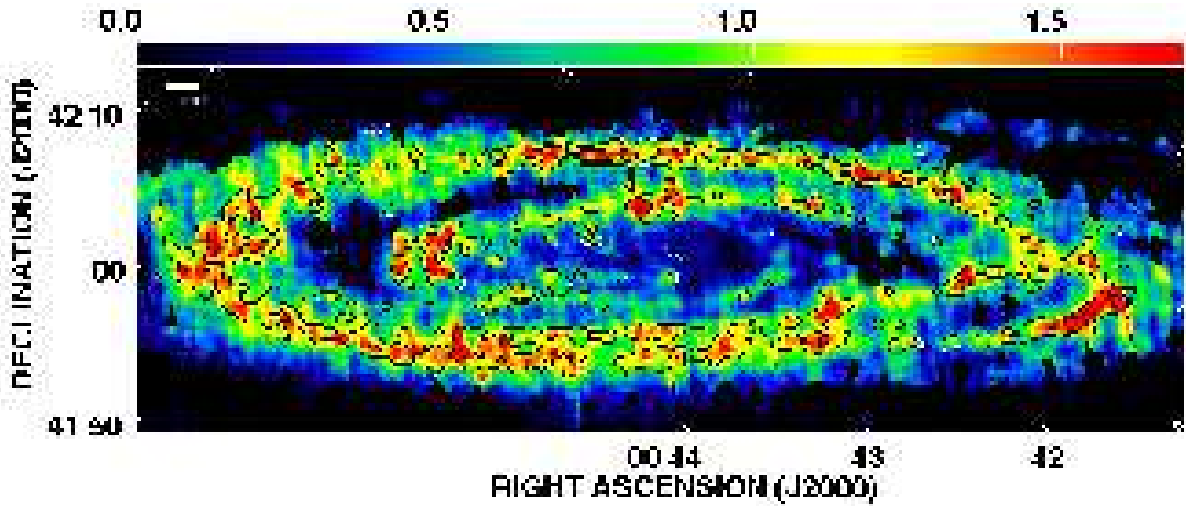}}
\caption[]{ Distribution of the dust optical depth at H$\alpha$ wavelength in
M~31. The bar at the top shows $\tau_{{\rm H}_{\alpha}}$. The angular
resolution of 45$\arcsec$ is shown in the lower right-hand corner of the map. The cross indicates the location of the center. Overlayed are contours of molecular gas column density N(H$_2$) with levels of 250 and 800\,$\times 10^{18}$\,mol.\,cm$^{-2}$. Note that maxima in  $\tau_{{\rm H}_{\alpha}}$ not always coincide with maxima in N(H$_2$). }
\end{center}
\label{fig:tau}
\end{figure*}
\begin{figure*}
%\resizebox{15cm}{!}{\includegraphics*{m31.Tdust.ring.east.ps}
%\includegraphics*{m31.Tdust.ring.west.ps}}
\resizebox{\hsize}{!}{\includegraphics*{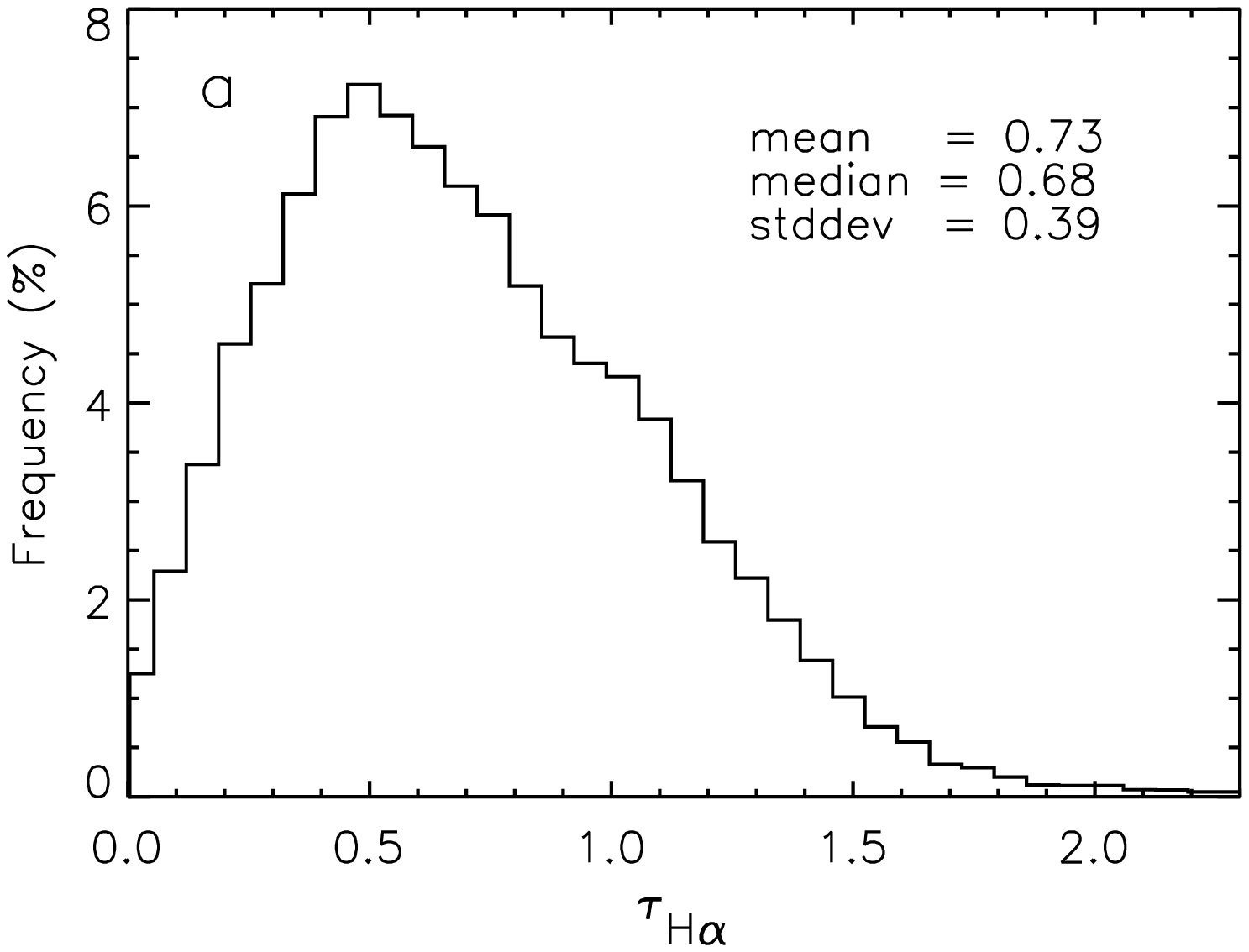}
\includegraphics*{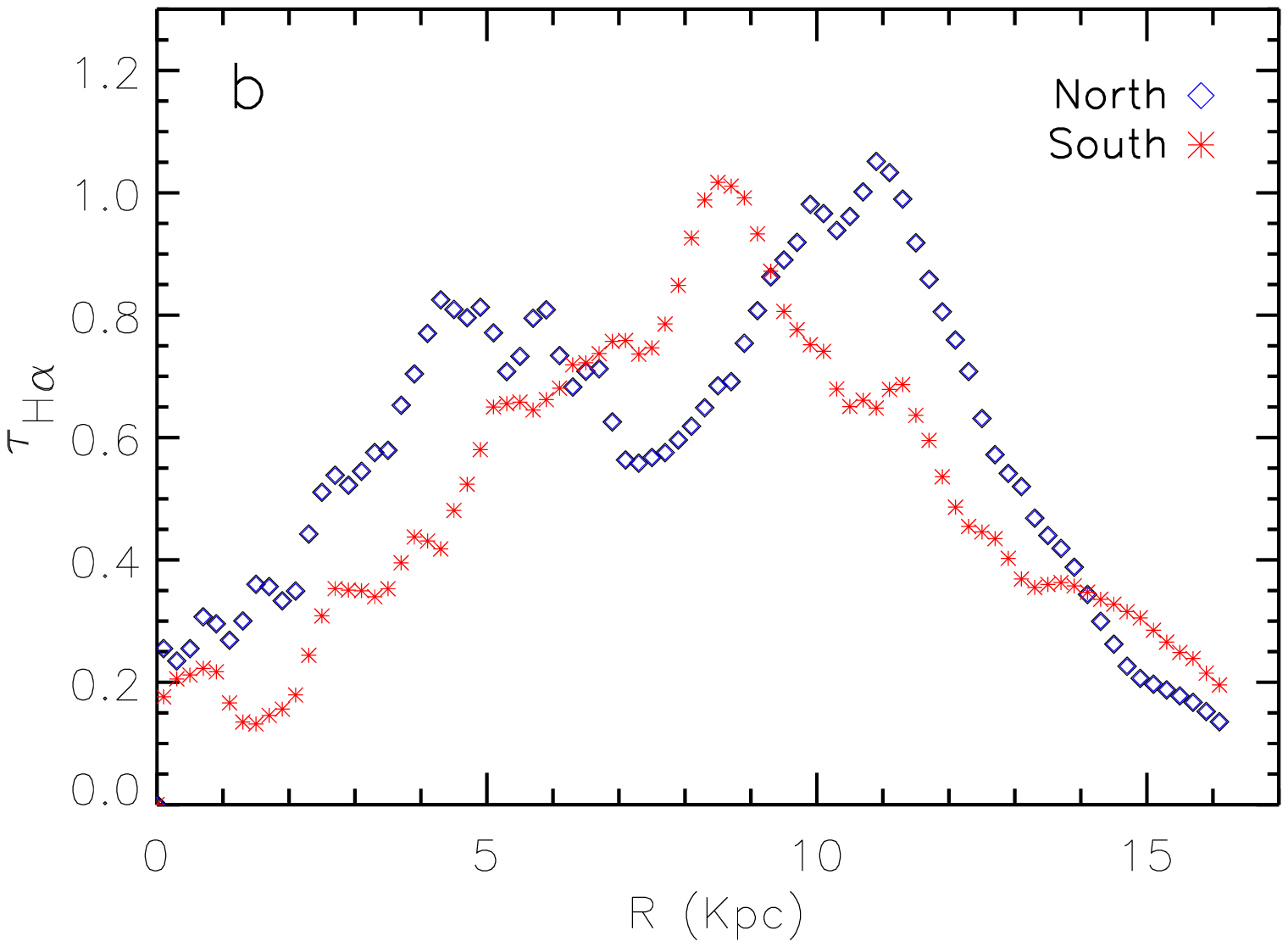}}
%\resizebox{7.3cm}{!}{\includegraphics*{m31.Tdust.ring.ps}}
\caption{{\it a}) Histogram of the dust optical depth shown in Fig.~3, {\it b}) radial distribution of the mean  optical depth at the H$\alpha$ wavelength in rings of width of 0.2\,kpc in the galactic plane in the north and south of M~31. The errors are smaller than the size of the symbols.}
\label{fig:tauhistring}
\end{figure*}
Near the center ($R\,<\,1$\,kpc),  $\tau_{\rm eff}$ varies between 0.03 and 0.13, 
corresponding to an extinction of $A({\rm H}\alpha)=1.086\,\times\,\tau_{\rm eff}\simeq$\,
0.03-0.14 mag.  At larger radii, the mean extinction increases, particularly in dense 
clouds and star forming regions, reaching a maximum of $A({\rm H}\alpha)\simeq$\,1.2\,mag 
at the densest dust cloud in the south-east of the `10\,kpc ring' (RA\,=\,00$^h$ 
41$^m$ 05.10$^s$ and DEC\,=\,+40$^{\circ}$ 38$\arcmin$ 17.73$\arcsec$). The range of 
extinction values agrees with that derived from the optical study of dust lanes by 
\cite{Walterbos_88}  and the photometric study of \cite{Williams}. 
\section{Radial distributions of dust and gas emission}
\subsection{Radial profiles}
\label{sec:profile}

In this section, we present the mean surface brightness along the line of sight
of dust and gas components as a function of galactocentric radius $R$. The surface
brightnesses are averaged in {\bf 200\,pc-wide} circular rings about the nucleus in the plane
of M~31. This is equivalent to averaging in elliptical rings of 53$\arcsec$ width in the plane of the sky.

For simplicity we used a constant inclination angle of 75$^{\circ}$ at all radii, 
appropriate for the emission at $R > 30\arcmin$ (6.8\,kpc), although in H$\alpha$ and HI
the inner regions are seen more face-on \citep{Ciardullo,Braun_91,Chemin}. However, using i\,=\,75$^{\circ}$ for $R < 30\arcmin$ instead of i\,=\,68$^{\circ}$ \citep[the area-weighted mean of the inclinations for the interval $R\,=\,1.9 - 6.8$\,kpc given by ][]{Chemin} does not change our results. The smaller inclination shifts the radial positions of the inner arms about 0.5\,kpc inwards but the general
shape of the profiles remains the same, and as all profiles change in a similar
way their inter-comparison is not affected. Furthermore, the results of the classical
correlations for $R < 30\arcmin$ presented in Sect.~6 are the same within the errors
for i\,=\,68$^{\circ}$ and i\,=\,75$^{\circ}$.    
\begin{figure}
\begin{center}
\resizebox{7.8cm}{!}{\includegraphics[angle=-90]{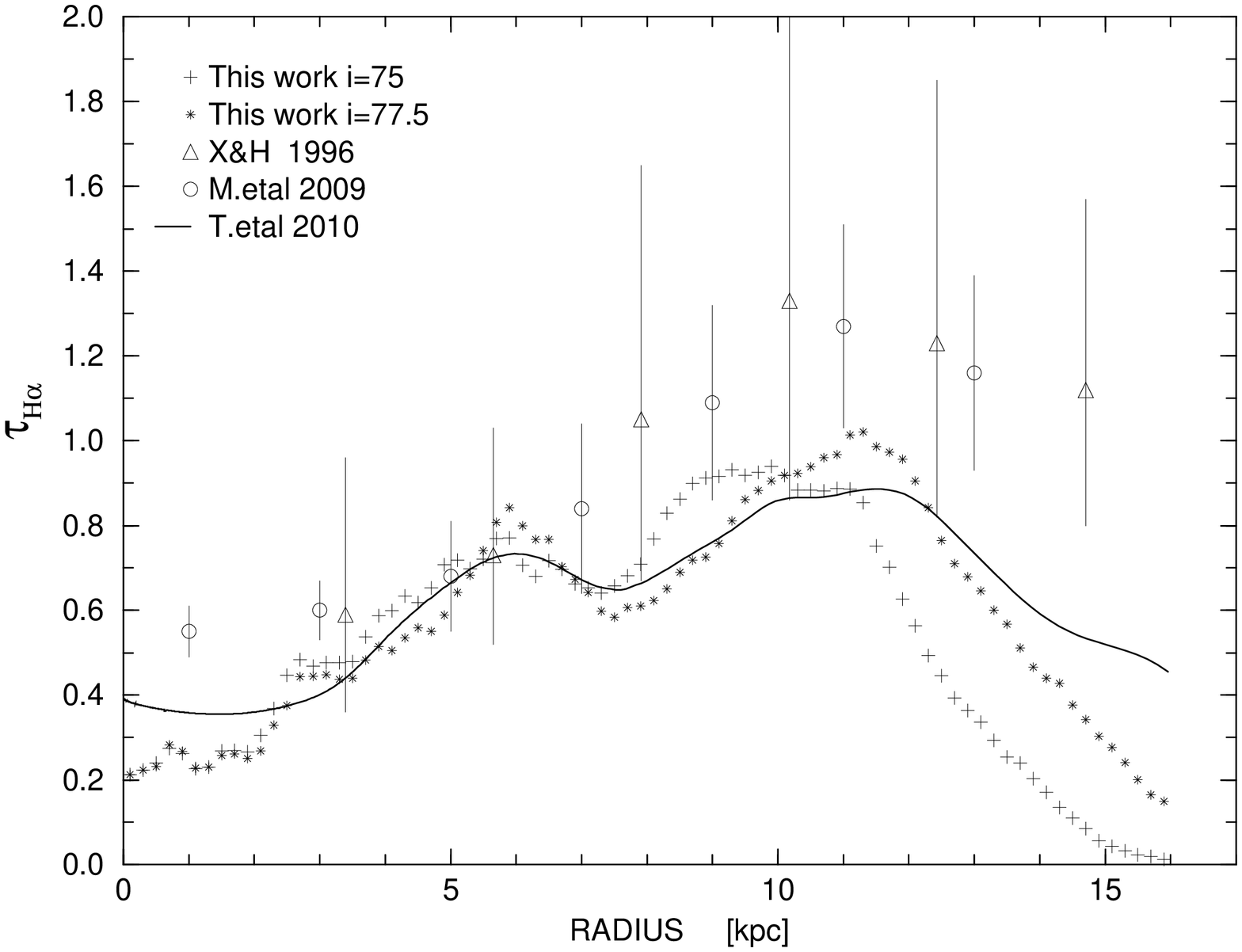}}
\caption{Radial variation of the (total) mean optical depth
in H$\alpha$ along the line of sight for the full area (north+south).
{\it Plusses}: our data averaged in {\bf 0.2\,kpc-wide} rings in the plane 
of M~31 using i\,=\,75$^{\circ}$; {\it stars}: same but with i\,=\,77.5$^{\circ}$ for 
comparison with other work; {\it triangles}: \cite{Xu_96},
averages in {\bf 2\,kpc-wide} rings with i\,=\,77$^{\circ}$ (scaled to D\,=\,780\,kpc); 
{\it circles}: \cite{Montalto}, averages in {\bf 2\,kpc-wide} rings
with i\,=\,77.6$^{\circ}$; {\it solid line}: \cite{Tempel}, semi-major
axis cut through model with i\,=\,77.5$^{\circ}$. The errors in our data 
and in the curve of \cite{Tempel} are about 10\% of the 
mean values. }
\label{fig:extinction}
\end{center}
\end{figure}

\begin{figure*}
\begin{center}
\resizebox{\hsize}{!}{\includegraphics*{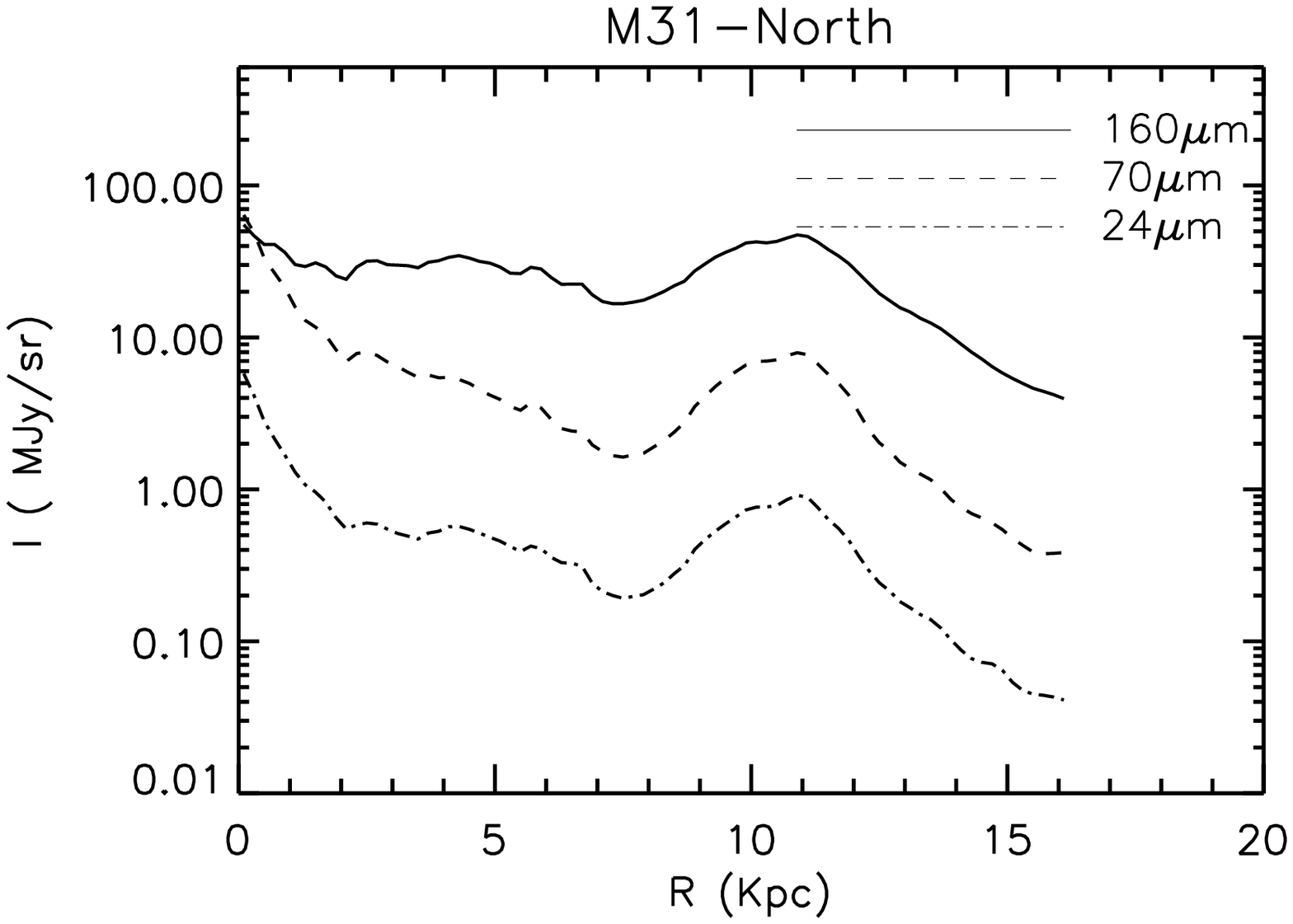}
\includegraphics*{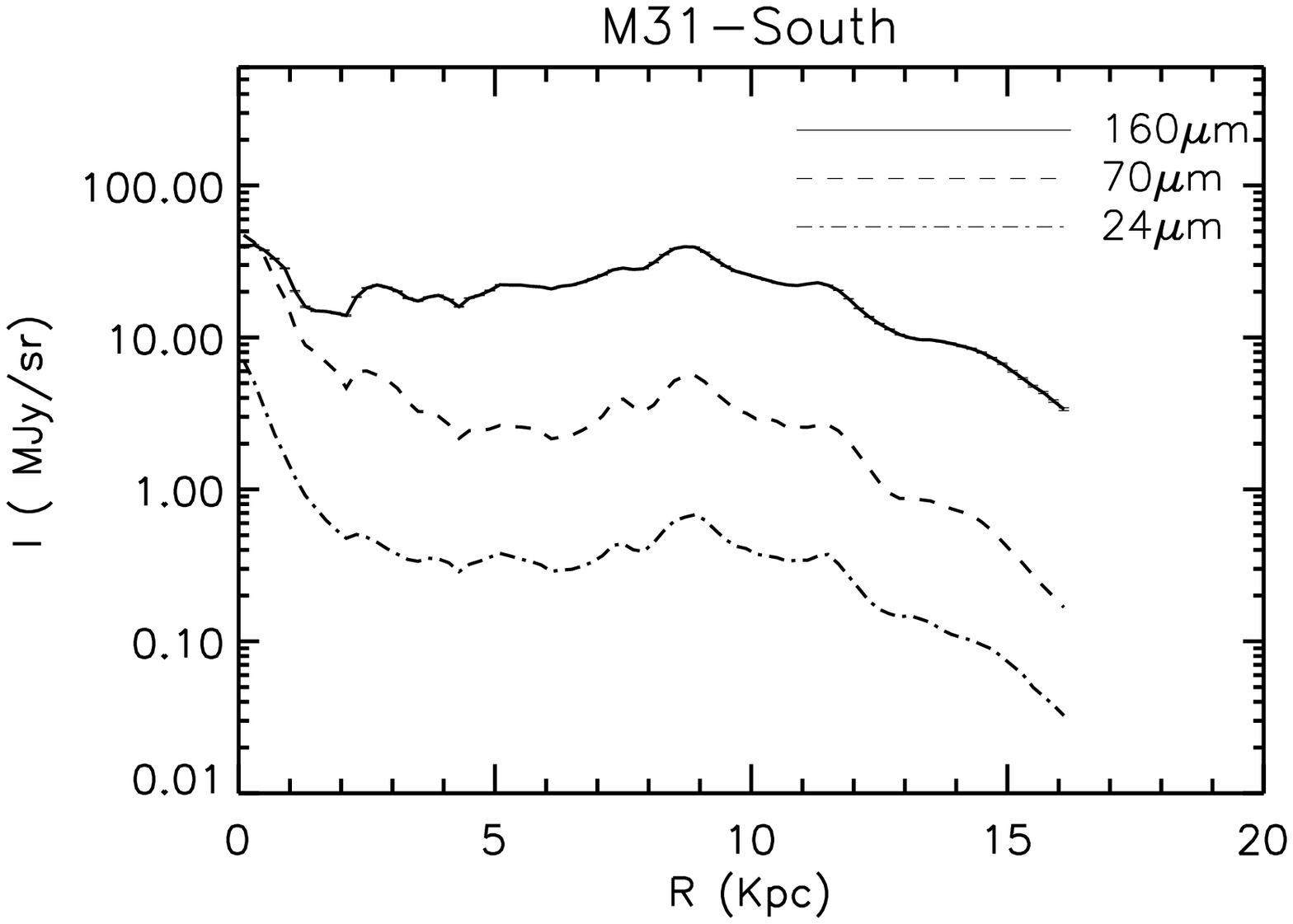}}
\resizebox{\hsize}{!}{\includegraphics*{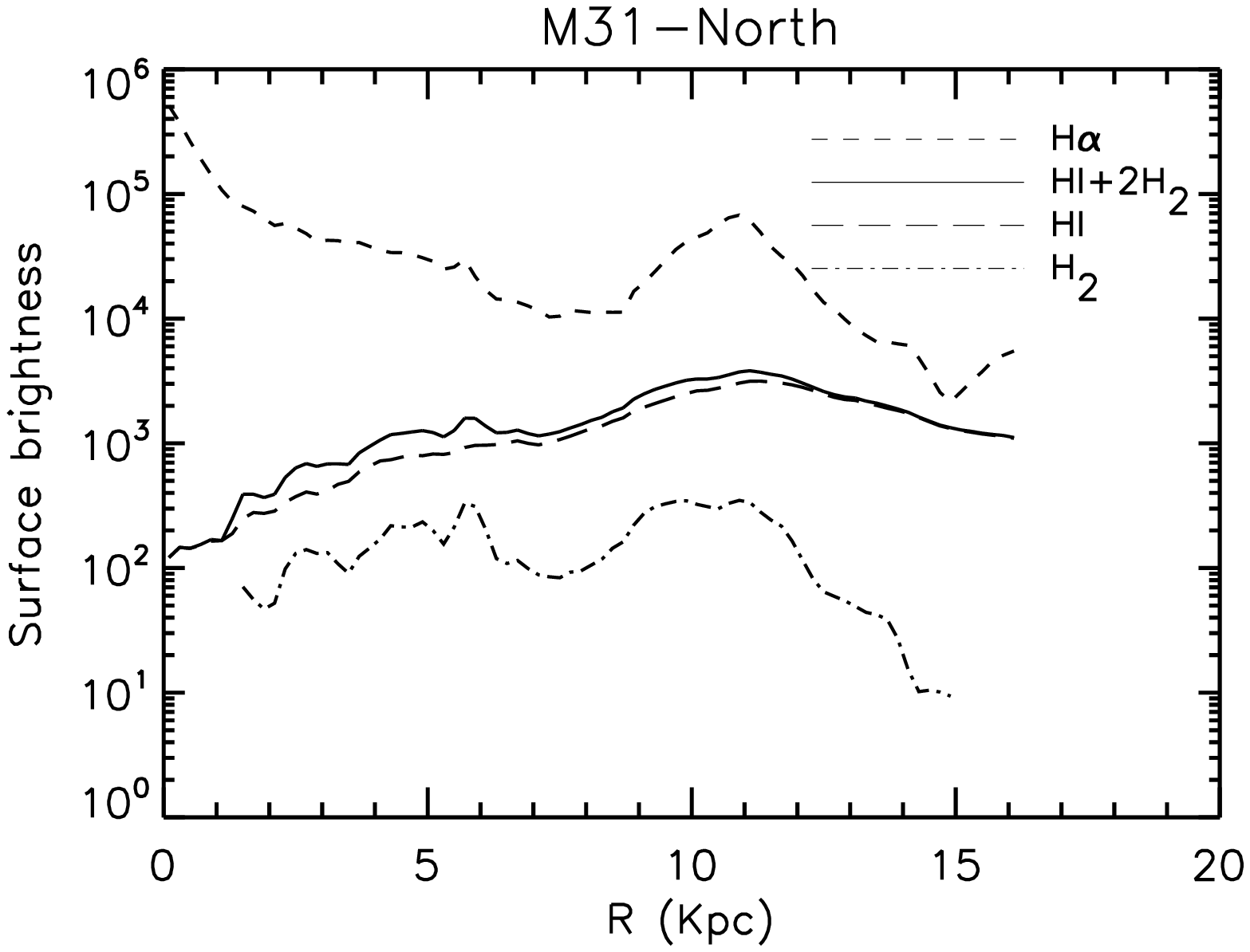}
\includegraphics*{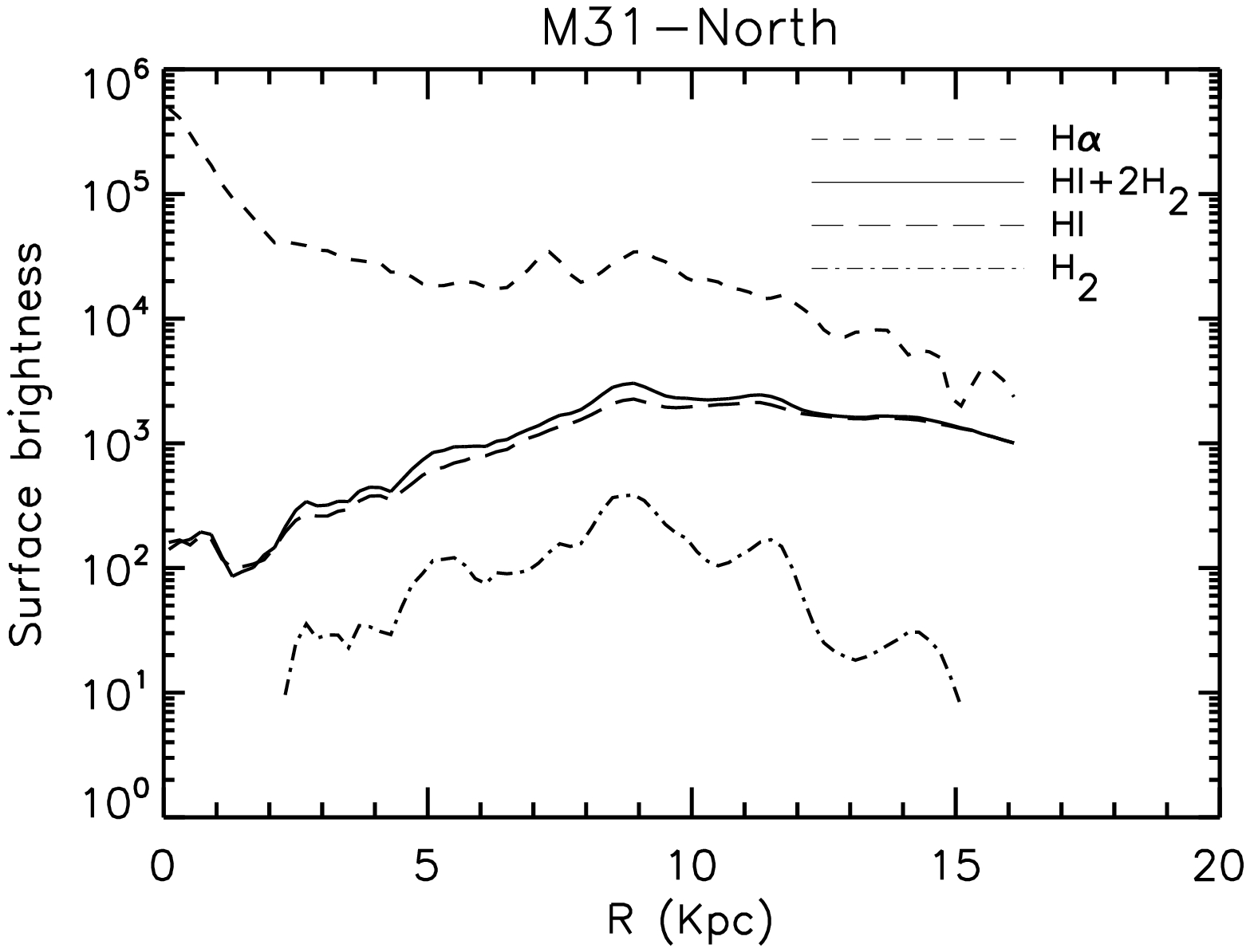}}
%\resizebox{8cm}{!}{\includegraphics*{surfbright.ps}}
\caption[]{{\it Top}: Radial profiles of the Spitzer IR emission from the northern ({\it left}) and the southern ({\it right}) halves of M~31. {\it Bottom}: Radial profiles of the surface densities of the atomic, molecular and total neutral gas together with that of the ionized gas (de-reddened H$\alpha$) for the northern ({\it left}) and southern ({\it right}) halves of M~31. The units are $10^{18}$\,atoms\,cm$^{-2}$ for  HI and HI+2H$_2$, $10^{18}$\,molecules\,cm$^{-2}$ for H$_2$, and   10$^{10}$\,erg\,s$^{-1}$\,cm$^{-2}$\,sr$^{-1}$ for  H$\alpha$ profiles. The profiles show intensities along the line of sight averaged in circular rings of 0.2\,kpc width in the plane of M~31 against the galactocentric radius. The errors are smaller than 5\% for all profiles, only for H$_2$ they increase from 10\% to 25\% at $R>12.3$\,kpc and in the inner arms for $R<4.5$\,kpc.   }
\label{fig:surfir}
\end{center}
\end{figure*}

Figure~\ref{fig:surfir} shows the mean IR intensities and the gas surface densities versus the galactocentric radius $R$ for the northern and southern halves
of M~31. The radial profiles of the IR emission at 24\,$\mu$m and 70\,$\mu$m are similar.  The 160\,$\mu$m emission, representing the colder dust emission, however, shows a generally flatter radial distribution than the 24\,$\mu$m and 70\,$\mu$m emission. In particular, the fast decrease of the 24\,$\mu$m and 70\,$\mu$m profiles from the center to  $R\,\simeq$\,2\,kpc does not occur at 160\,$\mu$m. This is in agreement with \cite{Haas} who concluded from their ISO 175\,$\mu$m map and IRAS data that the dust near the center is relatively warm. 
The fast central decrease of warmer dust emission may  be attributed to a decrease in the UV radiation field outside the nucleus, as a similar trend is seen in the GALEX UV profiles presented by \cite{Thilker_05}.
At all three IR wavelengths the arms are visible, even the weak inner arms.  The bright arms forming the `10\,kpc ring', are pronounced in  the north and followed by an exponential decrease toward larger radii.  

Although the general trend of the warm dust surface brightness
(at 24\,$\mu$m and 70\,$\mu$m) resembles more that of H$\alpha$ than those of
the neutral gas profiles (Fig.~\ref{fig:surfir}, lower panels), small variations
(e.g. in the inner 5\,kpc and for $R$\,=\,11-12\,kpc in the south)
follow those in the total gas distribution due to variations in the molecular gas.  Beyond about 5\,kpc, the radial profile of the cold dust (160\,$\mu$m) is similar to that
of the molecular gas, but with smoother variations. The minimum between 5~kpc and
10~kpc radius at 24\,$\mu$m and 70\,$\mu$m is less deep at 160\,$\mu$m and is missing
in the HI profile.

We obtained radial scale lengths between the maximum in the `10\,kpc
ring' and R=14.9~kpc for the northern ($l_{\rm N}$) and southern ($l_{\rm S}$) halves
of M~31 separately as well as for the total area ($l$). We fit an
exponential function of the form 
\begin{equation}
I(R)= I_{0}exp(-R/l),
\end{equation}
where $I_0$  is 
the intensity at $R$\,=\,10.9\,kpc for the total area and in the north, and 
$R$\,=\,8.9\,kpc in the south. The resulting scale lengths are listed in 
Table~\ref{table:slength}. 

In each half of M~31, the scale lengths of the warm dust emission 
are smaller than that of the cold dust. This confirms that the 
warm dust is mainly heated by the UV photons from the star forming 
regions in the `10\,kpc ring' and the cold dust mainly by the interstellar 
radiation field (ISRF) from old stars 
\citep{Xu_96}.

\begin{figure}
\begin{center}
\resizebox{7.5cm}{!}{\includegraphics*{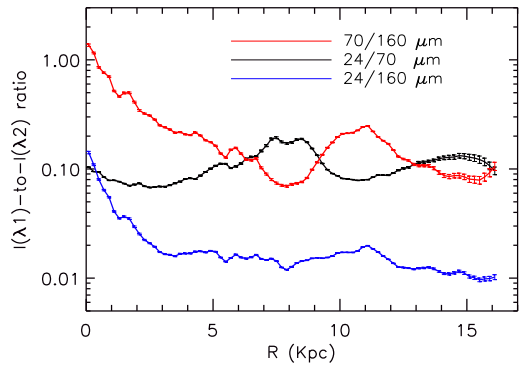}}
\caption[]{Ratio of the MIPS IR intensities against galactocentric radius in M~31.}
\label{fig:ratio}
\end{center}
\end{figure}

\begin{table}
\begin{center}
\label{table:scalelength}
\caption{Exponential scale lengths of dust and gas emissions from M~31. The scale lengths were calculated from $R$=\,10.9\,kpc to $R$=\,14.9\,kpc for the whole galaxy ($l$) and the northern half  ($l_{\rm N}$) and from $R$=\,8.9\,kpc to $R$=\,14.9\,kpc for the southern half  ($l_{\rm S}$). The scale length of $\tau_{\rm H\alpha}$ is also shown for comparison. }
\begin{tabular}{ l l l l} 
\hline
 &  $l_{\rm N}$\,(kpc) &  $l_{\rm S}$\,(kpc) & $l$\,(kpc)\\
\hline 
\hline
IR 160$\mu$m  & 1.86\,$\pm$\,0.06 & 3.87\,$\pm$\,0.18 & 2.29\,$\pm$\,0.11 \\
IR 70$\mu$m  & 1.38\,$\pm$\,0.08 & 3.07\,$\pm$\,0.17  & 1.66\,$\pm$\,0.09\\
IR 24$\mu$m  & 1.38\,$\pm$\,0.07 & 3.44\,$\pm$\,0.21  & 1.57\,$\pm$\,0.06 \\
H$_2$           & 1.14\,$\pm$\,0.07 & 2.40\,$\pm$\,0.22 &  1.27\,$\pm$\,0.11 \\
HI           & 4.54\,$\pm$\,0.24 & 15.90\,$\pm$\,1.70 & 5.90\,$\pm$\,0.20 \\
HI+2H$_2$    & 3.80\,$\pm$\,0.12 & 9.00\,$\pm$\,0.69  & 4.73\,$\pm$\,0.21 \\
H$\alpha$    & 1.08\,$\pm$\,0.03 & 2.92\,$\pm$\,0.10  & 1.39\,$\pm$\,0.04 \\
\hline
$\tau_{\rm H\alpha}$ & 2.48$\pm$0.07 & 5.06$\pm$0.22 & 3.17\,$\pm$\,0.15\\
\hline
\end{tabular}
\label{table:slength}
\end{center}
\end{table}
The scale lengths of the 24\,$\mu$m and  70\,$\mu$m emission are nearly the same and the 24\,$\mu$m-to-70\,$\mu$m intensity ratio (Fig.~\ref{fig:ratio}) hardly varies between $R\simeq10$\,kpc and $R\simeq15$\,kpc. This indicates a similar distribution of their origins.   Assuming that the main source of the 24\,$\mu$m emission is very small dust grains  and of the 70\,$\mu$m and 160\,$\mu$m emission is big grains, as argued by \cite{Walterbos_87}, a constant intensity ratio of the 24\,$\mu$m-to-70\,$\mu$m and  24\,$\mu$m-to-160\,$\mu$m emission suggests that the very small and big grains are well mixed in the interstellar medium. Other possible origins of the 24\,$\mu$m emission are stars with dust shells like evolved AGB stars or Carbon stars. Using the IRAS data,  \cite[][]{Soifer} attributed the 25\,$\mu$m emission from the bulge (central 8$\arcmin$) of M~31 to circumstellar dust emission from late-type stars.  In the disk of M~31, \cite{Walterbos_87} found no direct evidence for a contribution from stars with dust shells \citep[contrary to that in the Milky Way, ][]{Cox_86}.

The higher resolution and sensitivity of the MIPS IR intensity ratios (Fig.\ref{fig:ratio}), however, provide more information. Although at $R\,>\,3$\,kpc the variations in the IR intensity ratios  are not large,  their radial behavior is not the same. For instance, the 24-to-70\,$\mu$m intensity ratio peaks between 5\,kpc and 10\,kpc radius, whereas the 70-to-160\,$\mu$m intensity ratio peaks in the `10\,kpc ring'. The latter can be explained by the higher temperature of the dust heated by OB associations in the {\bf `10\,kpc ring'}. The fact that the 24-to-70\,$\mu$m intensity ratio is not enhanced in the {\bf `10\,kpc ring'} (and in the central region) shows the invalidity of this ratio for temperature  determination due to the important contribution from the very small grains.  On the other hand,  the enhancement of the 24-to-70\,$\mu$m in regions where there is no strong radiation field (between the arms) reveals  possibly different origins of the 24\,$\mu$m and 70\,$\mu$m emission.
% which plays a minor role (as the increase is not very strong).
The stellar origin, e.g. photosphere of cool stars or dust shell of the evolved stars, may provide the enhancement of the 24-to-70\,$\mu$m intensity ratio in the inter-arm region. In M~33,  \cite{Verley_09} attributed a similar enhancement of the diffuse 24\,$\mu$m emission  to dusty circumstellar shells of unresolved, evolved AGB stars. For M~31, this needs to be quantified through a more detailed study and modelling of the spectral energy distribution, which is beyond the scope of this paper.

%\begin{figure*}
%\begin{center}
%\resizebox{\hsize}{!}{\includegraphics*{surfbright_gas.east.ps}
%\includegraphics*{surfbright_gas.west.ps}}
%\caption[]{ }
%\end{center}
%\label{fig:surfgas}
%\end{figure*}

%Tables 1 and 2 summarize the characteristics and performance of data used in this work.

\begin{figure}
%\begin{center}
\resizebox{7cm}{!}{\includegraphics*{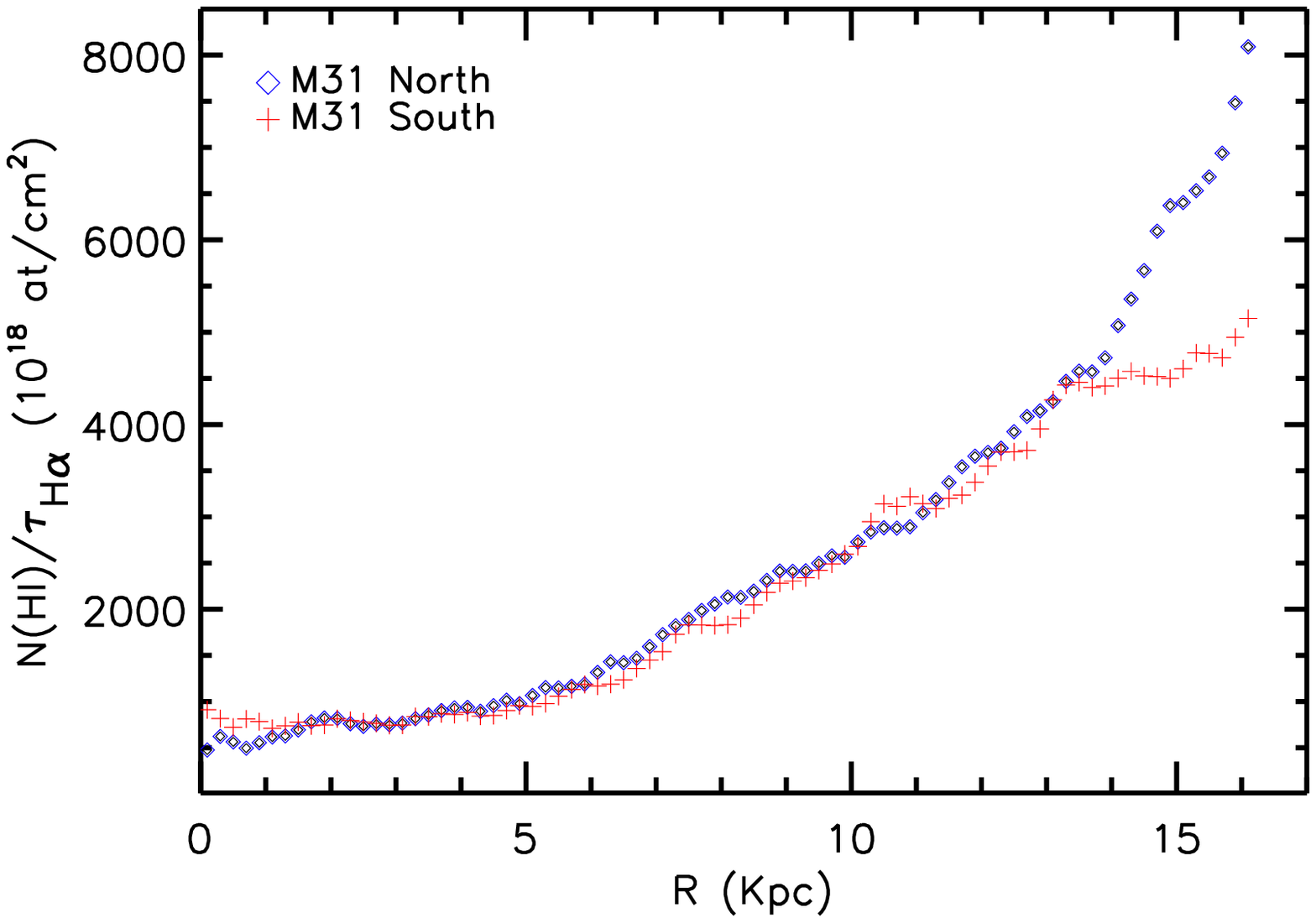}}
\resizebox{7cm}{!}{\includegraphics*{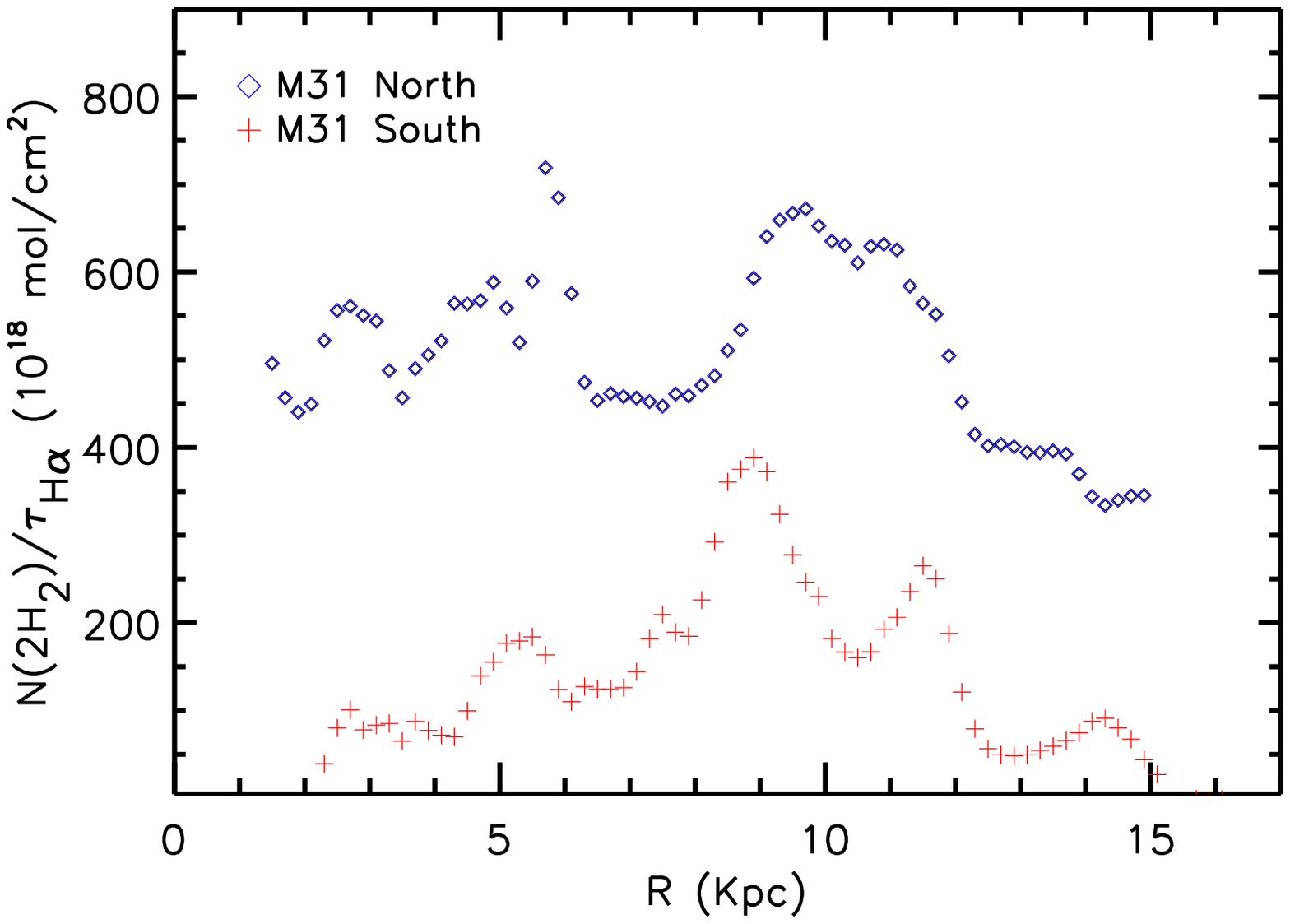}}
\resizebox{7cm}{!}{\includegraphics*{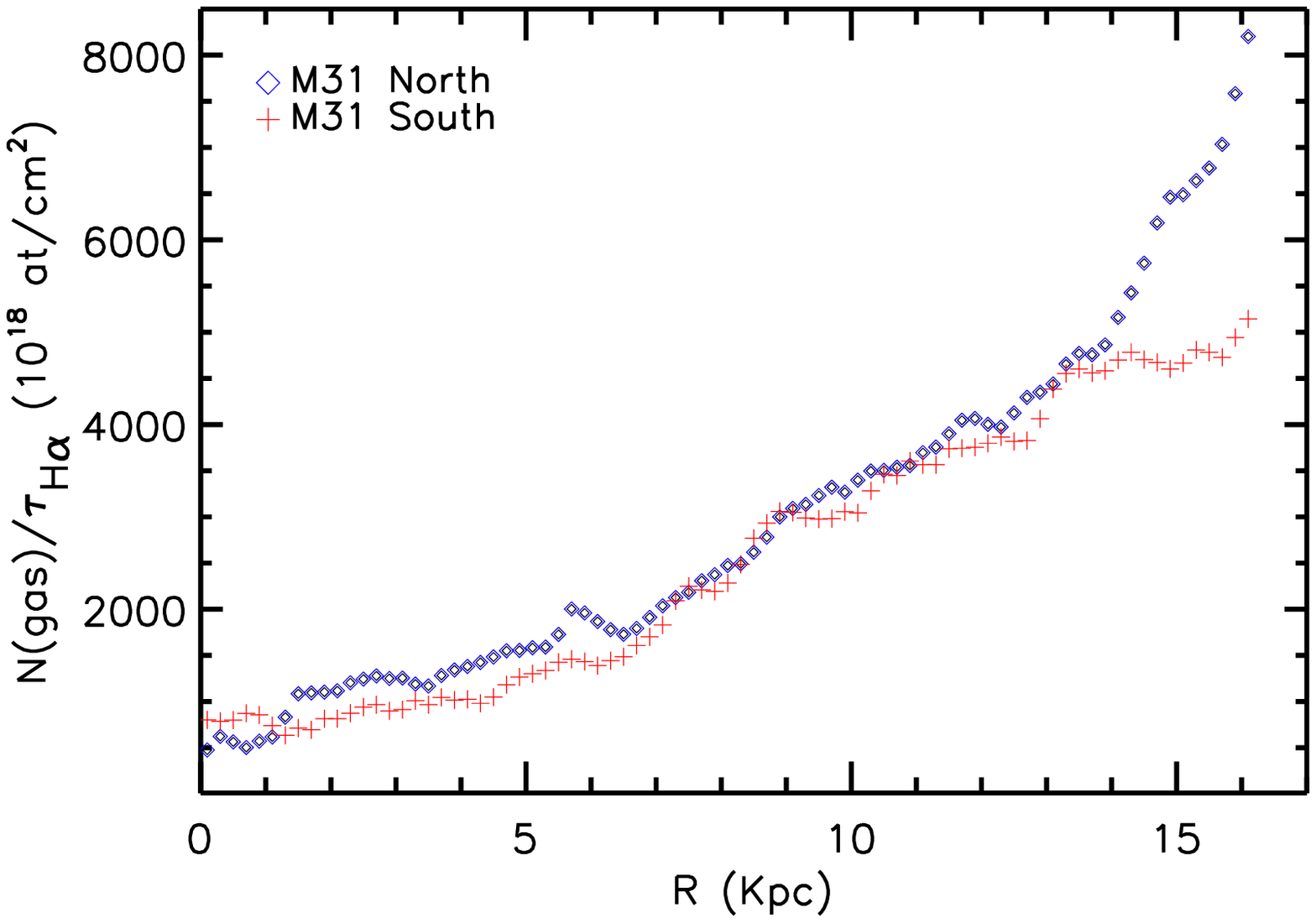}}
\caption[]{Radial profiles of the gas-to-dust ratios in M~31, the northern half and the southern half. {\it Top}:  N(HI)/$\tau_{{\rm H}\alpha}$, {\it middle}: N(2H$_2$)/$\tau_{{\rm H}\alpha}$ , {\it bottom}:  N(HI+2H$_2$)/$\tau_{{\rm H}\alpha}$. In the middle panel,  the northern profile is shifted by 300 units for clarity.  The errors are smaller than 5\% everywhere only for   N(2H$_2$)/$\tau_{{\rm H}\alpha}$ they increase from 10\% to 25\% at $R>12.3$\,kpc and in the inner arms for $R<4.5$\,kpc.  }
%\end{center}
\label{fig:gasdustratio}
\end{figure}

\subsection{Gas-to-dust ratio}
The gas-to-dust mass ratio and its variation across the galaxy can provide information about the metallicity distribution  \citep[e.g. ][]{Viallefond_82} and hence about the evolutionary history of the galaxy. 
The relative amount of dust and gas is expected to be correlated with the abundance of the heavy elements \citep{Draine_07}. 

A number of authors has studied the gas-to-dust ratio in M~31 by 
comparing HI column densities and optical or UV extinction \citep{Walterbos_88,
Bajaja_77,Xu_96,Nedialkov,Savcheva}. All authors found an increase of the atomic 
gas-to-dust ratio with radius. \cite{Walterbos_87} derived the HI gas-to-dust ratio 
using dust optical depth from IRAS 60\,$\mu$m and 100\,$\mu$m data. They found a
radial gradient that is 4-5 times larger than the abundance gradient
of \cite{Blair}. After adding the molecular and atomic gas
column densities, \cite{Nieten} obtained a strong radial
increase in the total gas-to-175\,$\mu$m intensity resulting from the
increase in the atomic gas-to-175\,$\mu$m intensity. As the dust 
optical depth is a better measure for the dust column density than 
the temperature-dependent dust emission, we re-investigated the 
gas-to-dust ratio in M~31 taking  advantage of the high resolution 
of the Spitzer MIPS data.

We calculated the radial profiles of the three gas-to-dust ratios   
from the mean column densities of N(HI), N(2H$_2$), N(HI+2H$_2$) and $\tau_{{\rm H}\alpha}$ in circular rings of 0.2\,kpc width in the plane of the galaxy. Figure \ref{fig:gasdustratio} (upper panel) shows that the atomic gas-to-dust ratio increases exponentially with 
radius by more than a factor of 10 from about $0.6\times10^{21}$ at\,cm$^{-2}$ at 
the center to about $6.5\times10^{21}$ at\,cm$^{-2}$  at $R$=\,15\,kpc. The increase is 
surprisingly smooth and, at least up to $R$=\,13\,kpc, nearly the same for 
the northern and southern half, indicating little variation between 
arm and inter-arm regions and within the arms. In contrast, the 
molecular gas-to-dust ratio (Fig.~\ref{fig:gasdustratio}, middle panel) does not increase systematically with radius but shows clear enhancements of a factor
2-3 in the spiral arms and the `10\,kpc ring'. The minima in the inter-arm 
regions are due to a stronger decrease in N(2H$_2$) than in $\tau_{{\rm H}\alpha}$. 
%Figure~\ref{fig:tau}
Figure~3 shows that  along the arms N(2H$_2$)/$\tau_{{\rm H}\alpha}$ also varies  significantly because maxima in H$_2$ emission and $\tau_{{\rm H}\alpha}$ are often not coincident. 
The variations in N(2H$_2$)/$\tau_{{\rm H}\alpha}$ are visible in the profile of the 
total gas-to-dust ratio (Fig.~\ref{fig:gasdustratio}, bottom panel) as weak enhancements 
at the positions of the arms near $R\,=\,6$\,kpc and $R\,=\,8-12$\,kpc. As the atomic 
gas is the dominant gas phase in M~31, dust mixed with HI gas largely
determines the optical depth. Inspection of the distribution of the
total gas-to-dust ratio across M~31 (not shown) reveals small-scale
variations along the arms of typically a factor of 2.

\begin{table*}
\begin{center}
\label{table4}
\caption{Exponential scale lengths $L$  and radial gradients of dust-to-gas ratios and the abundance [O/H] combined data from \cite{Blair} and \cite{Dennefeld}, where Ratio($R$)\,=\,C.\,exp$(-R/L)$. N(HI) and N(gas) are in 10$^{21}$\,at\,cm$^{-2}$. Errors are standard deviations. }
\begin{tabular}{ l l l l l } 
\hline
Ratio  & $R$ & \,\,\,\,\,\,\,\,\,\, \,C  &  $L$ & Gradient \\
   & (kpc) &  & (kpc) &(dex/kpc) \\
\hline 
\hline
$\tau_{{\mathrm H}\alpha}$/N(HI) & 0-16 &  1.95\,$\pm$\,0.08 & 6.2\,$\pm$\,0.1 & 0.070\,$\pm$\,0.001  \\
                                 & 5-15 &  1.97\,$\pm$\,0.08 & 6.1\,$\pm$\,0.2 & 0.071\,$\pm$\,0.001 \\
& &  &\\
$\tau_{{\rm H}\alpha}$/N(gas) & 0-16 &  1.35\,$\pm$\,0.05 &7.3\,$\pm$\,0.2 &  0.059\,$\pm$\,0.002 \\
                              & 5-15 &  1.27\,$\pm$\,0.06 & 7.4\,$\pm$\,0.2 & 0.059\,$\pm$\,0.002 \\
 & &  &\\
$[{\rm O/H}]\times 10^{-4}$    & 5-15  & 2.95\,$\pm$\,0.35 &  9.7\,$\pm$\,2.6 & 0.045\,$\pm$\,0.012  \\
\hline
\end{tabular}
%\label{table4}
\end{center}
\end{table*}
We conclude that the radial increase in the total gas-to-dust ratio of 
more than a factor 10 between the center and $R=\,15$\,kpc is entirely due to 
that of the atomic gas-to-dust ratio, whereas the molecular gas-to-dust 
ratio is only increased in the arms.  This confirms the conclusion of 
\cite{Nieten} based on the same gas data and the 175\,$\mu$m intensity. 

At which radius in M~31 would the gas-to-dust ratio observed in the solar
neighborhood occur? \cite{Bohlin} and \cite{Diplas_94}
derived N(HI)/E(B-V)=$4.8 \times 10^{21}$ at\,cm$^{-2}$\,mag$^{-1}$ and $(4.9 \pm 0.3) \times 10^{21}$ at\,cm$^{-2}$\,mag$^{-1}$, respectively, using the extinction towards large samples 
of stars to determine the color excess E(B-V). Since E(B-V)\,=\,A$_{\rm V}$/R$_{\rm V}$, 
where the visual extinction A$_{\rm V}$\,=\,1.234\,$\tau_{{\rm H}\alpha}$\,mag 
\citep[e.g. ][]{krugel} and the total/selective extinction R$_{\rm V}$=\,2.8\,$\pm$\,0.3 in M~31 \citep{Walterbos_88}, we have E(B-V)=\,0.44\,$\tau_{{\rm H}\alpha}$ mag$^{-1}$. Hence, a value of N(HI)/E(B-V)=\,$4.9 \times 10^{21}$ at\,cm$^{-2}$\,mag$^{-1}$ corresponds to N(HI)/$\tau_{{\rm H}\alpha}$\,=\,$2.2 \times 10^{21}$ at\,cm$^{-2}$, which occurs in M~31 near $R\,=\,$8.5\,kpc (Fig.~\ref{fig:gasdustratio}, top panel), just in the bright emission ring. The total gas-to-dust ratio near the sun of $5.8 \times 10^{21}$ at\,cm$^{-2}$\,mag$^{-1}$ \citep{Bohlin} corresponding to N(gas)/$\tau_{{\rm H}\alpha}$=\,$2.6 \times 10^{21}$ at\,cm$^{-2}$ occurs at nearly the same radius (Fig.~\ref{fig:gasdustratio}, bottom panel). Thus the gas-to-dust ratio near the sun is similar to that in the {\bf `10\,kpc ring'} in M~31, in agreement with earlier studies  \citep{Genderen,Walterbos_87}.

In contrast to Fig.~\ref{fig:gasdustratio}, we present in Fig.~9
the radial profiles of the dust-to-gas ratios, here for the total area in M~31. The two lower curves closely follow exponentials with scale lengths of 6.1$\pm$\,0.2\,kpc and 
7.4$\pm$\,0.2\,kpc for $\tau_{{\rm H}\alpha}$/N(HI) and $\tau_{{\rm H}\alpha}$/N(gas), respectively, between $R$=\,5\,kpc and $R$=\,15\,kpc (see Table 4). For nearly the same radial range ($R$=\,3-15\,kpc), \cite{Walterbos_87} derived a scale length
of $\tau_{100\mu{\rm m}}$/N(HI)$\simeq$\,4\,kpc from data near the major axis. 
\cite{Walterbos_88} obtained a scale length of A$_{\rm B}$/N(HI)$\simeq$\,9\,kpc 
for the inner and outer dust lanes, and the $\tau_{\rm V}$/N(HI) ratio of \cite{Xu_96}
for diffuse spiral arm regions also indicates a scale length of about 4\,kpc (all scale 
lengths were scaled to D=\,780\,kpc). Since our scale lengths 
are not restricted to specific areas, our results are more representative for the 
mean dust-to-gas ratios in the disk of M~31.

As dust consists of heavy elements and both dust and heavy elements are found
in star formation regions, the radial variations in the dust-to-gas ratio and the
metal abundance are expected to be similar (e.g. \citep{Hirashita_99,Hirashita_02}. This has
indeed been observed in several nearby galaxies \citep{Issa}. In M~31 the 
variation in the metallicity with radius is not well established. Measurements of the 
element abundance strongly depend on the empirical method and calibration applied 
\citep{Trundle}. \cite{Pagel} showed that the ([OII] +[OIII])/H$\beta$ ratio (the so 
called 'R$_{23}$') is a good probe of oxygen abundance and radial trends of this ratio 
have been studied in many nearby galaxies  \citep[e.g. ][]{Pagel_81,Evans,Henry,Garnett}. \cite{Blair} and \cite{Dennefeld} derived R$_{23}$ for HII regions in M~31. We combined their results and derived a scale length of log[O/H] of 9.7$\pm$\,2.6\,kpc  corresponding to a gradient of 0.045$\,\pm \,0.012$\,dex/kpc. Comparing four different calibrations, \cite{Trundle} derived gradients of 0.027 - 0.013\,dex/kpc using the 11 HII regions of \cite{Blair}, with Pagel's calibration giving 0.017\,$\pm$0.001\,dex/kpc. Since our value of the [O/H] gradient is based on 19 HII regions, we expect it to be more reliable than that of \cite{Trundle}. 

Table~4 shows that the radial gradient in $\tau_{{\rm H}\alpha}$/N(gas) best matches the 
metallicity gradient. In view of the large uncertainties, the gradients in the dust-to-gas
surface-density ratio and the oxygen abundance in M~31 may indeed be comparable. A much larger
sample of abundance measurements of HII regions is needed to verify this similarity. Our
result agrees with the approximately linear trend between gradients in dust-to-gas ratios
and [O/H] in nearby galaxies noted by \cite{Issa}.
\begin{figure}
\begin{center}
\resizebox{7.5cm}{!}{\includegraphics[angle=-90]{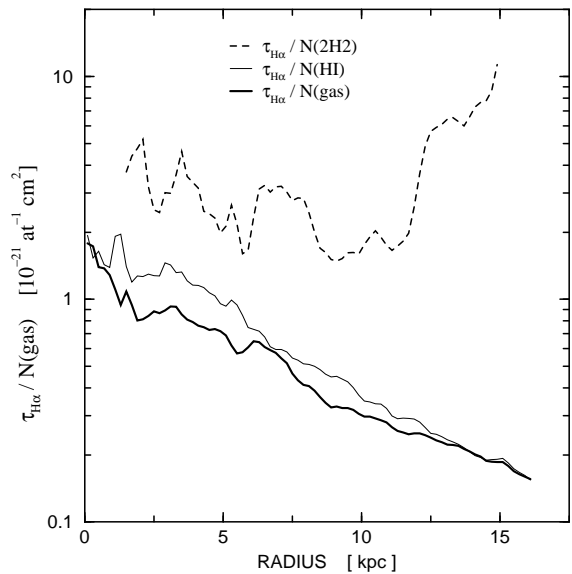}}
\caption[]{Dust-to-gas ratios as function of galactocentric radius
         for M~31, calculated from the radial 
         profiles of $\tau_{{\rm H}\alpha}$, N(HI), N(2H$_2$) and N(HI+2H$_2$)=~N(gas). } 
\end{center}
\label{fig:dusttogas}
\end{figure}

\section{Wavelet analysis of dust and gas emission}

\begin{figure*}
\begin{center}
\resizebox{13cm}{!}{\includegraphics*{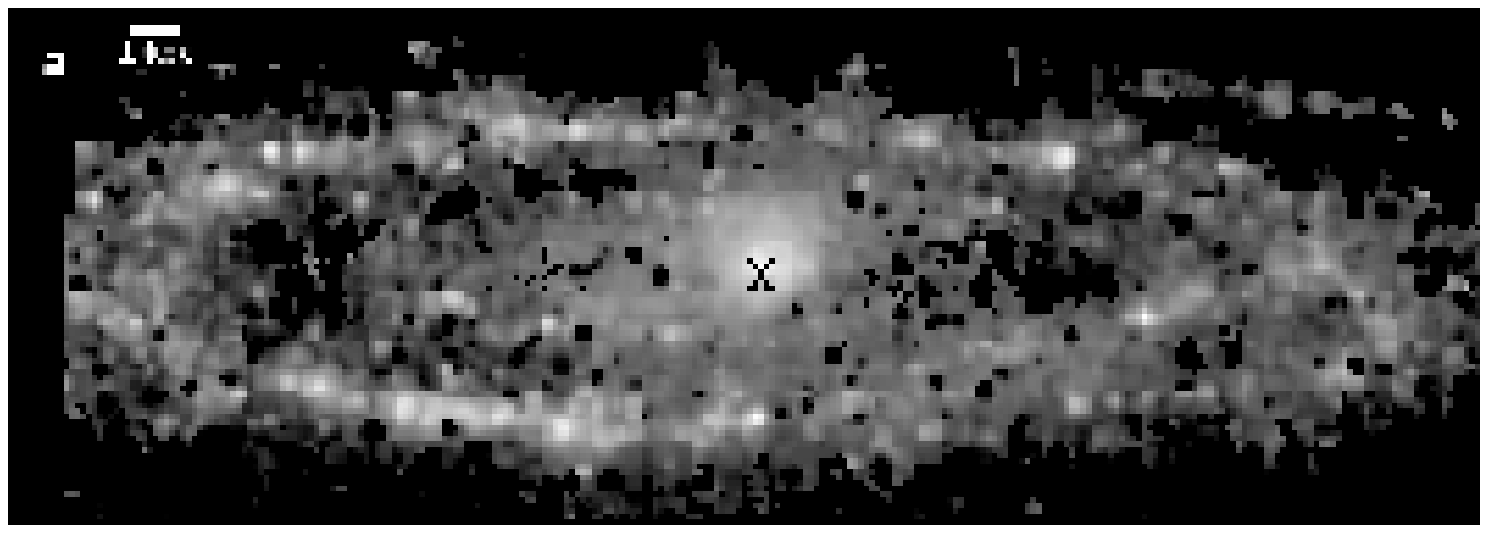}}
\resizebox{13cm}{!}{\includegraphics*{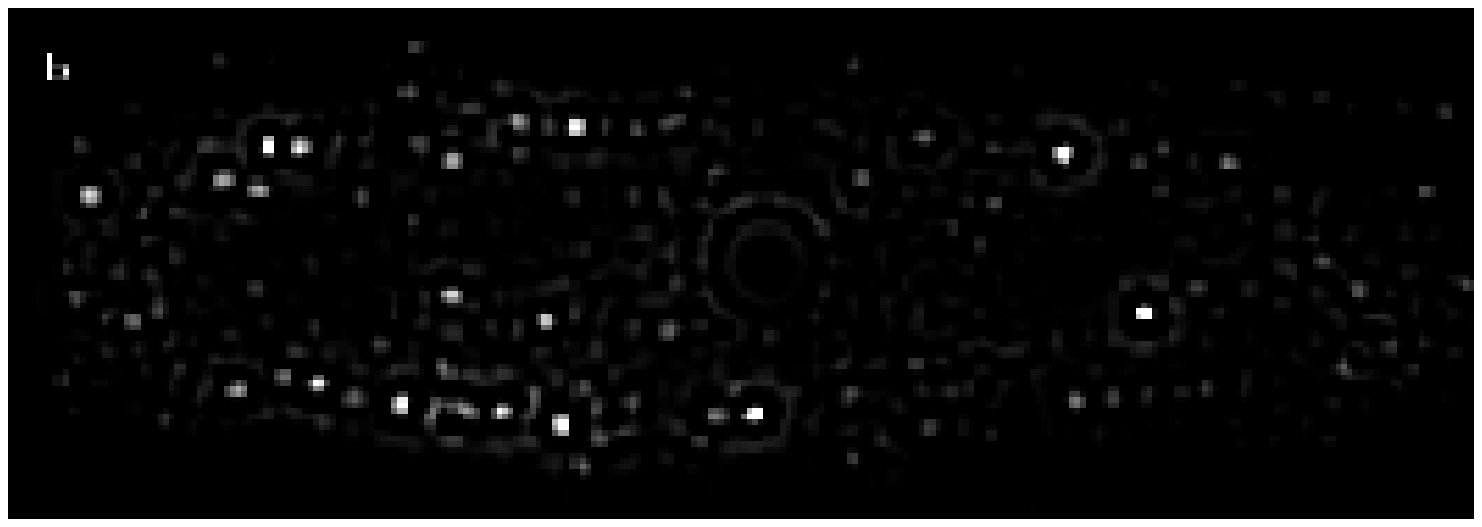}}
\resizebox{13cm}{!}{\includegraphics*{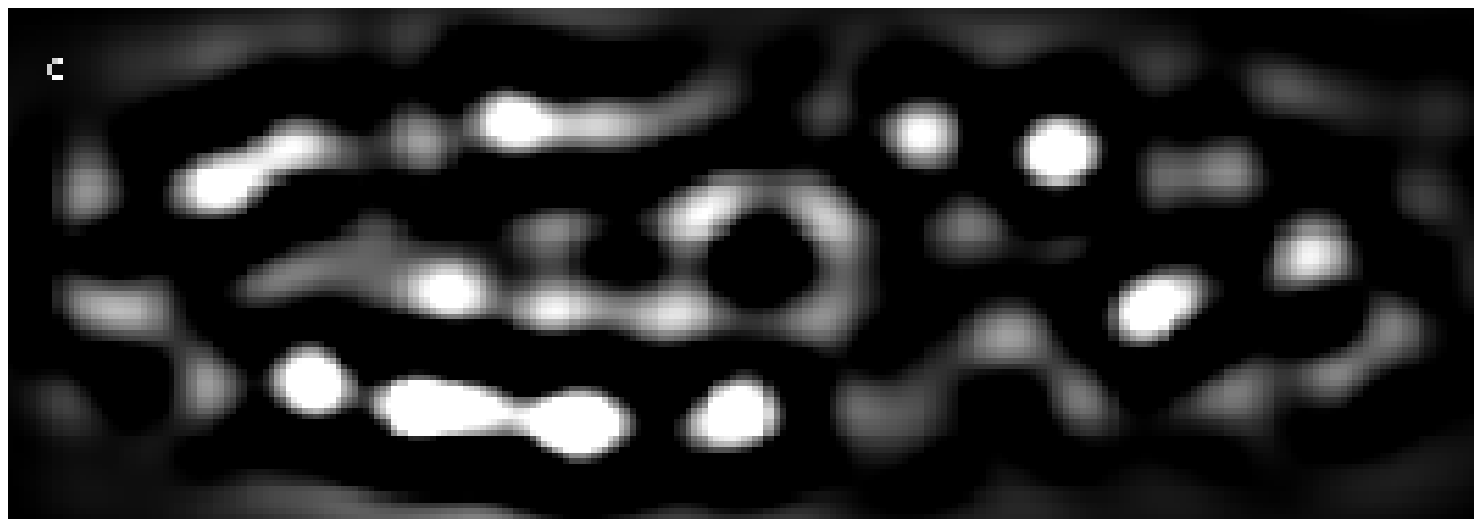}}
\resizebox{13cm}{!}{\includegraphics*{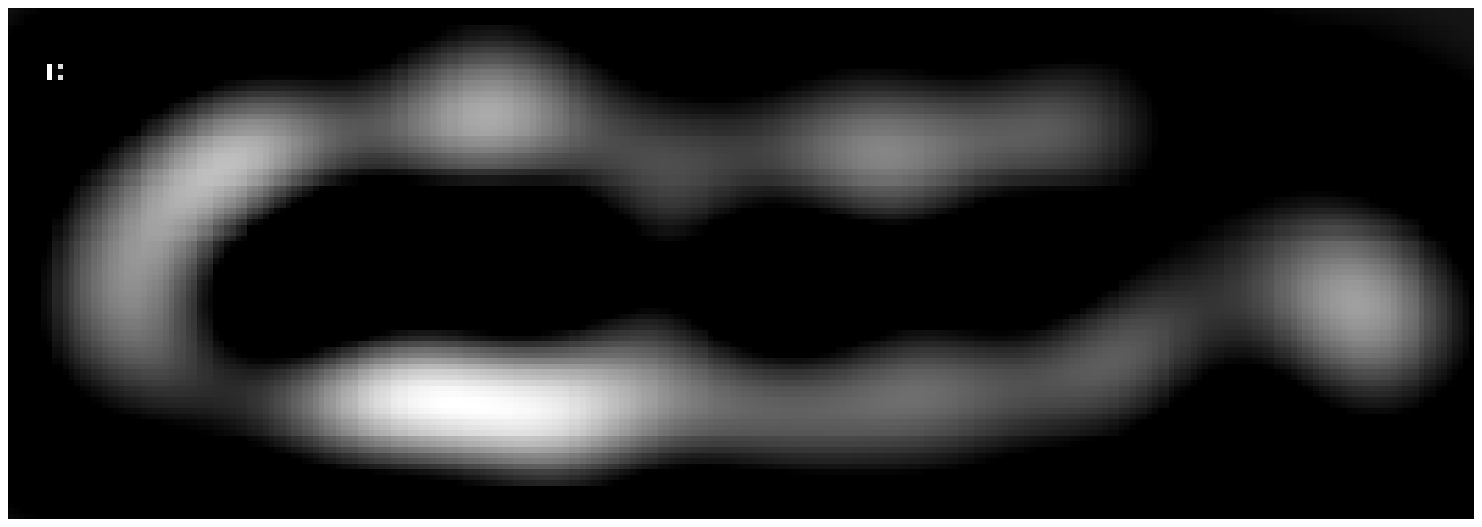}}
\caption[]{Distribution of the de-reddened H$\alpha$ emission ({\it a})  and  the wavelet decomposition for scales  0.4, 1.6, 4.0\,kpc ({\it b} to {\it d}). The central 2\,kpc was subtracted from the H$\alpha$ map before the decomposition. The cross in the H$\alpha$ map indicates the location of the center.  }
\label{fig:wave}
\end{center}
\end{figure*}

To investigate the  physical properties of different phases of the interstellar medium as a function of the size of emitting regions, wavelet transformation is an ideal tool. 
We use the {\it Pet Hat} wavelet \citep[see ][]{Frick_etal_01,Tabatabaei_1_07} to decompose the emissions of IR, HI, H$2$, HI+2H$2$, and de-reddened  H$\alpha$ into 10 spatial scales starting at 0.4\,kpc (about twice the resolution). 
The central 2\,kpc was subtracted  from all images before the wavelet transformation to prevent a strong influence of the nucleus on the results. As an example, we show the extinction-corrected H$\alpha$ map and the H$\alpha$ emission for 3 different scales in   Fig.~\ref{fig:wave}. On the scale of 0.4\,kpc, the distribution of HII complexes and large HII regions is borne out. The scale of 1.6\,kpc (the typical width of spiral arms) shows connected HII complexes along the arms, and on the scale of 4\,kpc we see the extended  emission from the '10\,kpc ring'. 

\subsection{Wavelet spectra}

The wavelet spectrum, {\it M(a)}, represents the distribution of the
emitting power as function of the scale $a$. The wavelet spectrum
will smoothly increase towards larger scales if most of the 
emission is coming from diffuse structures forming the largest
scales, here up to 25\,kpc. On the other hand, the spectrum will decrease
with increasing scale if compact structures are the dominant source
of emission. The spectra of the IR and gas emission are shown in 
Fig.~\ref{fig:wave1}.

All IR and gas spectra are intermediate between the two cases
described above. Only the spectra of the HI gas and the 160\,$\mu$m
emission generally increase with scale indicating the importance
of diffuse HI and cold dust emission. In addition, the HI spectrum 
exhibits a dominant scale at $a\,\simeq\,4$\,kpc corresponding to the width of 
the '10\,kpc ring', where strong diffuse emission occurs in interarm 
regions. The large width of the HI `ring' is also visible in the
radial profiles in Fig.~\ref{fig:surfir}. The dominant scale of the emission from
warm dust, molecular gas and H$\alpha$ is near 1\,kpc, where
complexes of giant molecular clouds and star forming regions show up. The IR spectra at 24\,$\mu$m and 70\,$\mu$m on scales $a<$\,6\,kpc look most similar indicating that the star forming regions are the main heating sources at both wavelengths.  On the other hand, the effect of the ISRF heating the cold dust is well indicated in the 160\,$\mu$m spectrum  where a general increase towards larger scales is found.   All spectra, apart from that of HI, show a minimum near $a$\,=\,6\,kpc corresponding to the large, weak interarm region inside the `10\,kpc ring'. The spectrum of H$\alpha$ is most similar to that of 70\,$\mu$m, which may explain why the H$\alpha$ emission correlates better with 70\,$\mu$m emission than with that at 24\,$\mu$m (see Sect. 6.2 and Table 6).

The spectrum of the H$\alpha$ emission is 
flat on small scales up to 1.6\,kpc, the width of the spiral arms 
in the H$\alpha$ map. This is understandable as the emission from 
very compact HII regions is unresolved at our resolution and not many large
HII complexes exist especially in the south (see the decomposed map in
Fig.~\ref{fig:wave}b of $a$=\,0.4\,kpc).

\subsection{Wavelet cross-correlations}

\begin{figure*}
\begin{center}
\resizebox{\hsize}{!}{\includegraphics*{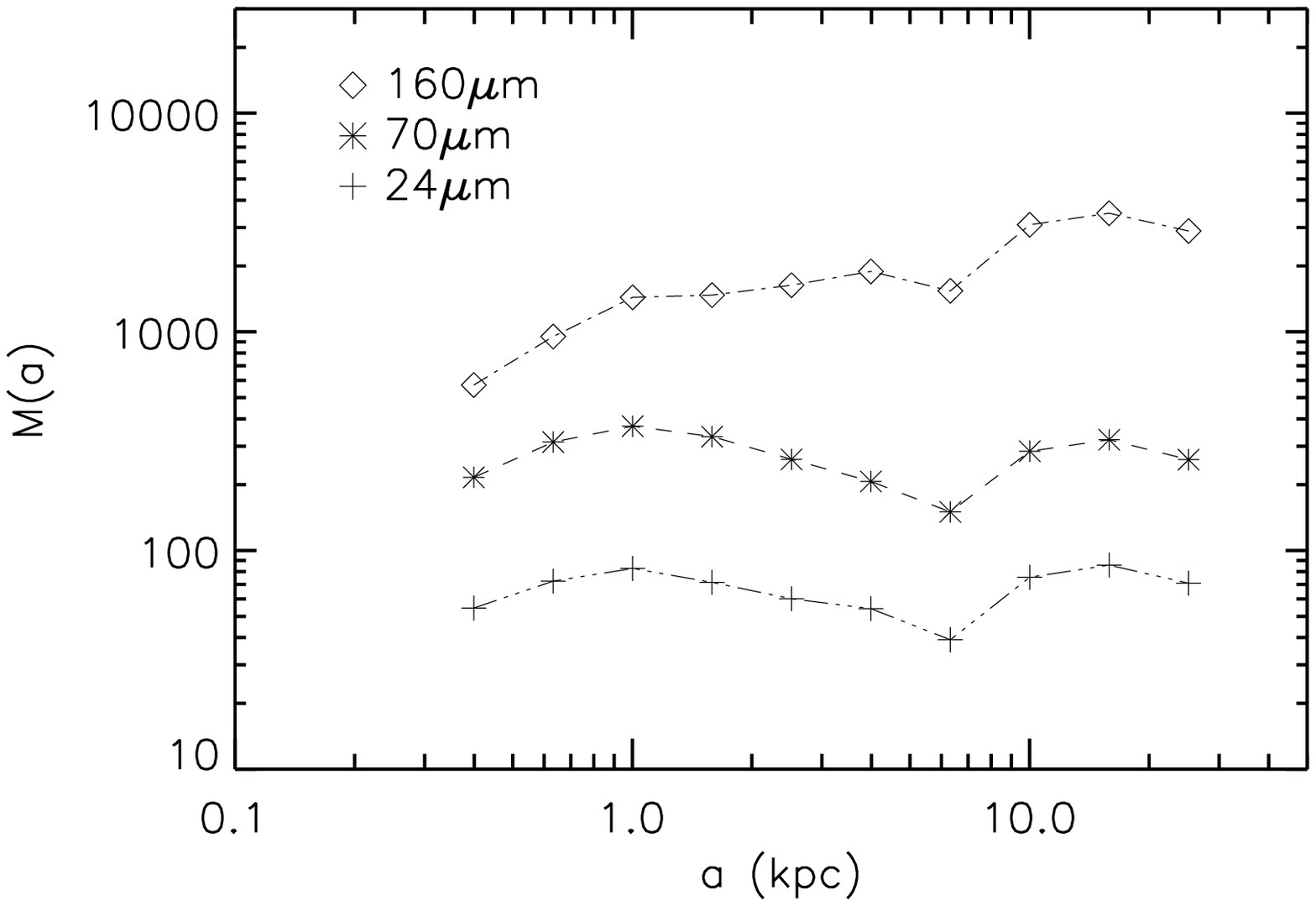}
\includegraphics*{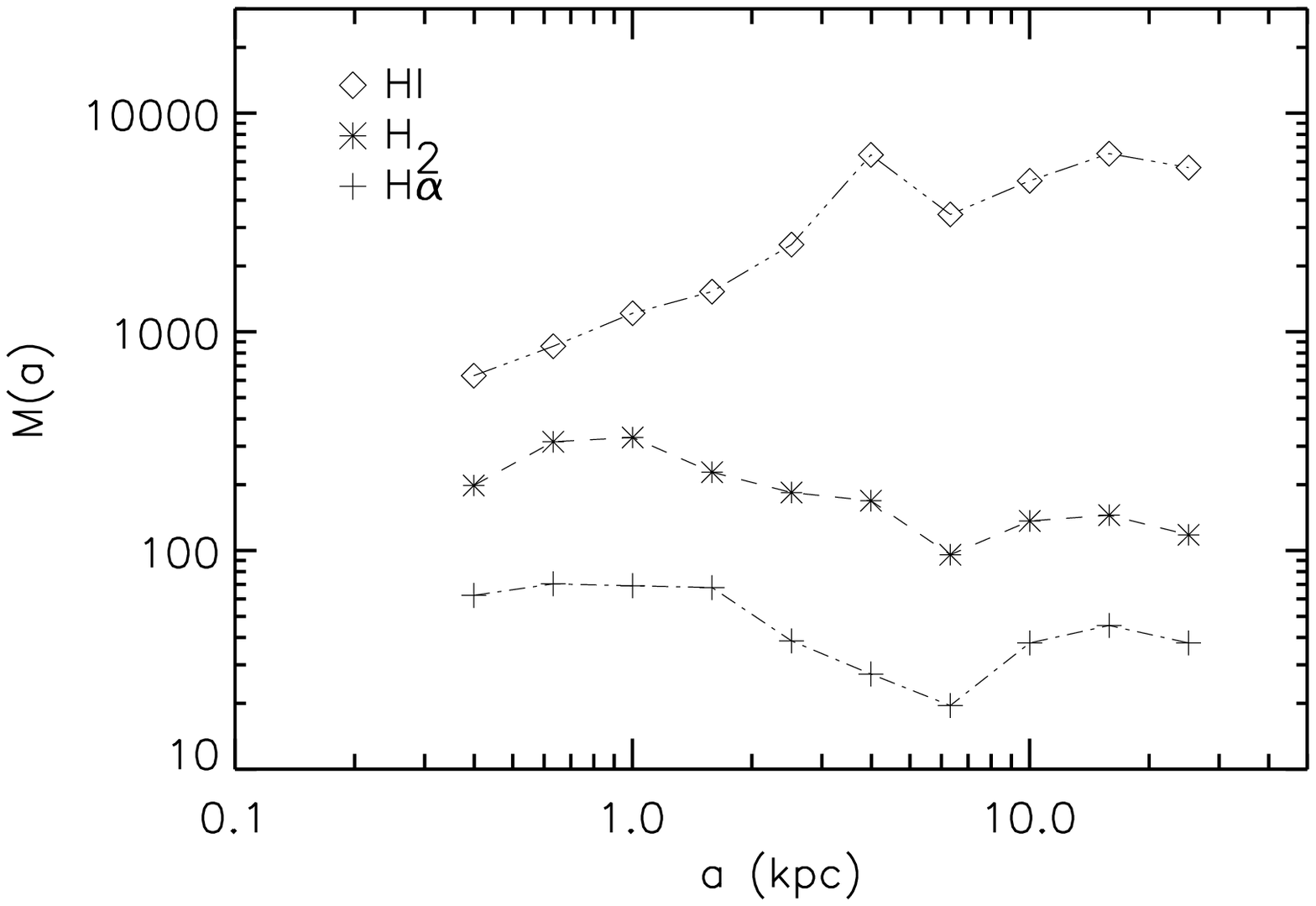}}
\caption[]{Wavelet spectra of MIPS IR ({\it left}) and gas ({\it right}) emission in M~31, shown in arbitrary units. The data points correspond to the scales 0.4, 0.6, 1.0, 1.6, 2.5, 4.0, 6.3, 10.0, 15.9, 25.1\,kpc. }
\label{fig:wave1}
\end{center}
\end{figure*}

\begin{figure*}
\begin{center}
\resizebox{\hsize}{!}{\includegraphics*{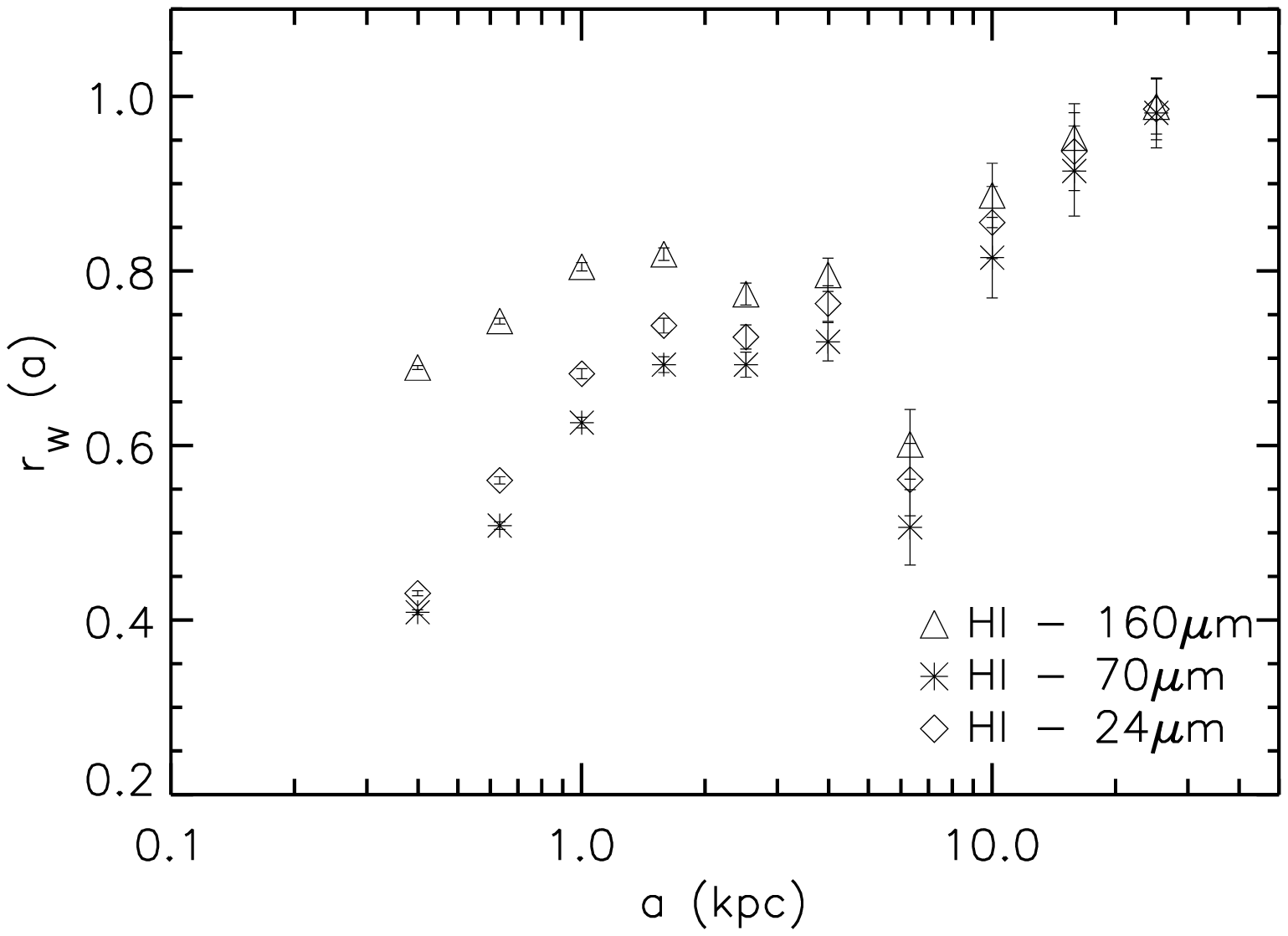}
\includegraphics*{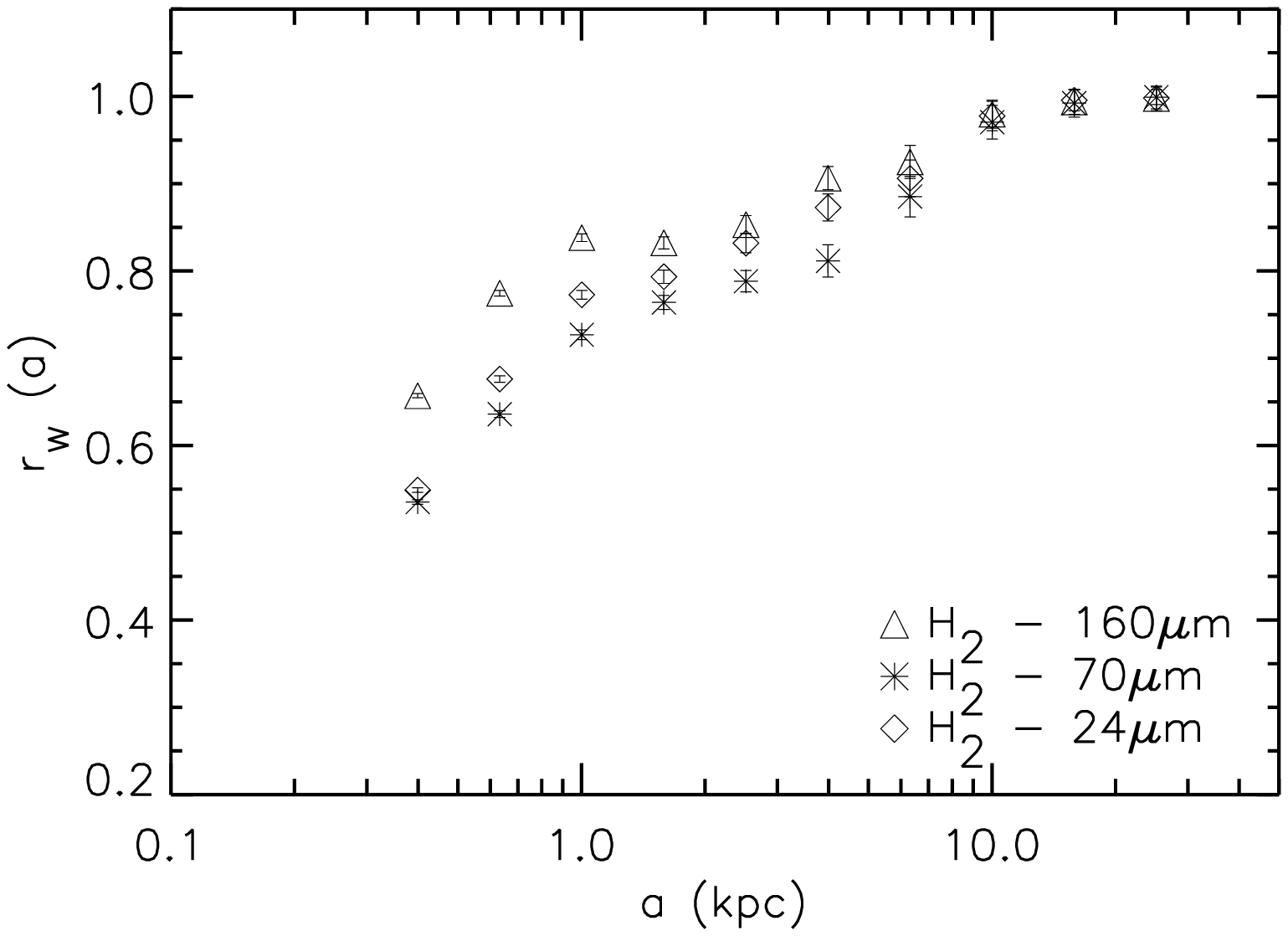}}
\resizebox{\hsize}{!}{\includegraphics*{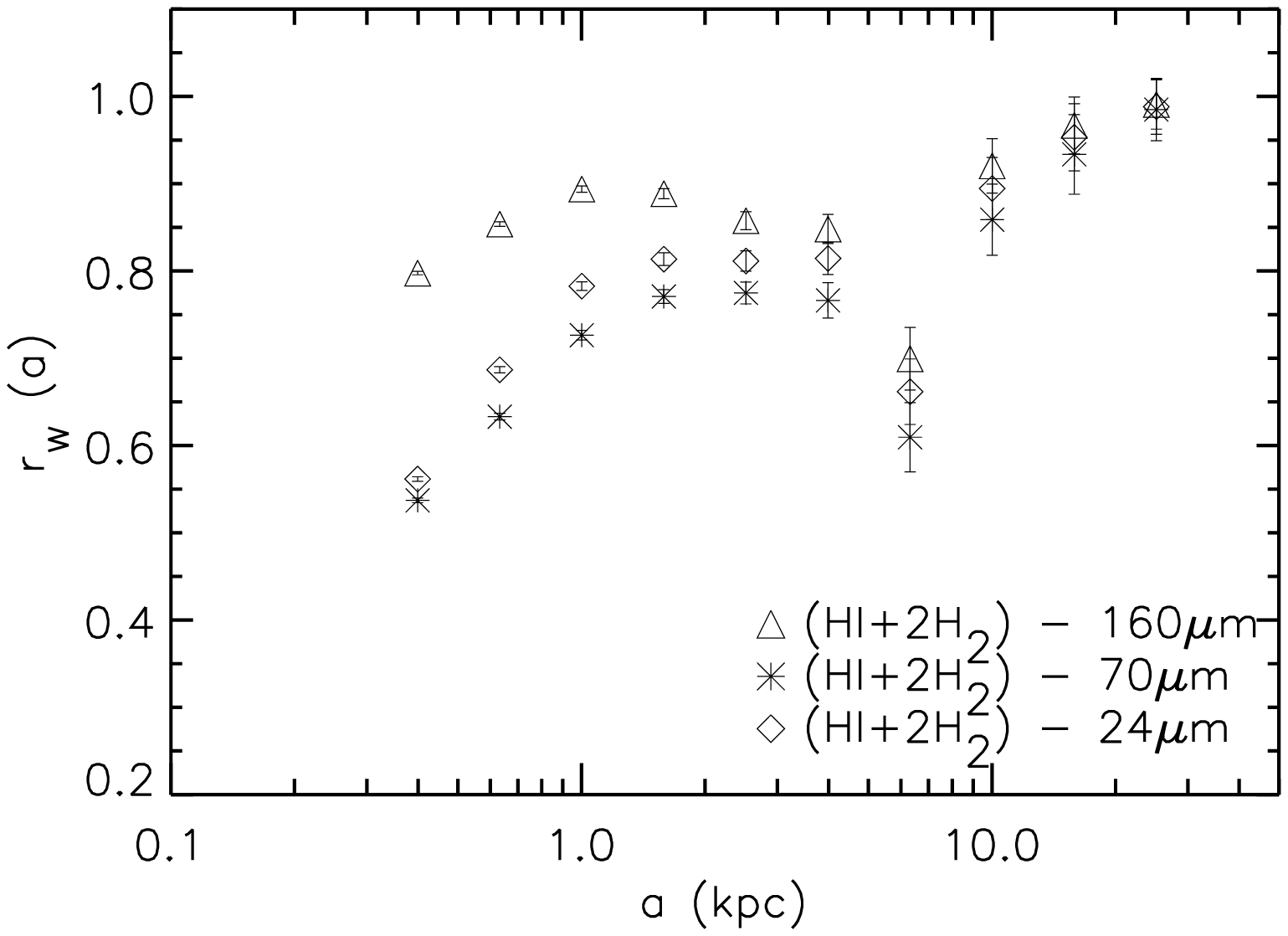} \includegraphics*{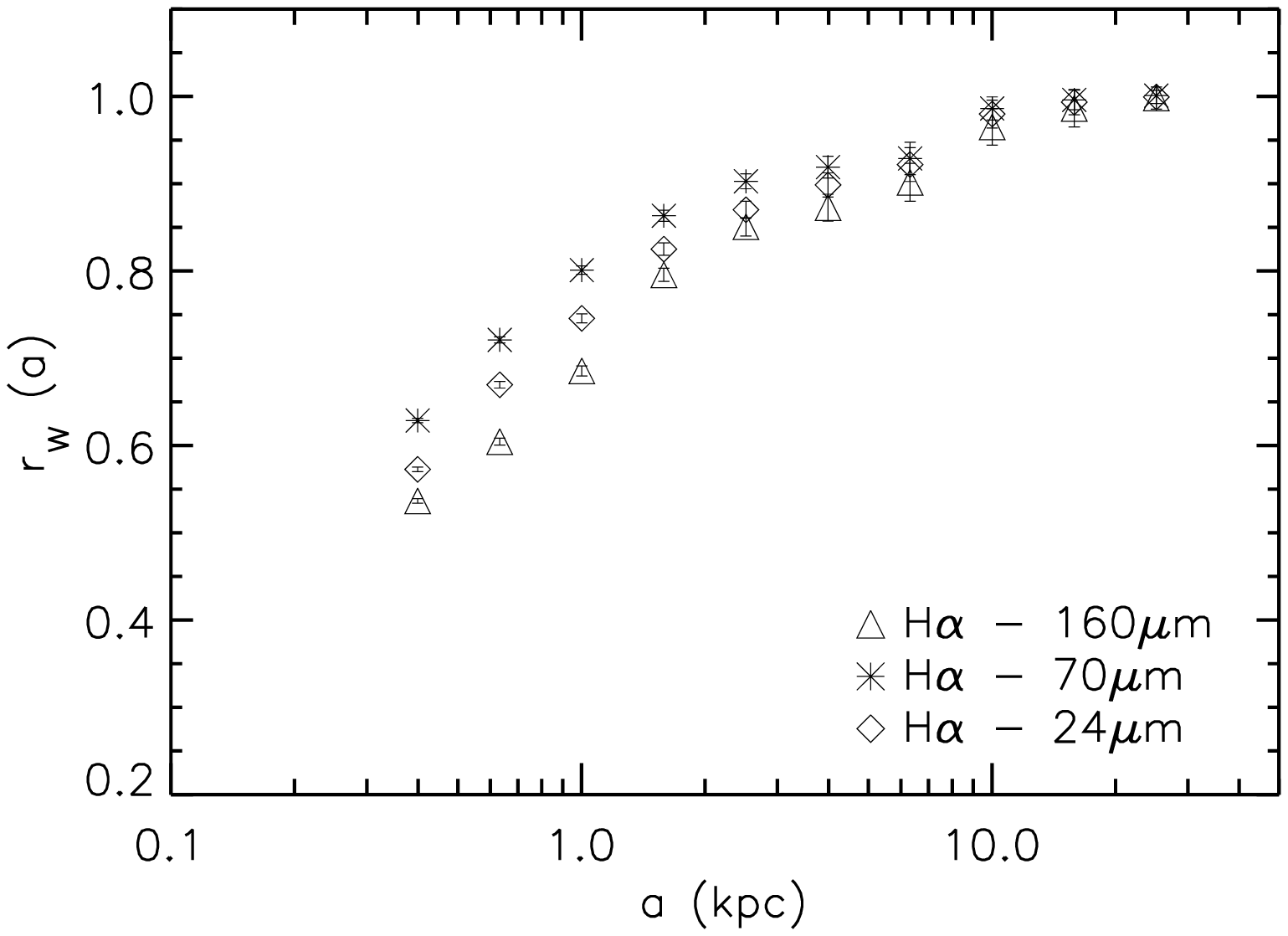}}
\caption[]{Wavelet cross correlations of atomic gas ({\it top-left}), molecular gas ({\it top-right}), and total neutral gas ({\it bottom-left}) with IR emission in M~31. The IR correlation with the ionized gas ({\it bottom-right}) is also shown. The data points correspond to the scales  0.4, 0.6, 1.0, 1.6, 2.5, 4.0, 6.3, 10.0, 15.9, and 25.1\,kpc. }
\label{fig:gas-IR}
\end{center}
\end{figure*}

We derive the cross-correlation coefficients, $r_w(a)$,
for different scales following \cite{Tabatabaei_1_07}. The correlation coefficients are plotted in terms of scale  in Fig.~\ref{fig:gas-IR}. They show that IR emission correlates 
with the emission from different gas phases on most scales. 
In all cases, 
emission from structures on scales larger than 10\,kpc are best 
correlated.  This corresponds to scales of the diameter of the 
'10\,kpc ring' and the over-all structure of the galaxy. On medium 
scales, the weakest correlation occurs between HI and dust emission
on $a=$\,6\,kpc.  This scale includes areas of  significant diffuse HI emission 
where the dust emission is weak  interior to the `10\,kpc ring' (compare also 
Fig.~\ref{fig:wave1}).    
On the smallest scale of 0.4\,kpc, the cold dust emission is best correlated with that of the total neutral gas, while the warm dust emission  at 70\,$\mu$m is best correlated with the ionized gas emission.
Note that on this scale, the 24\,$\mu$m and 70\,$\mu$m (warm dust) emissions hardly correlate with HI ($r_{w(a)}<0.5$) because only a small fraction of the HI emission occurs on this scale (see Fig.~\ref{fig:wave1}).  Furthermore, the  coefficients of the 70\,$\mu$m--H$\alpha$ correlation are higher than those of the 70\,$\mu$m--neutral gas correlation on scales $a\,<\,6.3$\,kpc.


\section{Classical correlations between dust and gas}

The wavelet cross-correlations for different scales in 
Fig.~\ref{fig:gas-IR} show on which scales the distributions of the 
various types of emission are significantly correlated. However, because the scale maps are normalized and information about absolute intensities is lost, they cannot be used to find quantitative relations between components of the ISM. Hence, to obtain numerical equations relating two distributions, we need classical correlations. Classical cross-correlations contain all scales that exist in
a distribution. For example, the high-intensity points of the H$\alpha$-70$\mu$m correlation in Fig.~\ref{fig:ha70} represent high-emission peaks on small scales in the spiral arms (compare
Fig.~\ref{fig:wave}b), whereas low-intensity points represent weak emission
around and between the arms on larger scales (compare Fig.~\ref{fig:wave}d).
The correlation coefficient of 79\% is a mean of all scales,
consistent with Fig.~\ref{fig:gas-IR}.

We made
pixel-to-pixel correlations between the distributions of
$\tau_{{\rm H}_{\alpha}}$ and H$_2$, HI, total gas as well as between 
de-reddened H$\alpha$ and 24\,$\mu$m, 70\,$\mu$m, and 160\,$\mu$m. We 
restricted the comparisons to radii where all data sets are complete, 
$R<50\arcmin$ (or 11.4\,kpc), and 
to intensities above 2\,$\times$\, rms noise. To reduce the influence 
of the gradient in the gas-to-dust ratio (see Sect. 4.2), 
we calculated correlations for two radial ranges: $0\arcmin<R<30\arcmin$ 
and $30\arcmin<R<50\arcmin$. We obtained sets of independent data points, 
i.e. a beam area overlap of $<5\%$, by choosing pixels spaced 
by more than 1.67$\times$ the beamwidth. Since the correlated variables 
are not directly depending on each other, we fitted a power law 
to the bisector in each case \citep{Isobe}. 

We also calculated the correlation coefficient, $r_c$, to show how well two 
components are correlated, and the student-t test to indicate the statistical 
significance of the fit. For a number of independent points of n\,$>100$, the fit 
is significant at the 3$\sigma$ level if $t>3$.  Errors in intercept $a_c$ 
and slope $b$ of the bisector are standard deviations (1\,$\sigma$).

\begin{figure}
\begin{center}
\resizebox{8cm}{!}{\includegraphics[angle=-90]{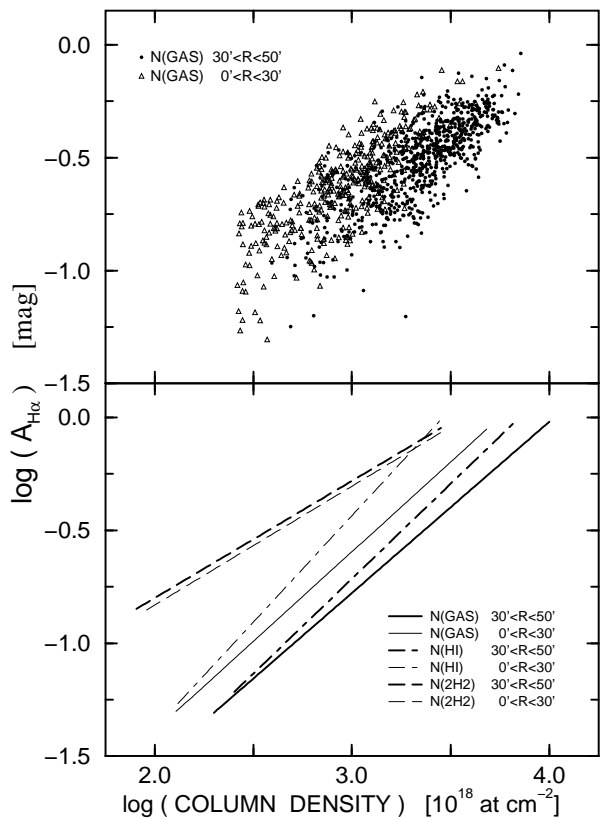}}
\caption[]{Classical cross-correlations between gas column densities and  dust extinction A$_{\rm H\alpha}$  in the radial ranges $0\arcmin<R<30\arcmin$ and $30\arcmin<R<50\arcmin$ (or 6.8~kpc~$<R<$~11.4~kpc). Only independent data points  (separated by 1.67$\times$ beamwidth) with values above 2$\times$ rms noise    were used.  {\it Top}. As an example, the scatter plot between N(gas) and A$_{\rm H\alpha}$. {\it Bottom}. Power-law fits to the various correlations given in  Table 5. Thick lines: $30\arcmin<R<50\arcmin$; thin lines: $0\arcmin<R<30\arcmin$. The shift between these intervals of the fits to the N(HI)-A$_{\rm H\alpha}$ and   N(gas)-A$_{\rm H\alpha}$ correlations is due to the radial increase in the atomic gas-to-dust ratio.}
\label{fig:class1}
\end{center}
\end{figure}

We first discuss the correlations between the neutral gas and dust extinction, scaled 
from $\tau_{{\rm H}_{\alpha}}$. Then we 
investigate the relationships between the emissions from dust 
and ionized gas. The results are 
given in Tables~\ref{table:dustgas1} and \ref{table:dustgas2}, and examples of correlation plots are shown in Figs.~\ref{fig:class1} to \ref{fig:sfr1}.

\subsection{Correlation between neutral gas and dust extinction}

In search of a general relationship between neutral gas and dust
extinction, a number of authors employed scatter plots between
gas column densities and extinction, optical depth or FIR surface
brightness \citep[e.g. ][]{Savage_78,Boulanger,Walterbos_88,Xu_96,Neininger_98,Nieten}. They obtained nearly linear relationships between these quantities. As the studies 
on M~31 have various shortcomings (lower limits for extinction, H$_2$ 
data not included and/or low angular resolution), we calculated 
classical correlations between the distribution of  dust extinction A$_{\rm H\alpha}$= 1.086\,$\tau_{\rm eff}$ and those of N(HI), N(2H$_2$) and N(gas) at our resolution of 45$\arcsec$. 
As the correlations are restricted to gas column densities above 2$\times$ 
the rms noise, values of $\tau_{{\rm H}_{\alpha}} <$ 0.04 are not included (see 
upper panel of Fig.~\ref{fig:class1}). The bisector fits given in Table 5 are 
plotted in the bottom panel of Fig.~\ref{fig:class1}. 

The relationships between A$_{\rm H\alpha}$ and N(2H$_2$) for the two radial
ranges are the same within errors, so the two areas can be combined.
With a correlation coefficient of $r_c \simeq$\,0.6, the correlation is not
very good, indicating that only a small part of the extinction is caused by dust in molecular clouds. This is not surprising
in view of the low molecular gas fraction in M~31 (see lower panels of 
Fig.~\ref{fig:surfir}) and the small area filling factor of the molecular gas compared to that of the atomic gas.
\begin{table*}
\begin{center}
%\label{table:5}
\caption{Power-law relations and correlation coefficients $r_c$ between  dust extinction and gas components. Ordinary least-squares fits of bisector log(Y)=$a_c+b$\,log(X) through n pairs of (logX, logY), where n is the number of independent points (Isobe et al. 1990); $t$ is the student-t test.}
\begin{tabular}{ l l l l l l l l} 
\hline
  X    &    Y &     $R$    &  \,\,\,\,\,\,     $a_c$  &\,\,\,\,\,\, $b$ &    n  & \,\,\,\,\,\, $r_c$  & $t$\\
 ($10^{18}$ at~cm$^{-2}$)& (mag) & ($\arcmin$) &&&&& \\
\hline 
\hline
N(2H$_2$) & A$_{\rm H\alpha}$ & 0-30 &   -1.88$\pm$0.06 &  0.53$\pm$0.03 &   207&   0.60$\pm$0.06 &   11\\
   &      & 30-50 &   -1.83$\pm$0.04 &  0.52$\pm$0.02 &   610  & 0.64$\pm$0.03 &   20 \\
        &     &   0-50 &   -1.85$\pm$0.03 &  0.52$\pm$0.01 &   817 &  0.63$\pm$0.03 &   23\\
& & & & & & &\\

N(HI) &  A$_{\rm H\alpha}$ &  0-30 &  -3.29$\pm$0.11 &  0.94$\pm$0.04 &   354 &  0.69$\pm$0.04&    18\\
      &        & 30-50 &  -3.26$\pm$0.07 &  0.84$\pm$0.02 &   768 &  0.72$\pm$0.03 &   29\\  
&&&&&&&\\   
N(gas) & A$_{\rm H\alpha}$ & 0-30 &   -2.97$\pm$0.08 &  0.79$\pm$0.03 &   350 &  0.80$\pm$0.03 &   25\\
      &     & 30-50 &   -3.05$\pm$0.06 &  0.76$\pm$0.02  &  766 &  0.77$\pm$0.02  &  33\\  
\hline
\end{tabular}
\label{table:dustgas1}
\end{center}
\end{table*}

The correlations between A$_{\rm H\alpha}$ and N(HI) are indeed better
($r_c \simeq\,0.7$) than those between A$_{\rm H\alpha}$ and N(2H$_2$), but the
relationships for the two radial intervals are not the same. Although
both are nearly linear (power-law exponent $b \simeq$\,0.9), their power laws are shifted (see Fig.~\ref{fig:class1}) in the sense that the values of A$_{\rm H\alpha}$ in $R=30\arcmin- 50\arcmin$ are about a factor of 2 lower than those inside $R=30\arcmin$.
This difference is caused by the radial decrease of $\tau_{{\rm H}_{\alpha}}$/N(HI) 
discussed in Sect. 4.2. The variation of this ratio within each of the
radial intervals contributes to the spread in the scatter plots and
reduces the correlation coefficients.

The correlations between total gas N(gas) and A$_{\rm H\alpha}$ are best
($r_c\simeq\,0.8$), as A$_{\rm H\alpha}$ represents dust mixed with both HI and H$_2$. They are close to linear ($b \simeq$\,0.8) and differ by nearly
a factor 2 in A$_{\rm H\alpha}$. The scatter plots for the two intervals are shown in the
upper panel of Fig.~\ref{fig:class1}. In linear plots both power-law fits are going 
through zero, suggesting that the dust causing the extinction and 
neutral gas are mixed down to very low densities.

Interestingly, extinction (or dust opacity) is proportional to the square root of N(2H$_2$), while it is about linearly related to the atomic gas density.  This is due to the quadratic dependence of N(2H$_2$) on N(HI) in M~31 observed by \citep[][]{Nieten}. This dependence is expected if in cool, dense, and dusty HI clouds the formation and destruction rates of H$_2$ are balanced \citep{Reach}.

\subsection{Correlation between ionized gas and dust}

Because the emission from ionized gas is a good tracer of the 
present-day star formation rate and massive stars both heat the 
dust and ionize the gas, a correlation between the emissions from
warm dust and ionized gas is expected. Relationships between the emission at 24\,$\mu$m
and Pa$\alpha$ or H$\alpha$ emission from HII regions in nearby galaxies as well as 
relationships between global luminosities of galaxies have been reported 
\citep[see ][ and references therein]{Kennicutt_09}. 

For M~31, the correlation between the emission from dust and ionized gas was first
tested by \cite{Hoernes_etal_98}, who found a good, nearly linear 
correlation between warm dust emission and free-free radio emission 
for the radial range $30\arcmin <R<90\arcmin$, using HIRAS data and multi-wavelength 
radio data. Here we correlate the extinction-corrected H$\alpha$
emission presented in Fig.~\ref{fig:wave}a with dust emission in the MIPS maps. 

\begin{figure}
\begin{center}
\resizebox{8cm}{!}{\includegraphics[angle=-90]{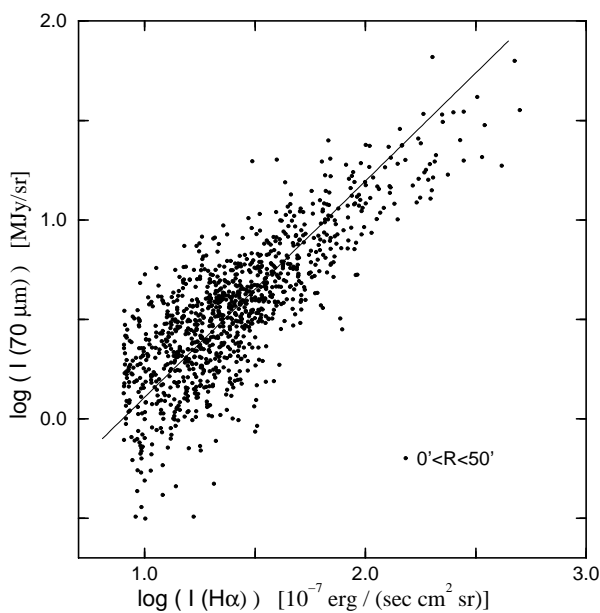}}
\caption[]{Scatter plot between the surface brightnesses of ionized gas
         and dust emission at 70\,$\mu$m for the radial range 0$\arcmin$-50$\arcmin$. Only independent data points (separated by 1.67$\times$ beamwidth) with values above 2$\times$ rms noise are included. The line shows the power-law fit given in Table 6,
         which has an exponent close to 1. In a linear frame, this fit 
         goes through the zero point of the plot.  }
\label{fig:ha70}
\end{center}
\end{figure}

The wavelet correlations in Fig.~\ref{fig:gas-IR} (bottom-right) show 
that in M~31 H$\alpha$ emission is best correlated with dust emission at 
70\,$\mu$m. This suggests that of the MIPS bands, the 70\,$\mu$m emission could best be used as the tracer of present-day star formation, making a numerical 
relation between the emissions at 70\,$\mu$m and H$\alpha$ of interest. 
Table~\ref{table:dustgas2} gives the bisector fits for the two radial ranges, which are 
very similar. Therefore, we present this correlation in Fig.~\ref{fig:ha70} for 
the entire radial range of $0\arcmin <R<50\arcmin$. The power-law fit for this radial 
interval is
\begin{eqnarray}
\nonumber
{\rm log}(I_{70})=(-0.98 \pm 0.02)+(1.09 \pm 0.02)\,{\rm log}(I_{{\rm H}\alpha}), \nonumber
\end{eqnarray} 
where $I_{70}$ is in MJy/sr and $I_{{\rm H}\alpha}$ in 10$^{-7}$\,erg~s$^{-1}$~cm$^{-2}$~sr$^{-1}$.
The correlation is quite good ($r_c \simeq 0.8$) and nearly linear. In a linear plot the power-law 
fit goes through zero suggesting that the correlation is also valid
for the lowest intensities. The good correlation indicates that the
heating sources that power dust emission at 70\,$\mu$m and ionize the gas
must indeed be largely the same. 
 
Naturally, H$\alpha$ emission is less correlated with the emission from cold dust at 160\,$\mu$m than with emission from warm dust seen at the shorter wavelengths. This is 
especially so at $R<30\arcmin$ where the radial profiles differ most (see Fig.~\ref{fig:surfir}). Moreover, the relation between the emission from cold dust and H$\alpha$ is non-linear (see the bisector slope $b$ in Table~\ref{table:dustgas2}).

Table~\ref{table:dustgas2}  shows that the correlations with 24\,$\mu$m  are slightly
worse than those with 70\,$\mu$m. In contrast, in M~33 the 24\,$\mu$m--H$\alpha$ correlation is better than the 70\,$\mu$m--H$\alpha$ correlation \citep{Tabatabaei_1_07}. This may suggest that in early-type galaxies like M~31 the contribution from evolved AGB stars to the 24\,$\mu$m
emission is larger than in late-type galaxies like M~33.
A significant stellar contribution to the 24\,$\mu$m emission from M~31 is also 
indicated by the enhancement of the 24\,$\mu$m-to-70\,$\mu$m intensity ratio in inter-arm regions
where the radiation field is weak (Fig.~\ref{fig:ratio}). 

Across M~31, the 24\,$\mu$m emission is linearly proportional to the extinction-corrected
H$\alpha$ emission ($b = 0.98 \pm 0.02$). A linear relationship was also found between the
luminosities at 24\,$\mu$m and extinction-corrected Pa$\alpha$ of HII regions in M~51 \citep{Calzetti_05} and between the luminosities at 24\,$\mu$m and extinction-corrected H$\alpha$
of HII regions in M~81 \citep{Perez}. Comparing the 24\,$\mu$m luminosities 
and corrected H$\alpha$ luminosities of HII regions in 6 nearby galaxies (including
M~51 and M~81), \cite{Relano_07} obtained a somewhat steeper power law with index
$1.21 \pm 0.01$, in agreement with the index for global luminosities of galaxies \citep[see
also ][]{Calzetti_07}. Thus, while the L$_{24}$ - L$_{\rm H\alpha}$ relationship
is linear within a single galaxy,  the relationships for HII regions in a sample of 
galaxies and for global luminosities are non-linear. According to \cite{Kennicutt_09}, the steepening is due to variations between galaxies in the contribution from 
evolved, non-ionizing stars to the heating of the dust that emits at 24\,$\mu$m.

\begin{table*}
\begin{center}
\caption{Power-law relations and correlation coefficients $r_c$  between  the emission from dust and ionized gas. Ordinary least-squares  fits of bisector log(Y)=$a_c+b$\,log(X) through n pairs of (logX, logY), where n is the number of independent points (Isobe et al. 1990); $t$ is the student-t test.}
\begin{tabular}{ l l l l l l l l}
\hline
  X    &    Y &     $R$   &  \,\,\,\,\,\,     $a_c$  & \,\,\,\,\,\,$b$ &    n  &  \,\,\,\,\,\,$r_c$  & $t$\\
($10^{-7}$ erg\,s$^{-1}$\,cm$^{-2}$\,sr$^{-1}$) & (MJy/sr) & ($\arcmin$)& & &  & &\\
\hline
\hline
 I$_{{\rm H}\alpha}$ &I$_{24\mu{\rm m}}$ &   0-30   & -1.76$\pm$0.05 &  0.94$\pm$0.03 &   417 &  0.79$\pm$0.03 &   27\\  
            &  & 30-50 &  -1.75$\pm$0.05  & 1.00$\pm$0.03  &  677  & 0.75$\pm$0.03 &   30\\     
             &  &  0-50  &  -1.75$\pm$0.05  &0.98$\pm$0.02  & 1094 &  0.76$\pm$0.02  &  38\\
& & & & & & &\\
I$_{{\rm H}\alpha}$ & I$_{70\mu{\rm m}}$ &   0-30   & -1.04$\pm$0.03 & 1.10$\pm$0.03  &  412  & 0.83$\pm$0.03 &   30\\
           &   &  30-50    &-0.96$\pm$0.03   &1.09$\pm$0.03  &  677  & 0.78$\pm$0.02  &  32  \\ 
            &    &  0-50    &-0.98$\pm$0.02  &1.09$\pm$0.02   & 1089  & 0.79$\pm$0.02   & 43 \\
& & & & & & &\\
 I$_{{\rm H}\alpha}$& I$_{160\mu{\rm m}}$ & 0-30 & 0.26$\pm$0.05 & 0.76$\pm$0.03 &  417 & 0.54$\pm$0.04 & 13\\
            &  &  30-50  &0.47$\pm$0.04  &0.72$\pm$0.02 &  678  & 0.73$\pm$0.03  &  28  \\ 
\hline
\end{tabular}
\label{table:dustgas2}
\end{center}
\end{table*}

\section{Star formation rate and efficiency}

Over the last 40 years many authors have studied the relationship 
between the rate of star formation and gas density in M~31 by 
comparing the number surface density of massive young stars or 
of HII regions with that of HI \citep{Berkhuijsen_77,Tenjes,Unwin,Nakai_84}. 
They found power-law exponents
near 2 as was also obtained for the solar neighborhood by \cite{Schmidt}, who first proposed this relationship with HI volume density. \cite{Kennicutt_98} showed that a similar relationship is expected between SFR and gas column densities. The early studies
suffered from the effects of dust absorption and could not consider
molecular gas \citep[apart from ][]{Tenjes}. As the necessary 
data are now available, we again address this issue. 

We compared the distribution of the H$\alpha$ emission corrected 
for dust attenuation (see Fig.~\ref{fig:wave}a) with those of HI, H$_2$ and total
gas. The corrected H$\alpha$ emission is a good measure for the
present-day star formation rate (SFR) that we first estimate for 
the total area observed using the relation of \citep{Kennicutt}:
\begin{equation}
{\rm SFR}({\rm M}_{\odot} {\rm yr}^{-1})=\, \frac{L_{\rm H_{\alpha}} ({\rm erg\,s^{-1}})}{1.26 \times 10^{41} }  \ ,
\end{equation} 
where $L_{\rm H_{\alpha}}$  is the ${\rm H_{\alpha}}$ luminosity. In an area of $110.0^{\arcmin} \times 38.5^{\arcmin}$ ($R<17$\,kpc), the luminosity of the de-reddened ${\rm H_{\alpha}}$ emission is  $L_{\rm H_{\alpha}}\,=\,4.75 \times\,10^{40}$\,erg\,s$^{-1}$ or $L_{\rm H_{\alpha}}\,=\,1.7 \times 10^7\,L_{\odot}$ for the distance 
to M~31 of 780\,kpc (see Table~1), giving SFR\,=\,0.38\,${\rm M}_{\odot} {\rm yr}^{-1}$. 
However, this value is rather uncertain for two reasons. First, the contribution from
the inner disk ($R <\,25\arcmin$, nearly 6\,kpc) is overestimated because in this area the number
of ionizing stars is low and the gas must be mainly heated by other sources (see Sect. 7.1).
Second, our H$\alpha$ map is limited to about 55$\arcmin$ (12.5\,kpc) along the major axis, so
some of the emission between $R$\,=\,12.5\,kpc and $R$\,=\,17\,kpc is missing. Subtracting the
luminosity from the area $R <$\,6\,kpc gives a lower limit to the SFR of 0.27\,M$_{\odot}$\,yr$^{-1}$ for the radial range 6\,$< R <$\,17\,kpc. Earlier estimates of the recent SFR for a larger
part of the disk indicated 0.35 - 1\,M$_{\odot}$\,yr$^{-1}$ \citep{Walterbos_94,Williams,Barmby}. Recently, \cite{Kang} derived a SFR of 0.43\,M$_{\odot}$\,yr$^{-1}$
(for metallicity 2.5\,$\times$\,Solar) from UV observations of young star forming regions
($<$10\,Myr) within 120$\arcmin$ from the center ($R <$\,27\,kpc). Their Fig.~13 suggests that about 20\% of this SFR is coming from $R >$ 17\,kpc and a negligible amount from $R <$\,6\,kpc.
So for the range 6\,kpc\,$< R <$\,17\,kpc they find a SFR of about 0.34 M$_{\odot}$\,yr$^{-1}$,
which is consistent with our lower limit of 0.27\,M$_{\odot}$\,yr$^{-1}$.

A SFR of 0.3\,M$_{\odot}$\,yr$^{-1}$ yields a mean face-on surface density of 
$\Sigma_{\rm SFR}$\,=\,0.4 M$_{\odot}$\,Gyr$^{-1}$\,pc$^{-2}$ between $R$\,=\,6\,kpc and $R$\,=\,17\,kpc. This is about 6 times lower than the value of $\Sigma_{\rm SFR}$=\,2.3\,M$_{\odot}$\,Gyr$^{-1}$\,pc$^{-2}$ that \cite{Verley_09} obtained for the disk of M~33 ($R <$\,7kpc), also using de-reddened H$\alpha$ data.
 
We can also calculate the star formation efficiency between $R$=\,6\,kpc and $R$=\,17\,kpc in M31. The total molecular gas mass in the entire area of $R<17$\,kpc  in M~31 is 
M(H$_2$)\,=\,3.6\,$\times\,10^8$\,M$_{\odot}$ \citep{Nieten}  and that in the area
6\,kpc\,$<\,R\,<$17\,kpc is M(H$_2$)\,=\,2.9\,$\times\,10^8$\,M$_{\odot}$. Hence the 
star formation efficiency SFE\,=\,SFR/M(H$_2$) between 6\,kpc and 17\,kpc radius is 
SFE\,=\,0.9\,Gyr$^{-1}$.  It is equivalent to a molecular depletion 
time scale of 1.1\,Gyr.  {\bf Hence, the disk of M~31 is about three times less
efficient in forming young massive stars than the northern part of the disk of M~33 \citep{Gardan}. }

\subsection{Star formation rate in the `10~kpc ring'}
The radial distributions of H$\alpha$ emission in the 
bottom panels of Fig.~\ref{fig:surfir} show a steep decrease from the center
to $R \simeq$\,2\,kpc, followed by a shallower decrease to a minimum
near $R$\,=\,6\,kpc. In the radial profile of $\Sigma_{\rm SFR}$, shown in the
upper panel of Fig.~\ref{fig:sfr2} for the total area, the inner arms at
$R$\,=\,2.5\,kpc and $R$\,=\,5.5\,kpc are only visible as little wiggles
superimposed onto a high background. Clearly, the starforming
regions in these arms hardly contribute to the ionization of the
gas at $R <$\,6\,kpc. \cite{Devereux_etal_94b} noted that at these
radii the H$\alpha$ emission is filamentary and unlike that in
starforming regions, and since not many young, massive, ionizing
stars are found interior to the {\bf `10\,kpc ring'} \citep{Berkhuijsen_89,Tenjes,Kang} the gas must be ionized by other sources. Naturally, the same holds for the
heating of the warm dust, the emission of which also strongly increases
towards the center. In an extensive discussion, \cite{Devereux_etal_94b} concluded that a collision with another galaxy in the past
may explain the ionization of the gas and the heating of the dust
as well as several other peculiarities (e.g. the double nucleus)
in the inner disk of M~31 \citep[see also ][]{Block}. We note that the UV emission may also
be influenced by this event, because it shows a similarly steep
increase towards the center as the H$\alpha$ emission. Furthermore,
all radial profiles that increase towards the center are anti-
correlated with the radial profiles of HI and total gas (see Fig.~\ref{fig:surfir},
bottom panels). This leads to apparent deviations in the Kennicutt-Schmidt
law in the inner disk if $\Sigma_{\rm SFR}$ is calculated from the usual
star formation tracers \citep[see ][]{Boissier,Yin}.  

If massive stars are not responsible for the ionization of the
gas and the heating of the dust, we can neither use the H$\alpha$
emission nor the infrared emission as tracers of present-day
star formation at $R <$\,6\,kpc in M~31, as was also pointed out by 
\cite{Devereux_etal_94b}. Therefore, we investigated the relationship 
between SFR and neutral gas only for the interval $30\arcmin <R< 50\arcmin$ 
($R$=\,6.8-11.4\,kpc) containing the '10\,kpc ring'.

\begin{figure*}
\begin{center}
\resizebox{\hsize}{!}{\includegraphics[angle=-90]{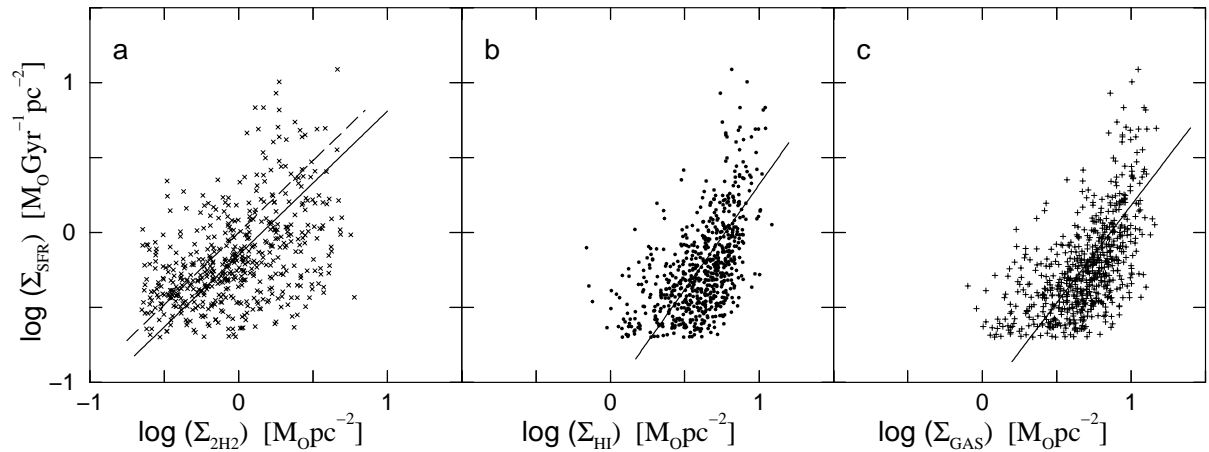}}
\caption[]{Scatter plots between the surface density of the star formation rate  and neutral gas surface densities for the radial interval $30\arcmin <R<50\arcmin$ (6.8~kpc~$<R<$~11.4~kpc). All surface densities are face-on values. Only independent data points (separated by 1.67$\times$ beamwidth) above 2$\times$ rms noise were used. {\bf a}. SFR versus molecular gas; {\bf b}. SFR versus atomic gas; {\bf c}. SFR versus total gas. Full lines indicate the power-law fits given in Table 6. The relationship in a. is linear and nearly the same as the average relationship for 7 bright galaxies (dashed line) derived by \cite{Bigiel_08}. Note that the cut off in  $\Sigma_{\rm HI}$ and $\Sigma_{\rm GAS}$ is near 10\,M$_{\odot}$~pc$^{-2}$. }
\label{fig:sfr1}
\end{center}
\end{figure*}

The correlation plots in Fig.~\ref{fig:sfr1} and the results in Table~\ref{table:sfr} 
show that  $\Sigma_{\rm SFR}$ is not well correlated with  the surface densities of 
either H$_2$, HI or total gas ($r_c \sim$ 0.45-0.59). In spite of this, the fitted 
bisectors are statistically  significant ($t>3$). Interestingly, we find a linear 
relationship between $\Sigma_{\rm SFR}$ and $\Sigma_{\rm 2H_2}$  
(exponent $b=0.96\pm$0.03), which closely agrees with the average 
relationship for 7 nearby galaxies, much brighter than M~31 (see 
Fig.~\ref{fig:sfr1}a), analyzed by \cite{Bigiel_08}. While in these galaxies 
molecular hydrogen is the dominant gas phase, most of the neutral 
gas in M~31 is atomic (compare Fig.~\ref{fig:sfr1}a,b). Hence, the surface density 
of SFR is linearly related to that of molecular gas, irrespective of 
the fraction of molecular gas or the absolute value of the total 
gas surface density in a galaxy. \cite{Bigiel_08} arrived at the 
same conclusion after comparing the galaxies in their sample. 

The correlation between $\Sigma_{\rm SFR}$ and total gas surface density is slightly better than that between $\Sigma_{\rm SFR}$  and  molecular gas surface density. The bisector fit in Table~7 yields the Kennicutt-Schmidt-law
\begin{equation}  
 \Sigma_{\rm SFR} = (0.076 \pm 0.005)\,\Sigma_{\rm GAS}^{1.30 \pm 0.05},  
\end{equation}
where $\Sigma_{\rm GAS}$ and $\Sigma_{\rm SFR}$ are in M$_{\odot}$\,pc$^{-2}$ and M$_{\odot}$\,Gyr$^{-1}$\,pc$^{-2}$, respectively.
The exponent of $1.30 \pm 0.05$ is well in the range of 
1.1-2.7 derived by \cite{Bigiel_08}. As a galaxy of low surface brightness, 
the SFRs in M~31 are correspondingly low. Our $\Sigma_{\rm SFR}$--$\Sigma_{\rm GAS}$
relationship nicely fits on the low-brightness extension of the compilation 
of available galaxy data in Fig.~15 of \cite{Bigiel_08},  formed by the
outer parts of their 7 galaxies and the global values for 20 galaxies of low
surface brightness.

Very recently, \cite{Braun_09} also studied the dependence of SFR
on gas density in M~31 using the new Westerbork HI survey and the 
CO survey of \cite{Nieten}. They estimated the SFR from the
surface brightnesses at IRAC 8\,$\mu$m, MIPS 24\,$\mu$m and GALEX FUV following
the procedure of \cite{Thilker_05}. Our Fig.~\ref{fig:sfr1}a is comparable
to the radial range 8-16\,kpc in their Fig.~20D  that shows the same range
in $\Sigma_{\rm SFR}$ as we find. Note that the molecular
gas densities of \cite{Braun_09} are a factor of 1.6 larger (+0.21 dex)
and have a larger dynamic range than our values due to differences
in scaling of the CO data, inclination, angular resolution and radial
range. Scaling our relationship to the assumptions of \cite{Braun_09}
gives  ${\rm log}(\Sigma_{\rm SFR})=\, -0.44\, +\, 0.96\,{\rm log}(\Sigma_{\rm 2H_2})$, which is in good agreement with their Fig.~20D. 

The dependencies of SFR surface density on total gas surface density in
Fig.~\ref{fig:sfr1}c and in Fig.~20E of \cite{Braun_09} have the same pear-like
shape characterized by a broadening towards lower  $\Sigma_{\rm SFR}$ and a rather sharp
cut-off near  $\Sigma_{\rm GAS}$\,=\,10\,M$_{\odot}$\,pc$^{-2}$. The cut-off comes from the
$\Sigma_{\rm SFR}$-$\Sigma_{\rm HI}$ relation (see Fig.~15b) and occurs at the same value as in the bright galaxies analyzed by \cite{Bigiel_08}, who interpreted the lack of higher surface mass densities as a saturation effect. \cite{Braun_09} show that in M~31 this truncation indeed 
vanishes after correcting the HI data for opacity, which could lead to somewhat 
steeper slopes in Figs.~15b and c.

\begin{table*}
\begin{center}
\caption{Kennicutt-Schmidt law in M~31 for $30\arcmin<R<50\arcmin$.  Ordinary-least squares fits of the bisector log($\Sigma_{\rm SFR}$)=$a_c+b$\,log($\Sigma$), where $\Sigma_{\rm SFR}$ is the face-on surface density of the star formation rate in M$_{\odot}$\,Gyr$^{-1}$\,pc$^{-2}$ and $\Sigma$ the face-on value of the gas surface density in M$_{\odot}$\,pc$^{-2}$; n is the number of independent points; $r_c$ is the correlation coefficient and $t$ the student-t test.}
\begin{tabular}{ l l l l l l l}
\hline
$\Sigma$(M$_{\odot}$~pc$^{-2}$)   & \,\,\,\,\,\,\,\,  $a_c$  &\,\,\,\,\,\, $b$ &    n  & \,\,\,\,\,\, $r_c$& $t$\\
\hline
\hline
$\Sigma_{\rm 2H_{2}}$ &  -0.15$\pm$0.01 &  0.96$\pm$0.03 &   537 &  0.45$\pm$0.04& 12 \\
$\Sigma_{\rm HI}$     &    -1.08$\pm$0.03 &  1.40$\pm$0.05  &  670  & 0.55$\pm$0.03& 17 \\
$\Sigma_{\rm GAS}$  &  -1.12$\pm$0.03 &  1.30$\pm$0.05 &  668 & 0.59$\pm$0.03& 19 \\
\hline
\end{tabular}
\label{table:sfr}
\end{center}
\end{table*}

 \subsection{Radial variations in the Kennicutt-Schmidt law}
\begin{figure*}
\begin{center}
\resizebox{13cm}{!}{\includegraphics[angle=-90]{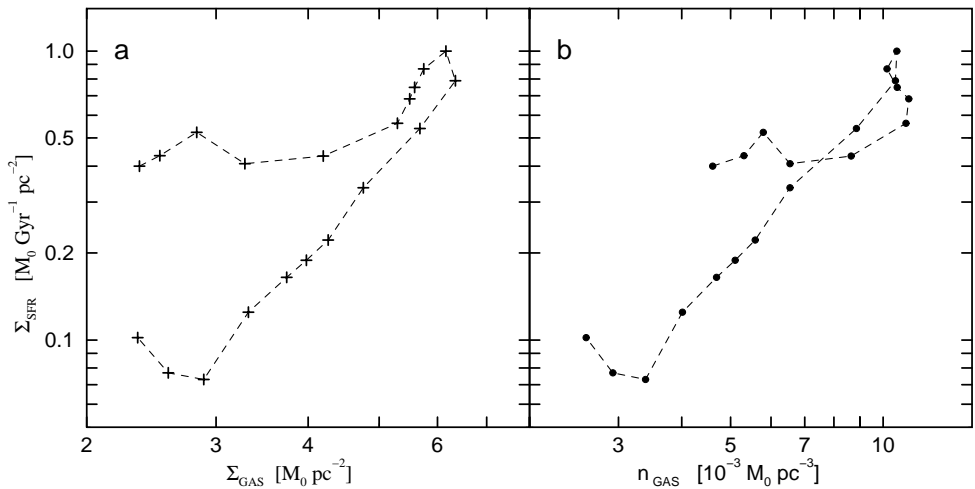}}
\caption[]{  Mean face-on values of $\Sigma_{\rm SFR}$, averaged in 0.5\,kpc-wide 
circular rings in the plane of M~31, plotted against the corresponding 
mean values of {\it (a)} gas surface density $\Sigma_{\rm GAS}$, and {\it (b)} gas volume
density $n_{\rm GAS}$. The upper left point is for the ring $R$\,=\,6.0-6.5\,kpc, the
maximum in $\Sigma_{\rm SFR}$ is in ring 10.5-11.0\,kpc and the minimum in ring 
$R$\,=\,14.5-15.0\,kpc. Typical errors are 0.01\,M$_{\odot}$\,Gyr$^{-1}$\,pc$^{-2}$ in $\Sigma_{\rm SFR}$, 0.02\,M$_{\odot}$\,pc$^{-2}$ in $\Sigma_{\rm GAS}$,  and 3\,$\times$\,10$^{-5}$\,M$_{\odot}$\,pc$^{-3}$  in $n_{\rm GAS}$. Points for $R>$\,12kpc suffer from 
missing data points near the major axis, the number of which increases 
with radius. }
\label{fig:KS}
\end{center}
\end{figure*}

In Fig.~\ref{fig:KS}a, we plot the mean values in {\bf 0.5\,kpc-wide} rings 
in the plane of M~31 of $\Sigma_{\rm SFR}$ against those of $\Sigma_{\rm GAS}$ {\bf from $R$\,=\,6\,kpc to $R$\,=\,16\,kpc}. The points form a big loop with a horizontal
branch for $R$\,=\,6 - 8.5\,kpc and a maximum $\Sigma_{\rm SFR}$ in the ring $R$\,=\,10.5 - 11.0\,kpc (see also Fig.~\ref{fig:sfr2}). This behavior was already noted by \cite{Berkhuijsen_77}
and \cite{Tenjes}, who used the number density of HII regions as
tracer of SFR and HI gas, and was recently confirmed by \cite{Boissier}
from GALEX UV data and total gas. Both \cite{Berkhuijsen_77} and \cite{Tenjes} showed that the differences between the slopes inside and outside
the maximum of the starforming ring is greatly reduced when the increase
in the scale height of the gas with increasing radius is taken into account. We calculated
the scale height, h, from the scale height of the HI gas given by Eq.~13 of
\cite{Braun_91}, scaled to D\,=\,780\,kpc, assumed half this value for that of the
H$_2$ gas, and a constant scale height for the ionizing stars. Fig.~\ref{fig:KS}b shows
$\Sigma_{\rm SFR}$ as a function of gas volume density $n_{\rm GAS}$\,= N(HI)/2h + N(2H$_2$)/h.
The points have moved towards each other, but the horizontal branch remained
and the behavior on the starforming ring has become more complicated.

The variations in slope in Fig.~\ref{fig:KS} are clear evidence for radial
variations in the index of the Kennicutt-Schmidt law. Such variations are 
not specific to M~31 as they are also seen in some of the galaxies
analyzed by \cite{Bigiel_08}. In order to quantify the variations,
we determined the bisectors in scatter plots for three circular rings:
$R$\,=\,7 - 9\,kpc, $R$\,=\,9 - 11\,kpc and $R$\,=\,11 - 13\,kpc, covering the horizontal
branch, the increasing part inside the maximum and the decreasing part
outside the maximum, respectively. The results are given in Table~8.
The index for the star fromation law for surface densities is unity
for the 7-9\,kpc ring and about 1.6 for the other two rings. Hence, the slope of $b$\,=\,1.30\,$\pm$\,0.05 obtained for the {\bf `10\,kpc ring'} ($R$\,=\,6.8 - 11.4\,kpc) in Sect.~7.1 represents the mean value of the first two rings considered here.
The scatter plots between $\Sigma_{\rm SFR}$ and gas volume density yield bisector 
slopes that are about 0.2 smaller than those for surface density. The
correlation coefficients are all close to $r_c = 0.59 \pm 0.03$ for the
{\bf `10\,kpc ring'} (see Table~7), indicating that even in 2\,kpc-wide rings
the intrinsic scatter is considerable. This implies that on scales
of a few hundred parsec significant variations in the index of the 
Kennicutt-Schmidt law and in the star formation efficiency occur.

\begin{table*}
\begin{center}
\caption{Kennicutt-Schmidt law in three radial intervals in M~31.
Ordinary least-squares fits of the bisector log($\Sigma_{\rm SFR}$)=$a_c\,+\,b$\,log(X),
where $\Sigma_{\rm SFR}$ is the face-on surface density of the star formation rate
in M$_{\odot}$\,Gyr$^{-1}$\,pc$^{-2}$, X\,=\,$\Sigma_{\rm GAS}$ is the face-on gas surface density in M$_{\odot}$\,pc$^{-2}$ and X\,=\,$n_{\rm GAS}$ is the gas volume density in at cm$^{-3}$. 
n is the number of independent points; $r_c$ is the correlation coefficient
and t the student-t test. }
\begin{tabular}{ l l l l l l l }
\hline
\,\,\,$R$(kpc) & \,\,\ X &\,\,\,\,\,\, $a_c$ &  \,\,\,\,\,\, $b$ &   n  & \,\,\,\,\,\, $r_c$& $t$\\
\hline
\hline
7-9 & $\Sigma_{\rm GAS}$  &  -0.90$\pm$0.04 &1.03$\pm$0.06 &  216 &  0.54$\pm$0.06 & 9 \\
9-11 &                  &  -1.43$\pm$0.06  & 1.67$\pm$0.08 &  356  & 0.63$\pm$0.04 & 15 \\
11-13  &                &  -1.46$\pm$0.06  &  1.55$\pm$0.08 & 297 & 0.62$\pm$0.05& 13 \\
&&&& & &\\
7-9  &  $n_{\rm GAS}$  &  0.18$\pm$0.04  &  0.88$\pm$ 0.06   & 218  &  0.51$\pm$0.06  &  9\\
9-11  &         &  0.45$\pm$0.04  &  1.50$\pm$0.07   & 356  &  0.62$\pm$0.04  & 15\\
11-13  &         &  0.35$\pm$0.04  &  1.35$\pm$0.07   & 297  &  0.62$\pm$0.05  & 14\\
\hline
\end{tabular}
\label{table:KS}
\end{center}
\end{table*}

\subsection{Radial variations of SFR and SFE}

\begin{figure}
\begin{center}
\resizebox{8cm}{!}{\includegraphics[angle=-90]{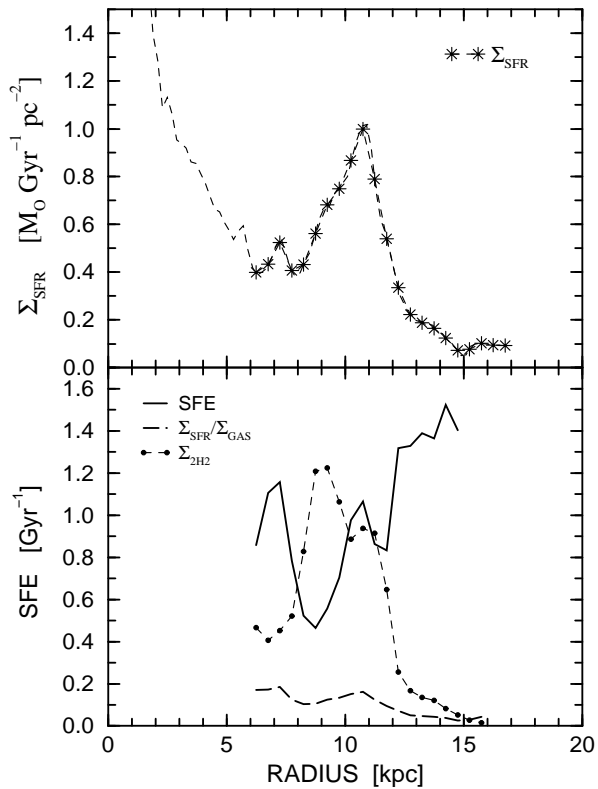}}
\caption[]{Radial variation of the face-on surface density of the star formation rate 
$\Sigma_{\rm SFR}$ and star formation efficiency SFE in M~31, averaged in 0.5\,kpc-wide 
rings in the plane of the galaxy. Beyond $R=$\,12\,kpc the data are not complete, 
because the observed area is limited along the major axis (see Fig.~10a). {\it Top}: 
Radial profile of  $\Sigma_{\rm SFR}$. {\it Bottom}: {\it full line} -  SFE\,=\,$\Sigma_{\rm SFR}$/$\Sigma_{\rm 2H_2}$; {\it long dashed line} - $\Sigma_{\rm SFR}$/$\Sigma_{\rm GAS}$; 
{\it dots} - radial profile of the molecular gas surface density  $\Sigma_{\rm 2H_2}$ seen face-on. 
Note that $\Sigma_{\rm SFR}$ and $\Sigma_{\rm 2H_2}$ do not peak at the same radius. 
Statistical errors in $\Sigma_{\rm SFR}$ and $\Sigma_{\rm 2H_2}$ are smaller than the symbols and those in SFE and $\Sigma_{\rm SFR}$/$\Sigma_{\rm GAS}$ are smaller than the thickness of the lines. Only beyond $R$\,=\,12\,kpc the error in SFE slowly increases to $0.3\,{\rm Gyr}^{-1}$  at $R$\,=\,15\,kpc.  }
\label{fig:sfr2}
\end{center}
\end{figure}

In Fig.~\ref{fig:sfr2} (upper panel) we present the radial profile of the SFR 
surface density between 6\,kpc and 17\,kpc, averaged in 0.5\,kpc-wide 
circular rings in the plane of M~31. The face-on values vary between about 
0.1 and 1\,M$_{\odot}$\, Gyr$^{-1}$\, pc$^{-2}$.  \cite{Boissier} and \cite{Braun_09}
obtained similar values for $\Sigma_{\rm SFR}$ in this radial range from GALEX UV data.
They are about 10 times smaller than the~surface densities of SFR between R\,=~1.5~kpc and R\,=~7\,kpc in the northern part of M~33 observed by \citep{Gardan}.

In the lower panel of Fig.~\ref{fig:sfr2} we show the radial profiles of the
surface density of the molecular gas and of the star formation 
efficiencies  SFE\,=\,$\Sigma_{\rm SFR}$/$\Sigma_{\rm 2H_2}$ and $\Sigma_{\rm SFR}$/$\Sigma_{\rm GAS}$. Although the maximum  $\Sigma_{\rm SFR}$ occurs on a relative maximum in the molecular gas density (in ring 10.5-11.0\,kpc),  $\Sigma_{\rm SFR}$ is only about 70\% of its maximum value where the molecular gas density is highest (in ring 9.0-9.5\,kpc). 
Consequently, SFE varies significantly with radius. Between $R=6$\,kpc and 
$R=15$\,kpc SFE fluctuates around a value of 0.9\,Gyr$^{-1}$, with a  minimum 
of  0.46\,$\pm$\,0.01\,Gyr$^{-1}$ near $R$=\,9\,kpc. Thus SFE is smallest where  $\Sigma_{\rm 2H_2}$
is highest! Up to $R$=12\,kpc the efficiency  $\Sigma_{\rm SFR}$/$\Sigma_{\rm GAS}$ shows the 
same trend as SFE.  The increase in SFE between 12~kpc and 15~kpc radius of 
a factor 1.5 results from the difference in radial scale lengths of  $\Sigma_{\rm SFR}$
(or H$\alpha$ emission) and the molecular gas density (see  Fig.~\ref{fig:surfir} and Table~3).
Interestingly, in M~33 \citep{Gardan} found a radial increase
in SFE of a factor 2 between 2~kpc and 6~kpc radius with similar
fluctuations around the mean as we observe in M~31, but the mean value
in M~31 is about three times lower than in M~33.  Furthermore, \cite{Leroy} found significant variations in the efficiency $\Sigma_{\rm SFR}/\Sigma_{\rm GAS}$ on a linear scale of 800\,pc in the sample of 12 spiral galaxies analyzed by them.

That large, small-scale variations in SFE exist in galaxies is also 
clear from the large spread in the scatter plots of  $\Sigma_{\rm SFR}$--$\Sigma_{\rm 2H_2}$ 
visible in Fig.~\ref{fig:sfr1}a and in several figures of \cite{Bigiel_08}. The 
same value of  $\Sigma_{\rm SFR}$ can occur in a range of $\Sigma_{\rm 2H_2}$ spanning more 
than a factor of 10.

We may conclude that neither the present-day star formation rate 
$\Sigma_{\rm SFR}$ nor the star formation efficiency SFE is well correlated 
with the molecular gas surface density. Hence, other factors than molecular
gas density must play an important role in the star formation process.
\cite{Bigiel_08} argue that local environmental circumstances
largely determine the SFE in spiral galaxies. These factors are 
extensively discussed by e.g. \cite{Leroy}.

\section{Summary}
%Using the highly sensitive and high resolution Spitzer MIPS data, we studied the dust characteristics and its relation to the other traces of the interstellar medium.  

In this paper, we studied the emission from dust, 
neutral gas and ionized gas in the disk of M~31, and the relationships 
between these components on various linear scales. 
We compared the Spitzer MIPS maps at 24\,$\mu$m, 70\,$\mu$m and 160\,$\mu$m \citep{Gordon_06} to the distributions of atomic gas seen in the HI line \citep{Brinks}, 
molecular gas as traced by the $^{12}$CO(1-0) line \citep{Nieten} and 
ionized gas observed in H$\alpha$ \citep{Devereux_etal_94b}. All data were smoothed to an angular resolution of 45$\arcsec$ corresponding to 170\,pc\,$\times$\,660\,pc in the plane of the galaxy.

For each of the dust and gas maps, we calculated the mean intensity 
distribution as a function of radius (Fig.~\ref{fig:surfir}), separately for the 
northern and the southern half of M~31. Using wavelet analysis, we decomposed the dust and gas distributions in spatial scales and calculated cross-correlations as a function of scale. We also used classical correlations to derive quantitative relations between the various dust and gas components.

Using the MIPS 70\,$\mu$m and 160\,$\mu$m maps, we derived the distributions of the  dust 
temperature and optical depth. The dust optical depth at the  H$\alpha$ wavelength was used 
to  a) investigate the dust-to-gas ratio, b) derive scaling relations between extinction and 
neutral gas emission, and c) de-redden the H$\alpha$ emission in order to estimate the 
recent star formation rate. We also presented the Kennicutt-Schmidt law indices obtained  for the bright emission ring {\bf near $R$=\,10\,kpc} in M~31. 
We summarize the main results and conclusions as follows.\\
\\
1. Dust temperature and opacity:\\
$\bullet$ The dust temperature steeply drops from about 30\,K in the center to about 
19\,K near $R=$\,4.5\,kpc, and stays between about 17~K and 20~K beyond this 
radius (Fig.~2). 
The mean dust temperature in the area studied is about 18.5~K. This is 
3~K less than the temperature obtained by \cite{Walterbos_87}
between the IRAS maps at 60\,$\mu$m and 100\,$\mu$m that both trace warmer dust 
than the MIPS maps at 70\,$\mu$m and 160\,$\mu$m used here.\\
$\bullet$ The dust optical depth at H$\alpha$ along the line of sight varies in 
a range between about 0.2 near the center and about 1 in the `10~kpc ring' (Fig.~4) with a mean value of 0.7$\pm$0.4 (the error is standard deviation) and a most probable value of $\simeq$\,0.5, indicating that M~31 is mostly optically thin to the H$\alpha$ emission. The total flux density of the H$\alpha$ emission increases by 30\% after correction for extinction.\\
\\
2. Radial distributions:\\
$\bullet$
The radial scale lengths between the maximum in the `10~kpc ring' and
$R=$~15~kpc of the warm dust are smaller than that of the cold dust, as is 
expected if the warm dust is mainly heated by UV photons from star
forming regions and cold dust by the ISRF. With the largest
scale length, atomic gas has the largest radial extent of the dust and gas 
components considered here.\\
$\bullet$ The radial gradient of the total gas-to-dust ratio is consistent
with that of the oxygen abundance in M~31. The gas-to-dust 
ratios observed in the solar neighborhood \citep{Bohlin} occur  
near $R=$~8.5~kpc in the disk of M~31 where N(gas)/$\tau_{{\rm H}\alpha}$=
$2.6 \times 10^{21}$ at\,cm$^{-2}$.\\
\\
3. Properties as a function of scale:\\
$\bullet$
Spatial scales larger than about 8~kpc contain most of the emitted power from 
the cold dust and the atomic gas, whereas the emissions from warm dust, 
molecular gas and ionized gas are dominated by scales near 1~kpc, 
typical for complexes of star forming regions and molecular clouds in spiral arms (Fig.~11). \\
$\bullet$ Dust emission is correlated ($r_w \ge 0.6$) with both neutral and ionized gas on scales~$>1$\,kpc. \\
$\bullet$ On scales\,$<1$\,kpc, ionized gas is best correlated with
warm dust and neutral gas (both HI and H$_2$) with cold dust. 
On the smallest scale of 0.4\,kpc, an HI--warm dust correlation 
hardly exists ($r_w \simeq 0.4$) because not much HI occurs on the 
scale of star forming regions (see Fig.~11).\\
\\
4. Relationships between gas and dust:\\
$\bullet$ 
H$\alpha$ emission is slightly better correlated with the
emission at 70\,$\mu$m than at 24\,$\mu$m (Fig.~13, Table 6), especially
on scales~$<$\,2 kpc (Fig.~12). As in M~33 the 24\,$\mu$m--H$\alpha$ correlation
is best, this suggests that in early-type galaxies like M~31 the contribution from 
evolved AGB stars to the 24\,$\mu$m emission is larger than in late-type galaxies like M~33.\\
$\bullet$ Dust extinction A$_{H\alpha}$ is not well correlated with N(2H$_2$)
indicating that dust mixed with molecular clouds does not
contribute much to the total extinction. Although the 
correlation with N(HI) is better, A$_{H\alpha}$ is best correlated 
with N(HI+2H$_2$).\\
$\bullet$ Dust opacity is proportional to the square root
of N(2H$_2$) but about linearly related to N(HI), as was also
found by \cite{Nieten} at 90$\arcsec$ resolution. This is an
indirect indication of a balance between the formation and
destruction rates of H$_2$ in cool, dusty HI clouds.\\
$\bullet$ In the central 2 kpc both the dust opacity and the HI 
column density are very low and the dust temperature is high.
This combination may explain the lack of H$_2$ in this region.\\
\\
5. SFR and SFE:\\
$\bullet$
The SFR in M~31 is low. The total SFR in the observed field between $R=$~6~kpc and $R=$~17~kpc 
is $0.27\,{\rm M}_{\odot} {\rm yr}^{-1}$ and the star formation efficiency is 0.9\,Gyr$^{-1}$, 
yielding a molecular depletion time scale of 1.1~Gyr. This is about three
times longer than observed in the northern part of M~33 (Gardan et al. 2007). 
The radial distribution of $\Sigma_{\rm SFR}$ in 0.5\,kpc-wide rings in the plane of 
the galaxy (Fig.~17) varies between about 0.1 and 1\,M$_{\odot}$\, Gyr$^{-1}$\, pc$^{-2}$, values that are about 10 times smaller than in  the northern part of M~33 \citep{Gardan}.  Between 
$R=$~6~kpc and $R=$~15~kpc, SFE varies between about 0.5\,Gyr$^{-1}$ and 1.5\,Gyr$^{-1}$, 
whereas the efficiency with respect to the total gas surface density 
slowly decreases from about 0.18\,Gyr$^{-1}$ to about 0.03\,Gyr$^{-1}$. \\ 
$\bullet$ SFR is not well correlated with  neutral gas and worst of all with molecular gas in the radial range $30\arcmin-50\arcmin$ containing the `10\,kpc ring' (Fig.~15, Table 7). In spite of this, the power-law fits are statistically significant. We find a linear
relationship between the surface densities of SFR and molecular gas
(power-law exponent 0.96~$\pm$~0.03),  and a power law with index 1.30~$\pm$~0.05 between the surface densities of SFR and total gas. These results agree with the average relationship
for 7 nearby galaxies much brighter than M~31 \citep{Bigiel_08}. While
in these galaxies molecular hydrogen is the dominant gas phase, most of 
the neutral gas in M~31 is atomic. Thus, the surface density of SFR depends
linearly on that of molecular gas irrespective of the fraction of molecular
gas or the absolute value of the total gas surface density in a galaxy.

Some important implications of this study are:
\begin{itemize}
\item[-] Precaution is required in using the total IR luminosity (TIR) as an indicator 
of recent SFR or to derive dust opacity for an early-type galaxy like M~31, 
because the cold dust is mainly heated by the ISRF and the warm dust emission 
at 24\,$\mu$m is partly due to evolved stars (especially in the bulge of the galaxy). 
 
\item[-] Neither the present-day SFR nor SFE is well correlated 
with the surface density of molecular gas or total gas. Therefore, other factors than 
gas density must play an important role in the process of star 
formation in M~31.
\end{itemize}

\begin{acknowledgements}
We are grateful to E. Kr\"ugel for valuable and stimulating comments. We thank K.M. Menten and R. Beck for comments and careful reading of the manuscript. The Spitzer MIPS data were kindly provided by Karl D. Gordon. E.Tempel kindly sent us a table of extinction values that we used
for Fig.~5. We thank an anonymous referee for extensive comments leading to improvements
in the manuscript. FT was supported through a stipend from the Max Planck Institute for Radio Astronomy (MPIfR).  

\end{acknowledgements}

\bibliography{s.bib} 

\begin{thebibliography}{85}
\expandafter\ifx\csname natexlab\endcsname\relax\def\natexlab#1{#1}\fi

\bibitem[{{Andriesse}(1974)}]{andriesse}
{Andriesse}, C.~D. 1974, \aap, 37, 257

\bibitem[{{Bajaja} \& {Gergely}(1977)}]{Bajaja_77}
{Bajaja}, E. \& {Gergely}, T.~E. 1977, \aap, 61, 229

\bibitem[{{Barmby} {et~al.}(2006){Barmby}, {Ashby}, {Bianchi}, {Engelbracht},
  {Gehrz}, {Gordon}, {Hinz}, {Huchra}, {Humphreys}, {Pahre},
  {P{\'e}rez-Gonz{\'a}lez}, {Polomski}, {Rieke}, {Thilker}, {Willner}, \&
  {Woodward}}]{Barmby}
{Barmby}, P., {Ashby}, M.~L.~N., {Bianchi}, L., {et~al.} 2006, \apjl, 650, L45

\bibitem[{{Berkhuijsen}(1977)}]{Berkhuijsen_77}
{Berkhuijsen}, E.~M. 1977, \aap, 57, 9

\bibitem[{{Berkhuijsen} \& {Humphreys}(1989)}]{Berkhuijsen_89}
{Berkhuijsen}, E.~M. \& {Humphreys}, R.~M. 1989, \aap, 214, 68

\bibitem[{{Bigiel} {et~al.}(2008){Bigiel}, {Leroy}, {Walter}, {Brinks}, {de
  Blok}, {Madore}, \& {Thornley}}]{Bigiel_08}
{Bigiel}, F., {Leroy}, A., {Walter}, F., {et~al.} 2008, \aj, 136, 2846

\bibitem[{{Blair} {et~al.}(1982){Blair}, {Kirshner}, \& {Chevalier}}]{Blair}
{Blair}, W.~P., {Kirshner}, R.~P., \& {Chevalier}, R.~A. 1982, \apj, 254, 50

\bibitem[{{Block} {et~al.}(2006){Block}, {Bournaud}, {Combes}, {Groess},
  {Barmby}, {Ashby}, {Fazio}, {Pahre}, \& {Willner}}]{Block}
{Block}, D.~L., {Bournaud}, F., {Combes}, F., {et~al.} 2006, \nat, 443, 832

\bibitem[{{Bohlin} {et~al.}(1978){Bohlin}, {Savage}, \& {Drake}}]{Bohlin}
{Bohlin}, R.~C., {Savage}, B.~D., \& {Drake}, J.~F. 1978, \apj, 224, 132

\bibitem[{{Boissier} {et~al.}(2007){Boissier}, {Gil de Paz}, {Boselli},
  {Madore}, {Buat}, {Cortese}, {Burgarella}, {Mu{\~n}oz-Mateos}, {Barlow},
  {Forster}, {Friedman}, {Martin}, {Morrissey}, {Neff}, {Schiminovich},
  {Seibert}, {Small}, {Wyder}, {Bianchi}, {Donas}, {Heckman}, {Lee},
  {Milliard}, {Rich}, {Szalay}, {Welsh}, \& {Yi}}]{Boissier}
{Boissier}, S., {Gil de Paz}, A., {Boselli}, A., {et~al.} 2007, \apjs, 173, 524

\bibitem[{{Boulanger} {et~al.}(1996){Boulanger}, {Abergel}, {Bernard},
  {Burton}, {Desert}, {Hartmann}, {Lagache}, \& {Puget}}]{Boulanger}
{Boulanger}, F., {Abergel}, A., {Bernard}, J.-P., {et~al.} 1996, \aap, 312, 256

\bibitem[{{Braun}(1990)}]{Braun_90}
{Braun}, R. 1990, \apjs, 72, 755

\bibitem[{{Braun}(1991)}]{Braun_91}
{Braun}, R. 1991, \apj, 372, 54

\bibitem[{{Braun} {et~al.}(2009){Braun}, {Thilker}, {Walterbos}, \&
  {Corbelli}}]{Braun_09}
{Braun}, R., {Thilker}, D.~A., {Walterbos}, R.~A.~M., \& {Corbelli}, E. 2009,
  \apj, 695, 937

\bibitem[{{Brinks} \& {Shane}(1984)}]{Brinks}
{Brinks}, E. \& {Shane}, W.~W. 1984, \aaps, 55, 179

\bibitem[{{Calzetti} {et~al.}(2007){Calzetti}, {Kennicutt}, {Engelbracht},
  {Leitherer}, {Draine}, {Kewley}, {Moustakas}, {Sosey}, {Dale}, {Gordon},
  {Helou}, {Hollenbach}, {Armus}, {Bendo}, {Bot}, {Buckalew}, {Jarrett}, {Li},
  {Meyer}, {Murphy}, {Prescott}, {Regan}, {Rieke}, {Roussel}, {Sheth}, {Smith},
  {Thornley}, \& {Walter}}]{Calzetti_07}
{Calzetti}, D., {Kennicutt}, R.~C., {Engelbracht}, C.~W., {et~al.} 2007, \apj,
  666, 870

\bibitem[{{Calzetti} {et~al.}(2005){Calzetti}, {Kennicutt}, {Bianchi},
  {Thilker}, {Dale}, {Engelbracht}, {Leitherer}, {Meyer}, {Sosey}, {Mutchler},
  {Regan}, {Thornley}, {Armus}, {Bendo}, {Boissier}, {Boselli}, {Draine},
  {Gordon}, {Helou}, {Hollenbach}, {Kewley}, {Madore}, {Martin}, {Murphy},
  {Rieke}, {Rieke}, {Roussel}, {Sheth}, {Smith}, {Walter}, {White}, {Yi},
  {Scoville}, {Polletta}, \& {Lindler}}]{Calzetti_05}
{Calzetti}, D., {Kennicutt}, Jr., R.~C., {Bianchi}, L., {et~al.} 2005, \apj,
  633, 871

\bibitem[{{Chemin} {et~al.}(2009){Chemin}, {Carignan}, \& {Foster}}]{Chemin}
{Chemin}, L., {Carignan}, C., \& {Foster}, T. 2009, \apj, 705, 1395

\bibitem[{{Ciardullo} {et~al.}(1988){Ciardullo}, {Rubin}, {Ford}, {Jacoby}, \&
  {Ford}}]{Ciardullo}
{Ciardullo}, R., {Rubin}, V.~C., {Ford}, Jr., W.~K., {Jacoby}, G.~H., \&
  {Ford}, H.~C. 1988, \aj, 95, 438

\bibitem[{{Cox} {et~al.}(1986){Cox}, {Kruegel}, \& {Mezger}}]{Cox_86}
{Cox}, P., {Kruegel}, E., \& {Mezger}, P.~G. 1986, \aap, 155, 380

\bibitem[{{Dennefeld} \& {Kunth}(1981)}]{Dennefeld}
{Dennefeld}, M. \& {Kunth}, D. 1981, \aj, 86, 989

\bibitem[{{Devereux} {et~al.}(1994){Devereux}, {Price}, {Wells}, \&
  {Duric}}]{Devereux_etal_94b}
{Devereux}, N.~A., {Price}, R., {Wells}, L.~A., \& {Duric}, N. 1994, \aj, 108,
  1667

\bibitem[{{Dickinson} {et~al.}(2003){Dickinson}, {Davies}, \&
  {Davis}}]{Dickinson}
{Dickinson}, C., {Davies}, R.~D., \& {Davis}, R.~J. 2003, \mnras, 341, 369

\bibitem[{{Diplas} \& {Savage}(1994)}]{Diplas_94}
{Diplas}, A. \& {Savage}, B.~D. 1994, \apj, 427, 274

\bibitem[{{Draine} {et~al.}(2007){Draine}, {Dale}, {Bendo}, {Gordon}, {Smith},
  {Armus}, {Engelbracht}, {Helou}, {Kennicutt}, {Li}, {Roussel}, {Walter},
  {Calzetti}, {Moustakas}, {Murphy}, {Rieke}, {Bot}, {Hollenbach}, {Sheth}, \&
  {Teplitz}}]{Draine_07}
{Draine}, B.~T., {Dale}, D.~A., {Bendo}, G., {et~al.} 2007, \apj, 663, 866

\bibitem[{{Draine} \& {Lee}(1984)}]{Draine}
{Draine}, B.~T. \& {Lee}, H.~M. 1984, \apj, 285, 89

\bibitem[{{Emerson}(1974)}]{Emerson_74}
{Emerson}, D.~T. 1974, \mnras, 169, 607

\bibitem[{{Evans}(1986)}]{Evans}
{Evans}, I.~N. 1986, \apj, 309, 544

\bibitem[{{Frick} {et~al.}(2001){Frick}, {Beck}, {Berkhuijsen}, \&
  {Patrickeyev}}]{Frick_etal_01}
{Frick}, P., {Beck}, R., {Berkhuijsen}, E.~M., \& {Patrickeyev}, I. 2001,
  \mnras, 327, 1145

\bibitem[{{Gardan} {et~al.}(2007){Gardan}, {Braine}, {Schuster}, {Brouillet},
  \& {Sievers}}]{Gardan}
{Gardan}, E., {Braine}, J., {Schuster}, K.~F., {Brouillet}, N., \& {Sievers},
  A. 2007, \aap, 473, 91

\bibitem[{{Garnett} {et~al.}(1997){Garnett}, {Shields}, {Skillman}, {Sagan}, \&
  {Dufour}}]{Garnett}
{Garnett}, D.~R., {Shields}, G.~A., {Skillman}, E.~D., {Sagan}, S.~P., \&
  {Dufour}, R.~J. 1997, \apj, 489, 63

\bibitem[{{Gordon} {et~al.}(2006){Gordon}, {Bailin}, {Engelbracht}, {Rieke},
  {Misselt}, {Latter}, {Young}, {Ashby}, {Barmby}, {Gibson}, {Hines}, {Hinz},
  {Krause}, {Levine}, {Marleau}, {Noriega-Crespo}, {Stolovy}, {Thilker}, \&
  {Werner}}]{Gordon_06}
{Gordon}, K.~D., {Bailin}, J., {Engelbracht}, C.~W., {et~al.} 2006, \apjl, 638,
  L87

\bibitem[{{Gordon} {et~al.}(2007){Gordon}, {Engelbracht}, {Fadda},
  {Stansberry}, {Wachter}, {Frayer}, {Rieke}, {Noriega-Crespo}, {Latter},
  {Young}, {Neugebauer}, {Balog}, {Beeman}, {Dole}, {Egami}, {Haller}, {Hines},
  {Kelly}, {Marleau}, {Misselt}, {Morrison}, {P{\'e}rez-Gonz{\'a}lez}, {Rho},
  \& {Wheaton}}]{Gordon_07}
{Gordon}, K.~D., {Engelbracht}, C.~W., {Fadda}, D., {et~al.} 2007, \pasp, 119,
  1019

\bibitem[{{Gordon} {et~al.}(2005){Gordon}, {Rieke}, {Engelbracht}, {Muzerolle},
  {Stansberry}, {Misselt}, {Morrison}, {Cadien}, {Young}, {Dole}, {Kelly},
  {Alonso-Herrero}, {Egami}, {Su}, {Papovich}, {Smith}, {Hines}, {Rieke},
  {Blaylock}, {P{\'e}rez-Gonz{\'a}lez}, {Le Floc'h}, {Hinz}, {Latter},
  {Hesselroth}, {Frayer}, {Noriega-Crespo}, {Masci}, {Padgett}, {Smylie}, \&
  {Haegel}}]{Gordon_05}
{Gordon}, K.~D., {Rieke}, G.~H., {Engelbracht}, C.~W., {et~al.} 2005, \pasp,
  117, 503

\bibitem[{{Haas} {et~al.}(1998){Haas}, {Lemke}, {Stickel}, {Hippelein},
  {Kunkel}, {Herbstmeier}, \& {Mattila}}]{Haas}
{Haas}, M., {Lemke}, D., {Stickel}, M., {et~al.} 1998, \aap, 338, L33

\bibitem[{{Henry} \& {Howard}(1995)}]{Henry}
{Henry}, R.~B.~C. \& {Howard}, J.~W. 1995, \apj, 438, 170

\bibitem[{{Hippelein} {et~al.}(2003){Hippelein}, {Haas}, {Tuffs}, {Lemke},
  {Stickel}, {Klaas}, \& {V{\"o}lk}}]{Hippelein}
{Hippelein}, H., {Haas}, M., {Tuffs}, R.~J., {et~al.} 2003, \aap, 407, 137

\bibitem[{{Hirashita}(1999)}]{Hirashita_99}
{Hirashita}, H. 1999, \apjl, 510, L99

\bibitem[{{Hirashita} {et~al.}(2002){Hirashita}, {Tajiri}, \&
  {Kamaya}}]{Hirashita_02}
{Hirashita}, H., {Tajiri}, Y.~Y., \& {Kamaya}, H. 2002, \aap, 388, 439

\bibitem[{{Hoernes } {et~al.}(1998){Hoernes }, {Berkhuijsen }, \& {Xu
  }}]{Hoernes_etal_98}
{Hoernes }, P., {Berkhuijsen }, E.~M., \& {Xu }, C. 1998, \aap, 334, 57

\bibitem[{{Isobe} {et~al.}(1990){Isobe}, {Feigelson}, {Akritas}, \&
  {Babu}}]{Isobe}
{Isobe}, T., {Feigelson}, E.~D., {Akritas}, M.~G., \& {Babu}, G.~J. 1990, \apj,
  364, 104

\bibitem[{{Issa} {et~al.}(1990){Issa}, {MacLaren}, \& {Wolfendale}}]{Issa}
{Issa}, M.~R., {MacLaren}, I., \& {Wolfendale}, A.~W. 1990, \aap, 236, 237

\bibitem[{{Kang} {et~al.}(2009){Kang}, {Bianchi}, \& {Rey}}]{Kang}
{Kang}, Y., {Bianchi}, L., \& {Rey}, S. 2009, \apj, 703, 614

\bibitem[{{Kennicutt} {et~al.}(2009){Kennicutt}, {Hao}, {Calzetti},
  {Moustakas}, {Dale}, {Bendo}, {Engelbracht}, {Johnson}, \&
  {Lee}}]{Kennicutt_09}
{Kennicutt}, R.~C., {Hao}, C., {Calzetti}, D., {et~al.} 2009, \apj, 703, 1672

\bibitem[{{Kennicutt}(1998{\natexlab{a}})}]{Kennicutt_98}
{Kennicutt}, Jr., R.~C. 1998{\natexlab{a}}, \araa, 36, 189

\bibitem[{{Kennicutt}(1998{\natexlab{b}})}]{Kennicutt}
{Kennicutt}, Jr., R.~C. 1998{\natexlab{b}}, \apj, 498, 541

\bibitem[{{Kennicutt} {et~al.}(2007){Kennicutt}, {Calzetti}, {Walter}, {Helou},
  {Hollenbach}, {Armus}, {Bendo}, {Dale}, {Draine}, {Engelbracht}, {Gordon},
  {Prescott}, {Regan}, {Thornley}, {Bot}, {Brinks}, {de Blok}, {de Mello},
  {Meyer}, {Moustakas}, {Murphy}, {Sheth}, \& {Smith}}]{Kennicutt_07}
{Kennicutt}, Jr., R.~C., {Calzetti}, D., {Walter}, F., {et~al.} 2007, \apj,
  671, 333

\bibitem[{{Kr\"ugel}(2003)}]{krugel}
{Kr\"ugel}, E. 2003, {The physics of interstellar dust} (The physics of
  interstellar dust, by Endrik Kr\"ugel.~IoP Series in astronomy and
  astrophysics, ISBN 0750308613.~Bristol, UK: The Institute of Physics, 2003.)

\bibitem[{{Kr{\"u}gel}(2009)}]{Kruegel_09}
{Kr{\"u}gel}, E. 2009, \aap, 493, 385

\bibitem[{{Leroy} {et~al.}(2008){Leroy}, {Walter}, {Brinks}, {Bigiel}, {de
  Blok}, {Madore}, \& {Thornley}}]{Leroy}
{Leroy}, A.~K., {Walter}, F., {Brinks}, E., {et~al.} 2008, \aj, 136, 2782

\bibitem[{{Magnier} {et~al.}(1997){Magnier}, {Hodge}, {Battinelli}, {Lewin}, \&
  {van Paradijs}}]{Magnier}
{Magnier}, E.~A., {Hodge}, P., {Battinelli}, P., {Lewin}, W.~H.~G., \& {van
  Paradijs}, J. 1997, \mnras, 292, 490

\bibitem[{{Montalto} {et~al.}(2009){Montalto}, {Seitz}, {Riffeser}, {Hopp},
  {Lee}, \& {Sch{\"o}nrich}}]{Montalto}
{Montalto}, M., {Seitz}, S., {Riffeser}, A., {et~al.} 2009, \aap, 507, 283

\bibitem[{{Nakai} \& {Sofue}(1982)}]{Nakai_82}
{Nakai}, N. \& {Sofue}, Y. 1982, \pasj, 34, 199

\bibitem[{{Nakai} \& {Sofue}(1984)}]{Nakai_84}
{Nakai}, N. \& {Sofue}, Y. 1984, \pasj, 36, 313

\bibitem[{{Nedialkov} {et~al.}(2000){Nedialkov}, {Berkhuijsen}, {Nieten}, \&
  {Haas}}]{Nedialkov}
{Nedialkov}, P., {Berkhuijsen}, E.~M., {Nieten}, C., \& {Haas}, M. 2000, in
  Proceedings 232. WE-Heraeus Seminar, ed. E.~M. {Berkhuijsen}, R.~{Beck}, \&
  R.~A.~M. {Walterbos}, 85--88

\bibitem[{{Neininger} {et~al.}(1998){Neininger}, {Gu{\'e}lin}, {Ungerechts},
  {Lucas}, \& {Wielebinski}}]{Neininger_98}
{Neininger}, N., {Gu{\'e}lin}, M., {Ungerechts}, H., {Lucas}, R., \&
  {Wielebinski}, R. 1998, \nat, 395, 871

\bibitem[{{Nieten} {et~al.}(2006){Nieten}, {Neininger}, {Gu{\'e}lin},
  {Ungerechts}, {Lucas}, {Berkhuijsen}, {Beck}, \& {Wielebinski}}]{Nieten}
{Nieten}, C., {Neininger}, N., {Gu{\'e}lin}, M., {et~al.} 2006, \aap, 453, 459

\bibitem[{{Pagel} \& {Edmunds}(1981)}]{Pagel_81}
{Pagel}, B.~E.~J. \& {Edmunds}, M.~G. 1981, \araa, 19, 77

\bibitem[{{Pagel} {et~al.}(1979){Pagel}, {Edmunds}, {Blackwell}, {Chun}, \&
  {Smith}}]{Pagel}
{Pagel}, B.~E.~J., {Edmunds}, M.~G., {Blackwell}, D.~E., {Chun}, M.~S., \&
  {Smith}, G. 1979, \mnras, 189, 95

\bibitem[{{P{\'e}rez-Gonz{\'a}lez} {et~al.}(2006){P{\'e}rez-Gonz{\'a}lez},
  {Kennicutt}, {Gordon}, {Misselt}, {Gil de Paz}, {Engelbracht}, {Rieke},
  {Bendo}, {Bianchi}, {Boissier}, {Calzetti}, {Dale}, {Draine}, {Jarrett},
  {Hollenbach}, \& {Prescott}}]{Perez}
{P{\'e}rez-Gonz{\'a}lez}, P.~G., {Kennicutt}, Jr., R.~C., {Gordon}, K.~D.,
  {et~al.} 2006, \apj, 648, 987

\bibitem[{{Reach} \& {Boulanger}(1998)}]{Reach}
{Reach}, W.~T. \& {Boulanger}, F. 1998, in Lecture Notes in Physics, Berlin
  Springer Verlag, Vol. 506, IAU Colloq. 166: The Local Bubble and Beyond, ed.
  {D.~Breitschwerdt, M.~J.~Freyberg, \& J.~Truemper}, 353

\bibitem[{{Rela{\~n}o} {et~al.}(2007){Rela{\~n}o}, {Lisenfeld},
  {P{\'e}rez-Gonz{\'a}lez}, {V{\'{\i}}lchez}, \& {Battaner}}]{Relano_07}
{Rela{\~n}o}, M., {Lisenfeld}, U., {P{\'e}rez-Gonz{\'a}lez}, P.~G.,
  {V{\'{\i}}lchez}, J.~M., \& {Battaner}, E. 2007, \apjl, 667, L141

\bibitem[{{Rieke} {et~al.}(2004){Rieke}, {Young}, {Engelbracht}, {Kelly},
  {Low}, {Haller}, {Beeman}, {Gordon}, {Stansberry}, {Misselt}, {Cadien},
  {Morrison}, {Rivlis}, {Latter}, {Noriega-Crespo}, {Padgett}, {Stapelfeldt},
  {Hines}, {Egami}, {Muzerolle}, {Alonso-Herrero}, {Blaylock}, {Dole}, {Hinz},
  {Le Floc'h}, {Papovich}, {P{\'e}rez-Gonz{\'a}lez}, {Smith}, {Su}, {Bennett},
  {Frayer}, {Henderson}, {Lu}, {Masci}, {Pesenson}, {Rebull}, {Rho}, {Keene},
  {Stolovy}, {Wachter}, {Wheaton}, {Werner}, \& {Richards}}]{Rieke}
{Rieke}, G.~H., {Young}, E.~T., {Engelbracht}, C.~W., {et~al.} 2004, \apjs,
  154, 25

\bibitem[{{Savage} {et~al.}(1978){Savage}, {Wesselius}, {Swings}, \&
  {The}}]{Savage_78}
{Savage}, B.~D., {Wesselius}, P.~R., {Swings}, J.~P., \& {The}, P.~S. 1978,
  \apj, 224, 149

\bibitem[{{Savcheva} \& {Tassev}(2002)}]{Savcheva}
{Savcheva}, A.~S. \& {Tassev}, S.~V. 2002, Publications de l'Observatoire
  Astronomique de Beograd, 73, 219

\bibitem[{{Schmidt}(1959)}]{Schmidt}
{Schmidt}, M. 1959, \apj, 129, 243

\bibitem[{{Soifer} {et~al.}(1986){Soifer}, {Rice}, {Mould}, {Gillett},
  {Rowan-Robinson}, \& {Habing}}]{Soifer}
{Soifer}, B.~T., {Rice}, W.~L., {Mould}, J.~R., {et~al.} 1986, \apj, 304, 651

\bibitem[{{Stanek} \& {Garnavich}(1998)}]{Stanek}
{Stanek}, K.~Z. \& {Garnavich}, P.~M. 1998, \apjl, 503, L131

\bibitem[{{Tabatabaei} {et~al.}(2007{\natexlab{a}}){Tabatabaei}, {Beck},
  {Krause}, {Berkhuijsen}, {Gehrz}, {Gordon}, {Hinz}, {Humphreys}, {McQuinn},
  {Polomski}, {Rieke}, \& {Woodward}}]{Tabatabaei_1_07}
{Tabatabaei}, F.~S., {Beck}, R., {Krause}, M., {et~al.} 2007{\natexlab{a}},
  \aap, 466, 509

\bibitem[{{Tabatabaei} {et~al.}(2007{\natexlab{b}}){Tabatabaei}, {Beck},
  {Kr{\"u}gel}, {Krause}, {Berkhuijsen}, {Gordon}, \&
  {Menten}}]{Tabatabaei_3_07}
{Tabatabaei}, F.~S., {Beck}, R., {Kr{\"u}gel}, E., {et~al.} 2007{\natexlab{b}},
  \aap, 475, 133

\bibitem[{{Tempel} {et~al.}(2010){Tempel}, {Tamm}, \& {Tenjes}}]{Tempel}
{Tempel}, E., {Tamm}, A., \& {Tenjes}, P. 2010, \aap, 509, 91

\bibitem[{{Tenjes} \& {Haud}(1991)}]{Tenjes}
{Tenjes}, P. \& {Haud}, U. 1991, \aap, 251, 11

\bibitem[{{Thilker} {et~al.}(2005){Thilker}, {Hoopes}, {Bianchi}, {Boissier},
  {Rich}, {Seibert}, {Friedman}, {Rey}, {Buat}, {Barlow}, {Byun}, {Donas},
  {Forster}, {Heckman}, {Jelinsky}, {Lee}, {Madore}, {Malina}, {Martin},
  {Milliard}, {Morrissey}, {Neff}, {Schiminovich}, {Siegmund}, {Small},
  {Szalay}, {Welsh}, \& {Wyder}}]{Thilker_05}
{Thilker}, D.~A., {Hoopes}, C.~G., {Bianchi}, L., {et~al.} 2005, \apjl, 619,
  L67

\bibitem[{{Trundle} {et~al.}(2002){Trundle}, {Dufton}, {Lennon}, {Smartt}, \&
  {Urbaneja}}]{Trundle}
{Trundle}, C., {Dufton}, P.~L., {Lennon}, D.~J., {Smartt}, S.~J., \&
  {Urbaneja}, M.~A. 2002, \aap, 395, 519

\bibitem[{{Unwin}(1980)}]{Unwin}
{Unwin}, S.~C. 1980, \mnras, 192, 243

\bibitem[{{van Genderen}(1973)}]{Genderen}
{van Genderen}, A.~M. 1973, \aap, 24, 47

\bibitem[{{Verley} {et~al.}(2009){Verley}, {Corbelli}, {Giovanardi}, \&
  {Hunt}}]{Verley_09}
{Verley}, S., {Corbelli}, E., {Giovanardi}, C., \& {Hunt}, L.~K. 2009, \aap,
  493, 453

\bibitem[{{Viallefond} {et~al.}(1982){Viallefond}, {Goss}, \&
  {Allen}}]{Viallefond_82}
{Viallefond}, F., {Goss}, W.~M., \& {Allen}, R.~J. 1982, \aap, 115, 373

\bibitem[{{Walterbos} \& {Braun}(1994)}]{Walterbos_94}
{Walterbos}, R.~A.~M. \& {Braun}, R. 1994, \apj, 431, 156

\bibitem[{{Walterbos} \& {Kennicutt}(1988)}]{Walterbos_88}
{Walterbos}, R.~A.~M. \& {Kennicutt}, Jr., R.~C. 1988, \aap, 198, 61

\bibitem[{{Walterbos} \& {Schwering}(1987)}]{Walterbos_87}
{Walterbos}, R.~A.~M. \& {Schwering}, P.~B.~W. 1987, \aap, 180, 27

\bibitem[{{Williams}(2003)}]{Williams}
{Williams}, B.~F. 2003, \aj, 126, 1312

\bibitem[{{Witt} \& {Gordon}(2000)}]{Witt}
{Witt}, A.~N. \& {Gordon}, K.~D. 2000, \apj, 528, 799

\bibitem[{{Xu} \& {Helou}(1996)}]{Xu_96}
{Xu}, C. \& {Helou}, G. 1996, \apj, 456, 163

\bibitem[{{Yin} {et~al.}(2009){Yin}, {Hou}, {Prantzos}, {Boissier}, {Chang},
  {Shen}, \& {Zhang}}]{Yin}
{Yin}, J., {Hou}, J.~L., {Prantzos}, N., {et~al.} 2009, \aap, 505, 497

\end{thebibliography}

\end{document}